\documentclass[final,onefignum,onetabnum]{siamart190516}

\ifpdf
\hypersetup{
  pdftitle={FLUPS: a Fourier-based Library of Unbounded Poisson Solvers},
  pdfauthor={Caprace, Gillis, Chatelain}
}
\fi

\usepackage[utf8]{inputenc}
\usepackage[english]{babel}
\usepackage{amsmath}
\usepackage{amsfonts}
\usepackage{amssymb}
\usepackage{amsopn}

\usepackage{fourier}
\usepackage[left=1.4cm,right=1.4cm,top=1.5cm,bottom=1.5cm]{geometry}

\usepackage{lineno}
\usepackage{hyperref}
\usepackage{setspace}
\usepackage{xcolor}
\usepackage{wrapfig}
\usepackage{graphicx}
\usepackage{pstool}
\usepackage{multirow}
\usepackage{tikz}% pour diagrammes
\usetikzlibrary{shapes.misc}
\usetikzlibrary{shapes.geometric}
\usepackage{setspace}
\usepackage{enumitem}

\usepackage{pgfplots}
\pgfplotsset{compat=newest}
\usepgfplotslibrary{groupplots}
\usepgfplotslibrary{dateplot}

\usepackage{subcaption}

\usepackage[numbers,sort&compress]{natbib}

\def\abs#1{\vert #1 \vert} % abs
\def\sabs#1{\vert #1 \vert} % small abs
\def\cC{\mathcal{C}}

\def\exp#1{\operatorname{exp} \left( #1 \right)} % exp

\def\erf#1{\operatorname{erf}\left(#1\right)}

\def\bf{\mathbf{f}}

\def\ic{\mathrm{i}}

\def\it#1{\textit{#1}} % italic
\def\iti{(\textsc{i})~}
\def\itii{(\textsc{ii})~}

\def\itiv{(\textsc{iv})~}

\def\bk{\mathbf{k}}

\def\bm{\mathbf{m}}

\def\O#1{\mathcal{O}\left( #1 \right)}
\def\bx{\mathbf{x}}

\def\bxi{\boldsymbol{\xi}}

\def\lapl{\nabla^{2}}

\newcommand{\be}{\begin{equation}}
\newcommand{\eec}{ \;\;,\end{equation}}
\newcommand{\ee}{ \end{equation}}
\newcommand{\eed}{\;\;.\end{equation}}
\newcommand{\bse}{\begin{subequations}}
\newcommand{\ese}{\end{subequations}}

\newcommand{\sect}{Section~\ref}

\newcommand{\app}{Appendix~\ref}

\newcommand{\fig}{\cref}

\newcommand{\eqqref}{\cref}

\newcommand{\eq}{\cref}
\newcommand{\fref}{\cref}

\definecolor{darkGreen}{RGB}{0,160,0}
\definecolor{darkBlue}{RGB}{51,51,204}
\definecolor{darkRed}{RGB}{160,0,0}
\definecolor{orange}{RGB}{255,120,0}
\definecolor{myGreen}{rgb}{0.0706, 0.5333, 0.1921}

\newcommand{\revthree}[1]{#1}
\newcommand{\revtwo}[1]{#1}
\newcommand{\revall}[1]{#1}
\newcommand{\typo}[1]{#1}

\definecolor{matlabBlue}{rgb}{0, 0.4470, 0.7410}
\definecolor{matlabOrange}{rgb}{0.8500, 0.3250, 0.0980}
\definecolor{matlabYellow}{rgb}{0.9290, 0.6940, 0.1250}
\definecolor{matlabPurple}{rgb}{0.4940, 0.1840, 0.5560}
\definecolor{matlabGreen}{rgb}{0.4660, 0.6740, 0.1880}
\definecolor{matlabRed}{rgb} {0.7765, 0.1569, 0.0902}
\definecolor{matlabBlack}{rgb}{0, 0, 0}

\definecolor{keyBlue}{rgb}{0.0392, 0.2274, 0.4235}
\definecolor{keyBlueTrans}{rgb}{0.4274, 0.4980, 0.5764}
\definecolor{keyGreen}{rgb}{0.05490, 0.38039, 0.003921}
\definecolor{keyGreenTrans}{rgb}{0.7019, 0.8313, 0.7176}
\definecolor{keyOrange}{rgb}{0.98039, 0.4980, 0.03529}
\definecolor{keyGray}{rgb}{0.50196, 0.50196, 0.50196}

\definecolor{pyBlue}{rgb}{0.11372,0.42352,0.67059}
\definecolor{pyBlueLight}{rgb}{0.6196,0.79215,0.88235}
\definecolor{pyOrange}{rgb}{1.0,0.4549,0.0627}
\definecolor{pyOrangeLight}{rgb}{0.9921568,0.643137,0.376470}
\definecolor{pyGreen}{rgb}{0.1529,0.5882,0.1529}
\definecolor{pyRed}{rgb}{0.81569,0.13725,0.14117}
\definecolor{pyPurple}{rgb}{0.53725,0.36078,0.7098}
\definecolor{myCacaDoie}{rgb}{0.580392156862745,0.43,0.27}
\definecolor{pyBrown}{rgb}{0.549019607843137,0.337254901960784,0.294117647058824}
\definecolor{pyRose}{rgb}{0.890196078431372,0.466666666666667,0.76078431372549}

\definecolor{pyOrange2}{rgb}{0.9921,0.5098,0.2078}
\definecolor{pyBlue2}{rgb}{0.3765,0.6431,0.8157}

\definecolor{pyBar1}{rgb}{0.1647,0.2745,0.2745}
\definecolor{pyBar2}{rgb}{0.0000,0.4588,0.4588}
\definecolor{pyBar3}{rgb}{0.4863,0.7804,0.9998}
\definecolor{pyBar4}{rgb}{0.4549,0.4549,0.4549}

\definecolor{keyGreen2}{rgb}{0.0627 , 0.4196 , 0.3882}%{16,107,99}
\definecolor{keyOrange2}{rgb}{0.9843 , 0.2941 , 0.2431}%{255,99,78}

\newcommand{\mymyLine}[3]{\raisebox{0.01pt}{\tikz{\draw[-,#1,#2,line width = #3 pt](0,0) -- (10pt,0);}}}

\newcommand{\lineSquare}[3]{
        \tikz{
            \draw[-,#1,solid,line width = #3 pt](0,0) -- (12.5pt,0);
            \filldraw[draw=#1,fill=#2, line width=#3 pt] (4.5pt,-2pt) rectangle (8.5pt,2pt);
        }
}

\newcommand{\ThinLineSquare}[2]{\lineSquare{#1}{#2}{0.5}}            
\newcommand{\ThickLineSquare}[2]{\lineSquare{#1}{#2}{1.0}}

\newcommand{\lineCircle}[3]{
        \tikz{
            \draw[-,#1,solid,line width = #3 pt](0,0) -- (13pt,0);
            \filldraw[draw=#1,fill=#2, line width=#3 pt] (6.5pt,0pt) circle (2pt);
        }
}
\newcommand{\ThinLineCircle}[2]{\lineCircle{#1}{#2}{0.5}}            
\newcommand{\ThickLineCircle}[2]{\lineCircle{#1}{#2}{1.0}}

\newcommand{\lineDiamondD}[3]{
        \tikz{
            \draw[-,#1,solid,line width = #3 pt](0,0)  -- (6pt,0) node[draw=#1, fill=#2, shape=diamond, line width=#3, inner sep=1.5pt]{} -- (13pt,0);
        }
}
\newcommand{\ThinLineDiamondD}[2]{\lineDiamondD{#1}{#2}{0.5}}

\tikzset{cross/.style={cross out, draw, minimum size=2*(#1-\pgflinewidth), inner sep=0pt, outer sep=0pt}}

\newcommand{\lineCross}[2]{
        \tikz{
            \draw[-,#1,solid,line width = #2 pt](0,0) -- (5pt,0) node[cross=3pt, line width=#2 pt,#1,inner sep=1.5pt]{} -- (10pt,0);
        }
}
\newcommand{\ThinLineCross}[1]{\lineCross{#1}{0.75}}

\newcommand{\lineTrianUp}[3]{
        \tikz{
            \draw[-,#1,solid,line width = #3 pt](0,0) -- (13pt,0);
	        \filldraw[draw=#1,fill=none, line width=#3 pt] (6.5pt,-.8pt) node[draw=#1, fill=#2, shape=regular polygon,regular polygon sides=3, line width=#3, inner sep=1.pt]{}  (6.5pt,1.2pt);
        }
}
\newcommand{\ThinLineTrianUp}[2]{\lineTrianUp{#1}{#2}{0.5}}

\graphicspath{{./figures/}}

\newsiamremark{remark}{Remark}
\newsiamremark{hypothesis}{Hypothesis}
\crefname{hypothesis}{Hypothesis}{Hypotheses}
\newsiamthm{claim}{Claim}

\headers{FLUPS}{Caprace, Gillis, Chatelain}

\title{FLUPS: a Fourier-based Library of Unbounded Poisson Solvers
\thanks{Submitted to the editors December 5, 2019. Corresponding authors: \email{denis-gabriel.caprace@uclouvain.be}, \email{thomas.gillis@uclouvain.be}. 
\funding{Denis-Gabriel Caprace, Thomas Gillis, Philippe Chatelain, 2019.}
}}

\author{Denis-Gabriel Caprace
\footnotemark[3] \thanks{These authors contributed equally to the work.} 
\and Thomas Gillis{\footnotemark[3]} {\footnotemark[2]}
\and Philippe Chatelain  
\thanks{Institute of Mechanics, Materials and Civil Engineering, Universit\'e catholique de Louvain, 1348 Louvain-la-Neuve, Belgium.}
}

\begin{document}

\maketitle

%% formating title

%\author[ucl]{Denis-Gabriel Caprace\corref{cor1}}
%\author[ucl]{Thomas Gillis\corref{cor1}}
%\author[ucl]{Philippe Chatelain}
%\cortext[cor1]{Corresponding authors\\Email address: \url{denis-gabriel.caprace@uclouvain.be} (Denis-Gabriel Caprace) and \url{thomas.gillis@uclouvain.be} (Thomas Gillis)}
%
%\address[ucl]{Institute of Mechanics, Materials and Civil Engineering, Université catholique de Louvain\\1348 Louvain-la-Neuve, Belgium}
%%\address[cse]{Computational Science and Engineering Laboratory, ETH Zürich\\CH-8092, Switzerland}

\begin{abstract}
% Fourier-based Library of Unbounded Poisson Solvers
%old
% A collection of distributed Poisson solvers for 2D and 3D homogeneous grids is presented, all gathered in a single, unique software. The Fourier-based Library of Unbounded Poisson Solvers  (\textit{FLUPS}) is designed to handle every possible combination of periodic, symmetric, half-unbounded and fully unbounded boundary conditions for the Poisson equation.
%new
A Fourier-based Library of Unbounded Poisson Solvers (\textit{FLUPS}) for 2D and 3D homogeneous distributed grids is presented. It is designed to handle every possible combination of periodic, symmetric, semi-unbounded and fully unbounded boundary conditions for the Poisson equation \revall{on rectangular domains with uniform resolution}.
FLUPS leverages a dedicated implementation of 3D Fourier transforms to solve the Poisson equation using Green's functions, in a fast and memory-efficient way. %\\
Several Green's functions are available, optionally with explicit regularization, \revthree{spectral truncation}, or using lattice Green\typo{'s} functions, and provide verified convergence orders from 2 to \revthree{spectral-like}.
The algorithm depends on the FFTW library to perform 1D transforms, while Message Passing Interface (MPI) communications enable the required remapping of data in memory. For the latter operation, a first available implementation resorts to the standard \textit{all-to-all} routines. A second implementation, featuring \textit{non-blocking} and persistent point-to-point communications, is however shown to be more efficient in a majority of cases and especially while taking advantage of the shared memory parallelism with OpenMP.
The scalability of the algorithm, aimed at massively parallel architectures, is demonstrated up to $73\,720$ cores. The results obtained with three different supercomputers show that the weak efficiency remains above $40\%$ and the strong efficiency above $30\%$ when the number of cores is multiplied by 16, for typical problems. These figures are slightly better than those expected from a third party 3D Fast Fourier Transform (FFT) tool, with which a 20\% longer execution time was also measured on average.
From the outside, the solving procedure is fully automated so that the user benefits from the optimal performances while not having to handle the complexity associated with memory management, data mapping and Fourier transform computation. \\
The parallel code is available under \typo{Apache license 2.0} at \url{github.com/vortexlab-uclouvain/flups}.
\end{abstract}



%\input{MR_IIM_Poisson_1_intro}
%!TEX root = mixPoissonSolver_main_siam.tex
%!TEX encoding = UTF-8 Unicode

\section{Introduction}%
\label{sec_intro}

%------------------------------------------------------------
% What is the problem
%-The Poisson equation is widespread among the computational physics domain and it has been subject a very broad study over the years.
The solution of %3D
Poisson problems is ubiquitous in computational physics as it concerns problems ranging from fluid dynamics and electromagnetism to particle physics. Consequently, there is a widespread need for dedicated, adapted  numerical solvers. When considering time-dependent applications such as fluid dynamics for example, this equation has to be solved many times per simulation, and the efficiency of solvers is hence also crucial.
%To date, numerous implementations exist to tackle (part of) the problem. Yet, none of them offer an integrated solution which combines the desired properties of versatility, performance, scalability, and portability.
% In this work, we consider the Poisson equation in a 3D parallelepiped-like domain with uniform resolution,
\revall{
In this work, we consider the Poisson equation in a 3D computational domain,}
\be
\nabla^2 \phi = f
\label{eq:intro:Poisson}
\eed
Often such problems are unbounded in nature, as they describe a potential field induced by sources located in a compact region of space. Nevertheless \revtwo{it is also common to exploit symmetries that may arise from the physical configuration of the domain, or periodicity, for the sake of modeling. In all cases, the geometry of the problem naturally drives the conditions that $f$ and $\phi$ need to satisfy at the computational domain boundaries.}
% Dirichlet and Neumann boundary conditions (BCs) may also appear from physical bounds to the domain . 
Therefore, \eqqref{eq:intro:Poisson} has to be complemented with a set of \revtwo{appropriate} boundary conditions (BCs) which, in each direction, specify the periodicity or an arbitrary combination of symmetries and (semi-)infinite direction.

%------------------------------------------------------------
% What has been done
\revall{Historically, three main families of approaches have emerged to efficiently solve \eqref{eq:intro:Poisson}: the Multigrid-based (MG) approaches
\cite{Trottenberg:2000}, the Fast Multipole Methods (FMM) \cite{Greengard:1997} 
and the Fast Fourier Transform (FFT)-based techniques \cite{Hockney:1965}.
% %,Chatelain:2010}
%
%domain decomposition and wavelet-based solver
%In the following, we address very briefly the main differences between the methods and 
In addition to the brief summary that we provide hereafter, we refer the reader to \cite{Gholami:2016} and references therein for a detailed description and comparison of these approaches.
In the MG approaches, \eqref{eq:intro:Poisson} is treated as a linear system usually emerging from a finite-difference or a finite-volume discretization, where several levels of resolution are recursively used to converge towards the solution. % improve the precision of the solution while limiting the cost of the associated iterative solver.
%Starting from a given grid level, the simplest implementation, the Geometric Multigrid, relies on a simple iterative solver (\textit{i.e.} a Gauss-Seidel or a Jacobi approach) to obtain an approximation of the solution. Coarser grids are recursively used to improve this approximation and decrease the cost of the iterative process. The generalization of this technique to more general matrix structures is denoted Algebraic Multigrid (AMG) and commonly used as a preconditioner for  the resolution of complex linear systems \cite{Trottenberg:2000}.
In the FMM and the FFT-based techniques,}
Green's functions are  employed to obtain the solution of \eqref{eq:intro:Poisson}, $\phi= G*f$, \typo{as the convolution of the right-hand side (RHS), $f$, with a precomputed  kernel, $G$}. The FMM treats near and far interactions separately, whereas the FFT framework does not require the distinction.
\revall{%Each of these three families presents peculiarities, 
Because of their specificities,  the choice of the most adapted approach for a given problem is mostly driven by three criteria: (i) the computational complexity; (ii) the characteristics of the source term, $f$, and its support; and (iii) the desired BCs.
}

% While considering unbounded/periodic problems, Green's functions are generally employed to obtain the solution of \eqqref{eq:intro:Poisson}, $\phi= G*f$, at the cost of a convolution between the right-hand-side, $f$, and a precomputed  kernel, $G$. 
% Two classes of methods exist, depending on how this convolution is computed: the Fast Multipole Method-based techniques (FMM) and the Fast Fourier Transform-based approaches (FFT). 
%
\revall{In terms of computational complexity,} for a domain with $N_p$ sources, tree-based algorithms (like the FMM) theoretically reduce the computational cost from $\mathcal{O}(N_p^2)$ (direct interactions) down to $\mathcal{O}(N_p)$ \cite{Gillman:2014}.  %However, the number of operations involved in the preparation of the tree and in the management of the data structure can substantially decrease this theoretical advantage.
\revall{Similarly, when the sources are uniformly distributed on a grid of size $N^3$, the multigrid approach %, as many other iterative solvers 
reaches a linear complexity, $\mathcal{O}\left(N^3 \right)$ (in 3D), where $N$ is the number of unknowns in each direction.}
However, \revall{in those conditions,} higher computational performances can be expected from an FFT-based algorithm which allows one to cast the convolution into a pointwise multiplication in the spectral space,
\be
\hat{\phi} = \hat{G} \hat{f}
\label{eq:Green_conv}
\eec
where $\hat{.}$ denotes the 3D Fourier transform.
\typo{The required forward and backward FFTs thus entail fast executions, in spite of their
complexity of $\mathcal{O}\left(N^3 \log(N)\right)$ operations per direction
}. %, where $N$ is the number of unknowns in each direction. 
\revall{Moreover, the three approaches can be compared in terms of their relative computational intensity \cite{Ibeid:2020}. While the multigrid methods present the lowest intensity, the FMM-based approaches are the most intense methods, as the computation of the multipole expansions requires many operations for fewer memory accesses.}
% \todo{need to talk about the complexity of MG then}.

%
%Yet, in practice, it often outperforms the tree algorithms in which the preparation and the management of the data structures involves heavy computations \typo{\cite{Dorschner:2020}} and only few can be precomputed.

%FFT-based techniques, which further requires the use of a uniform grid, have been repeatedly implemented and applied to Poisson's problem in the past decade \cite{Costa:2018,Qiang:2017,Anderson:2016,Fuka:2015}.
%legacy fishpack?
%Zheng:2016 (pourri)
%Jiang:2013 (pourri)
%Burdiardja:2011 (pourri, mauvais G000)

\revall{
Most importantly, one must also consider the adequacy of the method for the source term and the geometry of its support.
%If the source term encompasses a very large range of spatial scales, a multi-level computational approach is of interest and the MG approach is the commonly admitted choice. 
When $f$ is made of a distributed collection of compact sources, the FMM is most suited. On the other hand, FFT-based techniques, which require the use of a uniform grid, are best-in-class to handle evenly distributed and compactly supported source terms.}

\revall{
An additional advantage of the FMM over the others is that it naturally enables unbounded solutions.} 
The spectral decomposition inherent to the Fourier transforms implies assumptions on the boundary conditions: the \typo{discreet} Fourier transform (DFT) imposes periodic conditions, \typo{whereas} all the combinations of odd and even symmetries are \typo{accessible through} the various types of cosine and sine transforms (DCT and DST).
Still, fully unbounded directions can also be achieved in an FFT-based framework using zero-padding, as described in the domain doubling technique proposed by
Hockney and Eastwood \cite{Hockney:1988}.

\revall{
In the present work, we focus on the case of source terms compactly and homogeneously supported within a parallelepiped-like domain (see \cite{Fuka:2015,Anderson:2016,Qiang:2017,Costa:2018} for application examples). In that specific case, a uniform Cartesian mesh is well adapted} 
\revtwo{
%. While nothing prevents the use of MG or FMM-based methods, 
and the FFT-based methods were reported to exhibit the shortest time-to-solution \cite{Gholami:2016}.} \revthree{We here aim to present the efficient implementation of an FFT-based method, within an inclusive computational framework supporting various types of BCs.} 

\medskip

The use of Green's functions in a discrete physical domain implies a second classification \typo{among} the potential methods. In order to obtain the Green's function value on the grid, a first approach consists in sampling the Green's function corresponding to the continuous Poisson problem. % solution. %This function of course depends on the domain boundary conditions.
In that case, a specific treatment is required to handle the singularity of the Green\typo{'s} functions (which can thus not be evaluated in one or several locations). 
In this respect, \revthree{three} major techniques have been proposed:
\begin{enumerate}[label=\Roman*.]
    \item the replacement of the singular values with a cell-averaged equivalent \cite{Chatelain:2010};
    \item  the use of an explicit (high-order\typo{, spectral in the limit}) regularization of the Green\typo{'s} function which thus solves the regularized Poisson problem \cite{Hejlesen:2013,Spietz:2018,Hejlesen:2019};
    \item \revthree{the truncation of the Green\typo{'s} function in the physical space to a sphere of radius $R$ encompassing the source term. This truncated kernel method \cite{Vico:2016} leads to an analytical kernel which yields spectral-accurate results for isotropic problems. However, as detailed in \cite{Greengard:2018}, its generalization to anisotropic domains comes at the price of oversampling in the Fourier space, hence leading to a significant overhead, or the loss of the analytical expression.
    }
\end{enumerate}
%,Spietz:2018, Vico:2016
%
% A second approach consists in determining the fundamental solution of the discrete version of \eqqref{eq:intro:Poisson}, obtained using finite differences. This approach of so-called Lattice Green's Functions (LGF) has been initially proposed in 2D \cite{Martinsson:2002a,Gillman:2010} and further applied in 3D \cite{Liska:2014}; %The Lattice Green Functions (LGF) is hence obtained as a Green kernel for the unbounded Poisson equation, but  the corresponding expression requires 
% the derivation of the corresponding kernels involves the evaluation of an integral in infinite domain.
%
\revall{
A second approach consists in determining the fundamental solution of the discrete version of \eqqref{eq:intro:Poisson}, obtained using finite differences. The corresponding Lattice Green's Functions (LGFs, hereafter denoted type \itiv) were initially proposed in 2D \cite{Martinsson:2002a,Gillman:2010} and further applied in 3D \cite{Liska:2014}. %As no analytical expression is readily available,
The kernel computation involves the evaluation of an integral in infinite domain for close interactions, and requires the use of an expansion for further interactions \cite{Martinsson:2002a}. This splitting renders the LGF also naturally well suited for an implementation of the FMM approach.
%
%Finally, we would like to highlight that it is common to combine the FMM approach with FFT-based evaluation of the near-interactions \cite{Malhotra:2015,Dorschner:2020}.
%REV1: the LGF is really a bridge toward retaining the efficient point distribution of FMM but enabling it to be computed faster with FFTs (but not with their library)
In that respect, let us highlight that the computation of the interactions can be accelerated through the use of FFTs, as proposed in \cite{Liska:2014}, hence providing a hybrid implementation of the FMM, optionally also with multiresolution \cite{Dorschner:2020}.
}

%This treatment implies a second cle.g by replacing the singular values with cell-averaged values.
%Alternatively, a regularisation of the kernels can be introduced, as explained in \cite{Winckelmans:2004}, and further developed in \cite{Hejlesen:2013,Spietz:2018}. 
%While the first method was shown to exhibit a second order of convergence, the later can be used to obtain a method with an arbitrary order. In the limit, spectral convergence can even be reached, as shown in \cite{Vico:2016,Hejlesen:2019}, at the cost of evaluation of a Bessel integral function.

\medskip

Solving the Poisson equation \typo{is a} computationally intensive task and many \typo{previous} efforts have focused on parallel performance optimization,  a challenge due to the inherent elliptic character of the problem.
Because the distribution of the computation entails a subdivision of the domain, communications are needed between all the computational entities\typo{, which drive the overall computational performance of the algorithm.}
% These communication will drive the 
While FFT-based algorithms may rely on highly optimized (mostly serial) libraries, such as FFTW \cite{Frigo:2005}, %
their parallel 3D implementation for distributed architectures requires handling dedicated communication patterns (several of them are reviewed in \cite{Sunderland:2012}), which are subject to specific optimization.
 Among the numerous initiatives tackling 3D FFT computations, the renowned %I don't like ``very famous''
 P3DFFT library \cite{Pekurovsky:2012} takes advantage of the message passing interface (MPI) and the pencil (1D) decomposition of the computational domain. 
As highlighted \typo{in} \cite{Sbalzarini:2006a}, the 1D pencil decomposition indeed %exhibits a better scalability 
waives some partition-size limitations on massively parallel architectures.
A pencil-based transform is able to exploit $N^2$ cores at most (for a cubic domain of $N^3$ unknowns), but requires the transposition of data, which can be efficiently implemented as in \cite{Bowman:2015}. 
By contrast, a transform that relies on a 2D slab decomposition would be limited to the use of $N$ cores. 
Moreover, when used for 3D transforms, the slab decomposition features one actual \textit{all-to-all} communication as the slabs need to be transformed into pencils in the third direction: one MPI rank will thus communicate with all the $N_{rank}$ other ranks.
With the pencil decomposition, the communication can be arranged such that one rank only communicates twice with $\sqrt{N_{rank}}$ other ranks, which will reduce the communication latency on very large partitions. 
%As opposed to the decomposition of the domain in 2D slab, the pencil decomposition indeed enables the use of a large number of cores for the computation of FFTs, direction per direction, 
%
The same ideas were exploited in other packages such as PFFT \cite{pippig:2013}, FFTK \cite{Chatterjee:2018} and MPI4PYFFT \cite{Dalcin:2019} (see also references therein), with small variations and improvements on the parallel implementation. 
%CAUTION: DALCIN uses  MPI_Alltoallw with datatypes, and says that he is faster because he doesnt need to copy to buffer himself... well ok but I presume out non-blocking implementation beats him.
For the above-mentioned solutions, the scalability of the algorithm was tested on \textit{Bluegene} and \textit{Cray XT/XC} architectures, for some up to 64k cores and more, often showing similar performances.
Concurrently, hardware acceleration applied to Poisson's problem has also been considered \cite{Jodra:2017}.
%threaded: \cite{Mininni:2011}

%\todo{speak about the slabs?}
%The pencil decomposition enables the use of a large number of cores for the computation of FFTs, direction per direction, but requires the transposition of data. 
%The optimal treatment of the latter operation is problem-dependent, and the best performances will rely on a trade-off between the use of shared memory parallelism and inter-process communication  \cite{Bowman:2015}.

% What is missing
Although the above-mentioned efforts have brought several substantial advances to the solving methodology and to its numerical implementation (specifically related to 3D FFTs), 
an open-source solution dedicated to solving Poisson's equation with any type of BCs \revall{on a parallelepiped and uniform Cartesian mesh}, that would encompass all these improvements, is still missing to date. 
In particular, one might wish for a solution that would hide the complexity of the FFT-based implementation from the user---whose focus should be on the boundary value problem only---but that would still benefit from the above improvements.
Furthermore, such an integrated solution would push the optimization of the whole algorithm one step further, by tailoring the management of memory and data to the Poisson problem on a given mesh configuration, and by advantageously sharing the computational resources over the whole computational process.

Moreover, techniques to address certain combinations of BCs are also missing. For instance, one would be interested in computing the solution of \eqqref{eq:intro:Poisson} in a domain which is unbounded on one side, and symmetric on the other side. This corresponds to semi-infinite problems such as the boundary layer on a flat plate in fluid mechanics. 
The treatment of such a configuration in a \typo{FFT}-based Poisson solver was never exposed.

%What do we propose?
This work presents the concepts that have supported the development of the Fourier-based Library of Unbounded Poisson Solvers (\emph{FLUPS}). FLUPS provides an integrated solution for a variety of Poisson's problems with high computational throughput.
The library, written in C++ with an API in C, supports any arbitrary combination of BCs on 2D and 3D uniform grids.
The parallel implementation is based on the MPI standard and can also take advantage of shared-memory parallelism with OpenMP, depending on the size of the problem and the computational resources available.
\revthree{FLUPS is thus intended as an open-source, state-of-the-art, and highly optimized implementation of a Poisson solver over uniform grids,}
 available at \url{github.com/vortexlab-uclouvain/flups}.

%Structure of the present work
In this article, the general scheme for solving Poisson's equation in 3D with Green's functions is first explained.
Details are given about the various types of BCs, and the consequences in the algorithm in terms of Green's functions and FFT computations.
In particular, a novel method for the treatment of a semi-infinite direction with symmetry at one end is proposed.
Then, a verification and validation of {FLUPS} is presented, focusing on the accuracy of the Poisson solver, and on the weak and strong scalability of the algorithm up to \typo{$73\,720$} processors, measured on three different massively parallel architectures.

%!TEX root = mixPoissonSolver_main_siam.tex
%!TEX encoding = UTF-8 Unicode

\section{Methodology of FLUPS in 3D}%
\label{sec_methodo}

%\todo{do we speak about 2D specificity? noup on my opinion}

% move elsewhere??
FLUPS is capable of handling scalar and vector fields and, by default, assumes that data are given in a ``cell-centered'' framework; {i.e.} their numerical values correspond to the evaluation of the fields at the center of the grid cells.
 Each field must be attached to a topology object, which describes the size and the memory arrangement of the field, as will be detailed in \sect{sec:implementation_note}.
% later!!
%Most importantly, every topology owns its specific Fastest Rotating Index (FRI), i.e. an integer indicating the corresponding direction ($X=0$, $Y=1$ or $Z=2$) of two neighbour data in memory. For a given FRI, $ax_0$, the decomposition in memory follows the natural order of the indexing, i.e. $ax_1 = \left( ax_0 + 1 \right) \% 3$ and $ax_2 = \left( ax_0 + 1 \right) \% 3$.
%
The initialization of FLUPS always starts with the precomputation of the communication patterns and with the \typo{planning} of the FFTs, which depend on the user-specified BCs and on the topology of the RHS field, $f$.
The evaluation of the user-specified Green's function is also performed during the setup of the library, and the Green's function in the full spectral space $\hat{G}$ is eventually obtained \typo{and stored}.
Then, for any given RHS, the solution is obtained by first computing $\hat{f}$, {i.e.} the forward 3D FFT of $f$, by multiplying it with $\hat{G}$, and by computing the backward transform of the solution.
Each of the following sections focuses on the specific stages of this process.

%------------------------------------------------------------------
\subsection{Evaluation of Green's function for spectral and unbounded directions}
\typo{Following the numbering of the introduction,} the Green's functions available in FLUPS are \iti the singular expressions with replacement of the singularities by cell-average values as in \cite{Chatelain:2010}; \itii those of \cite{Hejlesen:2013,Spietz:2018} with explicit regularization up to the order \revthree{$m=10$} \revthree{ and their spectral generalization presented in \cite{Hejlesen:2019} }; and \itiv the lattice Green's functions of \cite{Liska:2014}.
%The extension of the library to type \itiii Green's function is possible, but was not envisioned at first.
The analytical expressions of the Green's functions in cases \typo{\iti, \itii and \itiv} are \typo{summarized} in \cref{appendix:greenfunctions}. 
%Both follow the same procedure to obtain the expression of the Green's function $G$.
%Their expression depend on the BCs of the domain.
%
Depending on the BCs, additional processing (including FFTs) is required to obtain $\hat{G}$.

\subsubsection{Fully unbounded problems}

For fully unbounded problems the Green's function is expressed in the physical space.
Indeed, the singular Green\typo{'s} functions of type \iti are obtained as the solution of
\be
\lapl G = \delta(\bx),
\label{eq:Green_Poisson_3dirunb_delt}
\ee
and the regularized type \itii Green's functions are obtained from
\be
\lapl G_m = \zeta_{\varepsilon,m}(\bx)
\label{eq:Green_Poisson_3dirunb_sig}
\eec
%; the LGF of (iii) are the solution of
%\be
%\lapl_h G_h = \delta(\bx)
%\label{eq:Green_Poisson_3dirunb_sig}
%\eed
where the analytical expression of $G$ is known as a function of the spatial coordinates.
The domain doubling technique must be used to obtain the result in the \typo{Fourier space} (thus involving a 3D FFT); see \sect{sec_domain_doubling}. 
In \eq{eq:Green_Poisson_3dirunb_sig}, $\zeta_{\varepsilon,m}$ is a regularization kernel\typo{, as in} \cite{Hejlesen:2013}. It is here assimilated to a Gaussian mollification of order $m$ and of characteristic smoothing length $\varepsilon = 2h$
 (see \app{sect:app:gaussian} for their mathematical expressions). 
%
%For case (iii), the lattice Green's function is obtained as the solution of 
%\be
%\lapl_h G = \delta(\bx)
%\label{eq:Green_Poisson_3dirunb_lgf}
%\eec
%where $\lapl_h$ stands for a discretized Laplacian operator. Most of the works in the literature have opted for second order finite differences. The Green's function values are then obtained at the price of the numerical computation of an integral, in addition to an approximation for long range interactions \cite{Liska:2014} (see \todo{Appendix} for more details).
%

\subsubsection{Fully spectral problems}

Conversely, if all directions are spectral ({i.e.,} \typo{with periodic or symmetric BCs),} %the Fourier transform can be readily applied, and} 
the Green's function is readily expressed in the full-spectral space.
For instance, \eqqref{eq:Green_Poisson_3dirunb_sig} is expressed in the Fourier space as $ \left( \ic \bk\right) \cdot \left( \ic \bk\right)  \; \hat{G} = \hat{\zeta}_{\varepsilon,m}(\bk)$, which yields for $\hat{G}$
\be
\hat{G}_m = -\frac{1}{\bk\cdot\bk} \hat{\zeta}_{\varepsilon,m}(\bk)
\eec
\typo{where $\bk = \left[ k_x \;,\; k_y \;,\; k_z \right]$, with $k_i \in \left[ -\pi/h_i \;;\; \pi/h_i \right]$.}
No FFTs are hence required to evaluate the Green's function. %The same rational applies to families (i) and (iii). %Similarly to the unbounded case, the spectral version of it is obtained using $\zeta= \delta(\bx)$ and the regularized version using a characteristic smoothing length $\sigma$, $\xi(\bx) = \zeta_\sigma(\bx)$.

\subsubsection{Partially unbounded problems}

For domains combining one (resp., two) spectral direction and two (resp., one) infinite (or semi-infinite) directions, Green's functions are obtained from the solution of a Helmholtz equation. 
Indeed, the Laplacian operator is split between the unbounded and the periodic/symmetric direction, and one considers the partial Fourier transform of \eq{eq:intro:Poisson} along the latter direction,  which hence becomes spectral \cite{Chatelain:2010}.
%This spectral direction is expressed as a function of the corresponding wave number $k$, while the unbounded directions are taken as a function of the real spatial coordinates, hence defining a Helmholtz equation. 
This results in a Helmholtz equation, which is expressed as a function of the wave number $k$ in the spectral direction and as a function of the spatial coordinates in the unbounded directions.
For example, considering two unbounded directions $X,Y$ and a periodic direction $Z$, the regularized Green\typo{'s} function is obtained as the solution of
\be
%\lapl_{x,y} \tilde{G}_{k_z} - k_z^2 \tilde{G}_{k_z} = \hat{\zeta}_\varepsilon^m(k_z)
\lapl_{x,y} \tilde{G}(x,y,k_z) - k_z^2 \tilde{G}(x,y,k_z) = \hat{\zeta}_{\varepsilon,m}(k_z)
\eec
where $\tilde{.}$ denotes the partially transformed function. %The expression of $\tilde{G}_{k_z}$ is known analytically as a function of $x,y$ and $k_z$.
The remaining spatial directions are then to be transformed using FFTs and the domain doubling technique.

%------------------------------------------------------------------
\subsection{3D FFT transform and convolution}
Numerically, the 3D FFT is obtained as the succession of 1D FFTs in all three directions, and each transform type is driven by the BCs in the considered direction.
For example, starting from $f(x,y,z)$ in spatial coordinates and seeking for the fully periodic transform, one thus obtains successively $\tilde{f}(x,y,k_z)$ as the result of the first DFT in $z$, ${\tilde{f}}(k_x,y,k_z)$ as the result of the second DFT in $x$, and finally $\hat{f}(k_x,k_y,k_z)$ as the result in all-spectral coordinates.
The type of BC dictates the order of the transforms, and hence the order of treatment of the directions, so as to minimize the number of computational operations (see \sect{sec:implementation_note}).

\medskip
%Two different techniques are available for the domain decomposition among processors: the slab-based and the pencil-based decomposition. 
%MOVED TO INTRO: As highlighted by \cite{Sbalzarini:2006a,Pekurovsky:2019}, the pencil decomposition provides a better scalability on massively parallel architectures as it is able to exploit $N^2$ cores at most (for a cubic domain of $N^3$ unknowns), while the slab decomposition is limited to the use of $N$ cores.  Moreover, the slab decomposition features one all-to-all communication, i.e. one MPI-rank communicates with $N_{rank}$ other ranks, while the pencil decomposition only requires one rank to communicate twice with $\sqrt{N_{rank}}$ other ranks.
%
%For all those reasons, we chose to apply the pencil-decomposition approach, yet the extension to slab-decomposition is straightforward.
In a fashion much akin to the P3DFFT library, FLUPS uses pencil decompositions of the domain. From one FFT operation to another, a change in topology is thus required---here called \textit{topology switch}.
The latter involves a very large number of communications, and a significant effort was dedicated to reducing the associated latency in FLUPS (see \sect{sec:implementation_note}).
Additionally, since all 1D FFT operations (here conducted using the FFTW library \cite{Frigo:2005}) require the contiguity of data in memory along the transformed direction, a memory transpose is also necessary to reorder the data along each pencil direction. Notice that non-unit stride 1D FFTs are also implemented in FFTW, but their use would here lead to a slower execution.

In the following subsections, the treatment of the transform in each direction is described, with a focus on the particularities of the various types of BCs. Moreover, the operations on the field $f$ and on the field $G$ are detailed when appropriate.
%As $f$ is real-valued in the spatial coordinates, FLUPS also manages the type (real or complex) of the data and selects the appropriate kind of transform.
The final result of the 3D FFT hence consists in a succession of these treatments for all the directions, which are processed separately as 1D problems.

%Finally, let us highlight that FLUPS assumes that all fields ($\phi,f$) have ``cell-centered'' values.

%Therefore, the Green's function is here understood as in a vertex-centered formalism, in such a way that the convolution of $f$ with $G$ indeed captures the self-interaction on each grid point, hence adding complexity in the treatment of the unbounded and spectral directions, as discussed hereunder.

%---------------------------------------------------
\subsubsection{Spectral direction}
The spectral (symmetric or periodic) directions are the most straightforward.
With periodic BCs, the forward Fourier transform of $f$ is readily obtained using a Direct Fourier Transform (DFT).
%
%According to [The documentation of FFTW?],  
%the periodic transform in the $j^{th}$ direction is implemented with a Direct Fourier Transform (DFT)% and yields
%\begin{align}
%\tilde{f}_k &= \sum_{j=0}^{N-1} f \left(x_j \right) \left[ \cos \left( \frac{2 \pi}{N} j  k \right) - i \sin \left( 2 \pi \frac{j}{N} k \right) \right]\\
%&= \sum_{j=0}^{N-1} f \left( x_j \right) e^{- \left( 2 \pi \frac{j}{N} k \right) }
%\end{align}
%where N is the number of grid point in the $j^{th}$ direction.
%
For symmetric directions, and accounting for the fact that fields $f$ and $\phi$ are ``cell-centered'',
symmetry conditions are implicitly satisfied at the cell edges using
\begin{itemize}
\item the type-$II$ DST for odd-odd cases,
\item the type-$IV$ DST for odd-even cases,
\item the type-$IV$ DCT for even-odd cases,
\item the type-$II$ DCT for even-even cases.
\end{itemize}
The latter are equivalent to, yet much more efficient than, the use a DFT on a domain of size $2N$ (for odd-odd and even-even cases) or $4N$ (for crossed cases), where the symmetries would be explicitly enforced by a copy of the data into a symmetric image in the extended domain. We refer the reader to \cite{Frigo:2005} and to the documentation of the FFTW package\footnote{The documentation of the FFTW library is available online at \url{www.fftw.org/fftw3_doc}.} for complete definitions.
Notice that the backward transform of $\hat{\phi}$ is performed with the corresponding DFT/DCT/DST type, as the solution field enjoys the same symmetries \revtwo{as the RHS}.
%\revtwo{We stress that, since we consider unbounded and periodicity conditions% as per eq.(1.1)
% , the solution $\phi$ and the right hand side, $f$, share to the same BCs.}

%
%The type-$II$ DCT is defined as
%\be
%%\hat{f}_k = 2 \sum_{j=1}^{n-1} f(x_j) \cos\left[ \pi \left(j+1/2\right) k / \left(n\right) \right]
%\tilde{f}_k = 2 \sum_{j=1}^{n-1} f(x_j) \cos\left[ \pi \left(j+1/2\right) k / \left(n\right) \right]
%\label{eq:DCT2}
%\eed
%
%The type-$II$ DST is defined as
%\be
%\tilde{f}_k = 2 \sum_{j=1}^{n-1} f(x_j) \sin\left[ \pi \left(j+1/2\right) \left(k+1\right) / \left(n\right) \right]
%\label{eq:DST2}
%\eed
%\todo{Give expressions of all DCT/DST... mmmh no + plots?}

%\todo{More comment on R2R, R2C, C2C here? and the corresponding input/output sizes?}

\medskip
THe Green's function needs not undergo an FFT in the spectral directions as the analytical expression is already expressed as a function of the spectral coordinates for both periodic and symmetric BCs.
However, the wave number $k$ used for its evaluation \typo{(i.e. the wave number of the current direction)} must correspond to the modes resulting from the transform of the source field.
Considering the definition of the various types of transforms,
\begin{itemize}
\item the type-$II$ DCT is given by
\be
\tilde{f} = 2 \sum_{j=0}^{N-1} f(x_j) \cos\left[ \pi \left(j+\frac12\right) \frac{\textrm{k}}{N} \right]
\label{eq:DCT2}
\eec
where the $\textrm{k}{\textrm{th}}$ entry in the result of the DCT corresponds to the mode $k$. The flip-flop mode, $\textrm{k}=N$, is trivially null and the Green's function $\tilde{G}$ is thus evaluated between $k=0$ and $k=N-1$;
\item the type-$II$ DST is given by
\be
\tilde{f} = 2 \sum_{j=0}^{N-1} f(x_j) \sin\left[ \pi \left(j+\frac12\right) \frac{\textrm{k}+1}{N} \right]
\label{eq:DST2}
\eec
where the $\textrm{k}{\textrm{th}}$ entry in the result of the DST corresponds to the mode $k+1$ since the mode $k=0$ is trivially null. $\tilde{G}$ is thus evaluated between $k=1$ and $k=N$;
%Therefore the mode $k=0$ is discarded in the Green's function;
%
\item the type-$IV$ DCT corresponds to
\be
%\hat{f}_k = 2 \sum_{j=1}^{n-1} f(x_j) \cos\left[ \pi \left(j+1/2\right) k / \left(n\right) \right]
\tilde{f} = 2 \sum_{j=0}^{N-1} f(x_j) \cos\left[ \left(\textrm{k}+\frac12\right) \pi  \frac{\left(j+\frac12\right)}{N} \right]
\label{eq:DCT4}
\eec
and the first entry $\tilde{f}(\textrm{k}=0)$ is actually the global mode $k=1/2$. $\tilde{G}$  is thus correspondingly evaluated with an offset of $\frac12$. The same conclusion holds for the type-$IV$ DST.
\end{itemize}

%Hence for $n$ information, it will return $n$ real information. One has to note that a flip-flop mode is by definition null. Adding this information, we retrieve the $n+1$ information generated by the Green's DCT.
%and the type-$II$ DST,
% we observe that 
%\todo{this is not clear}
%Yet, FLUPS assumes that all fields have  values. 
%It must be noticed, however, that the Green's functions must be given in a ``vertex-centered'' framework, such that the self-interaction is located in $x=y=z=0$ in order to correctly enforce the symmetries at the edge of the domain. This leads to different expressions
%
%
%This mismatch introduces some additionnal complexity and is discussed hereunder when needed.
%
%Yet, as the field is cell-centered, a special treatment of the first or last mode has to be considered for the symmetric cases. The first mode is zero by construction if a DST is used and the last mode is zero if a DCT is used. Hence the Green's function has to be computed at the correct modes depending on the considered BCs.

% Finally, it is
\typo{Notice that} the forward transform of $f$ in periodic directions uses a real to complex or a complex to complex \typo{transform}, depending on the input data type. The backward transform of $\hat{\phi}$ makes use of the corresponding \typo{inverse operation}.

%---------------------------------------------------
\subsubsection{Fully unbounded direction}
\label{sec_domain_doubling}
\begin{figure}[!ht]
\begin{minipage}{0.2\textwidth}
\begin{minipage}{\textwidth}
\includegraphics[width=\textwidth]{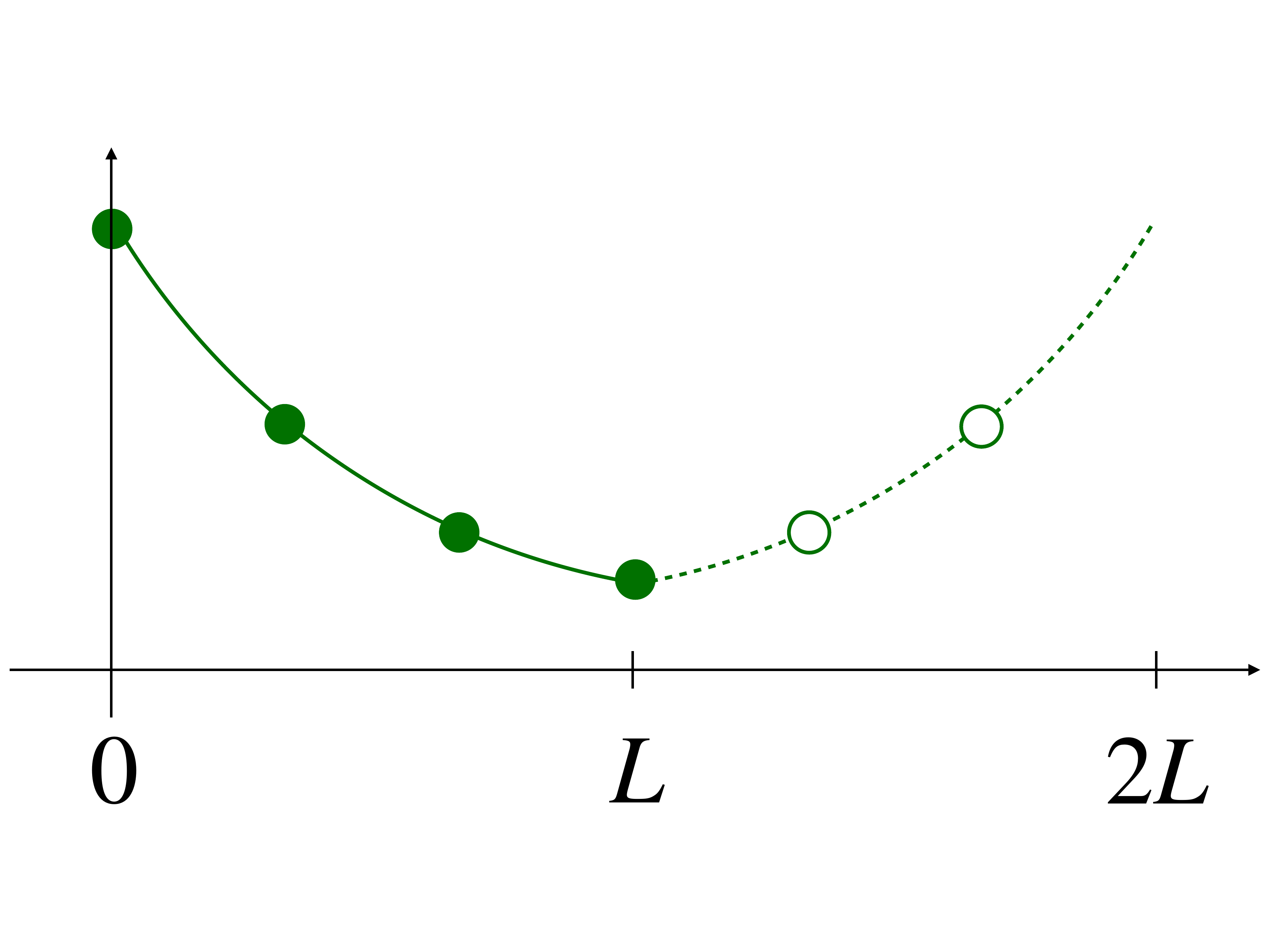}
\subcaption{Green's function $G$}
\label{fig:BC:full_unb_green}
\end{minipage}
\begin{minipage}{\textwidth}
\includegraphics[width=\textwidth]{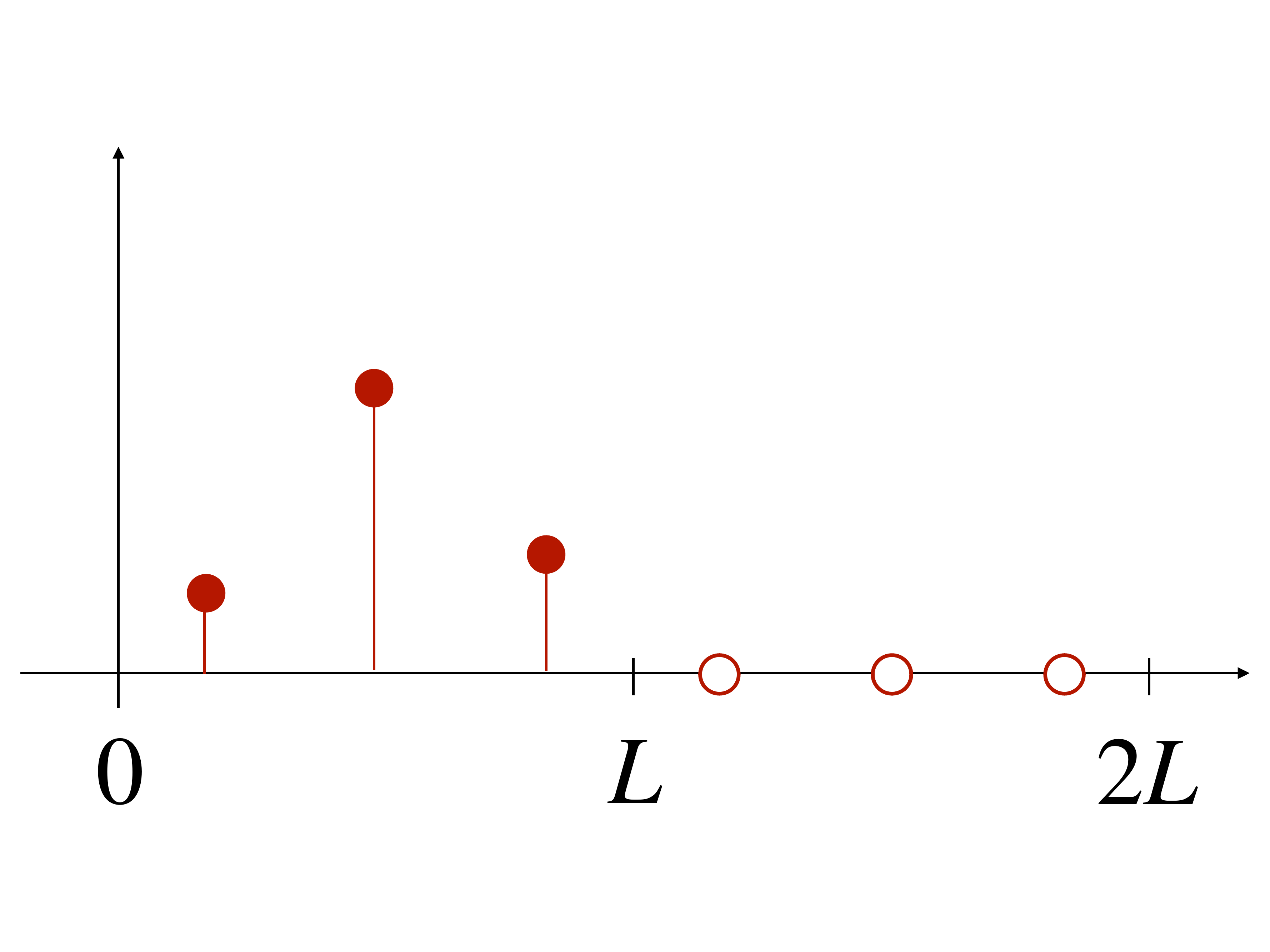}
\subcaption{ right hand side $f$}
\label{fig:BC:full_unb_src}
\end{minipage}
\end{minipage}%
\hspace{0.05\textwidth}%
\begin{minipage}{0.25\textwidth}
\includegraphics[width=\textwidth]{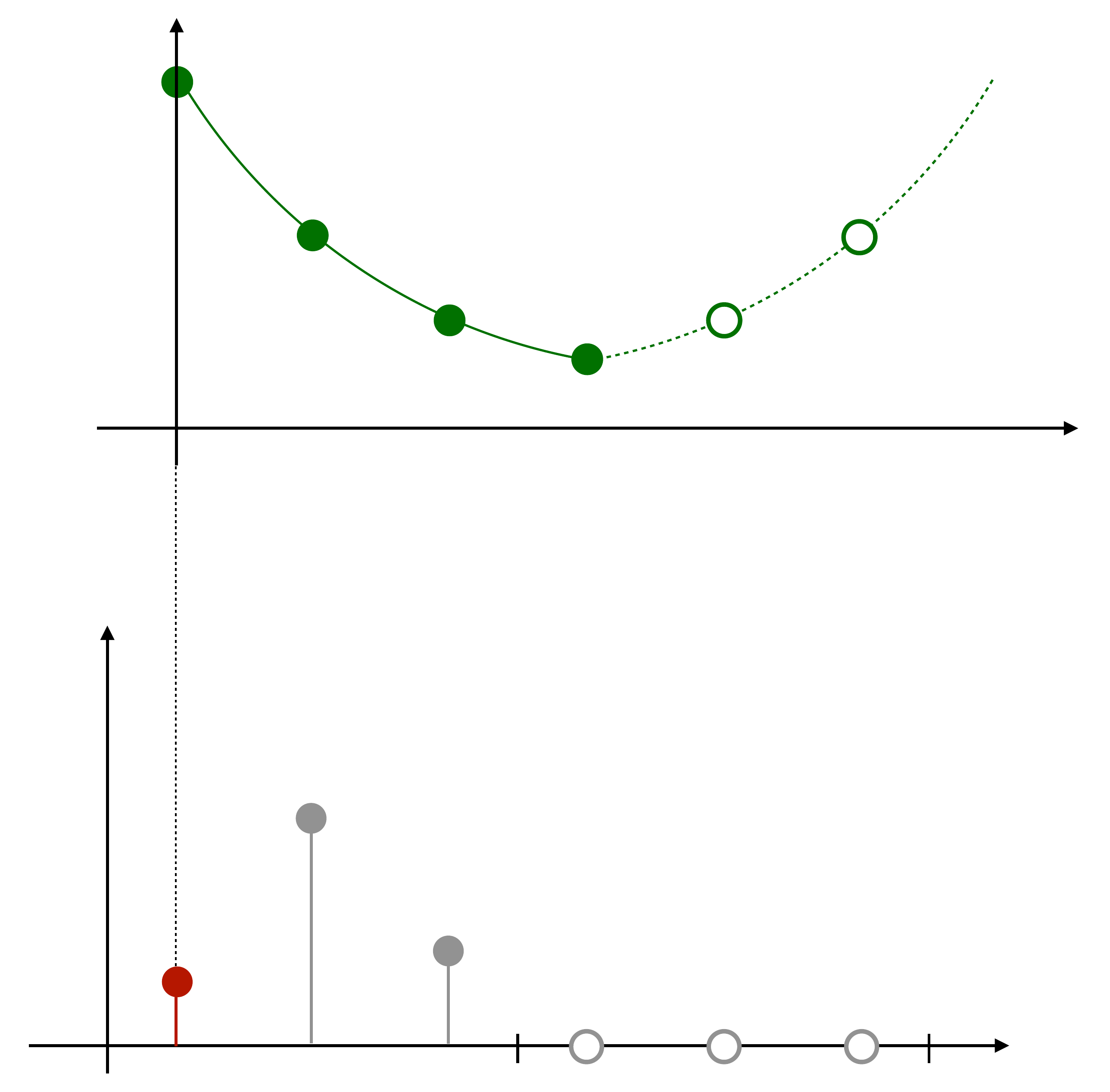}
\subcaption{convolution term $j=0$}
\label{fig:BC:full_unb_conv1}
\end{minipage}%
\begin{minipage}{0.25\textwidth}
\includegraphics[width=\textwidth]{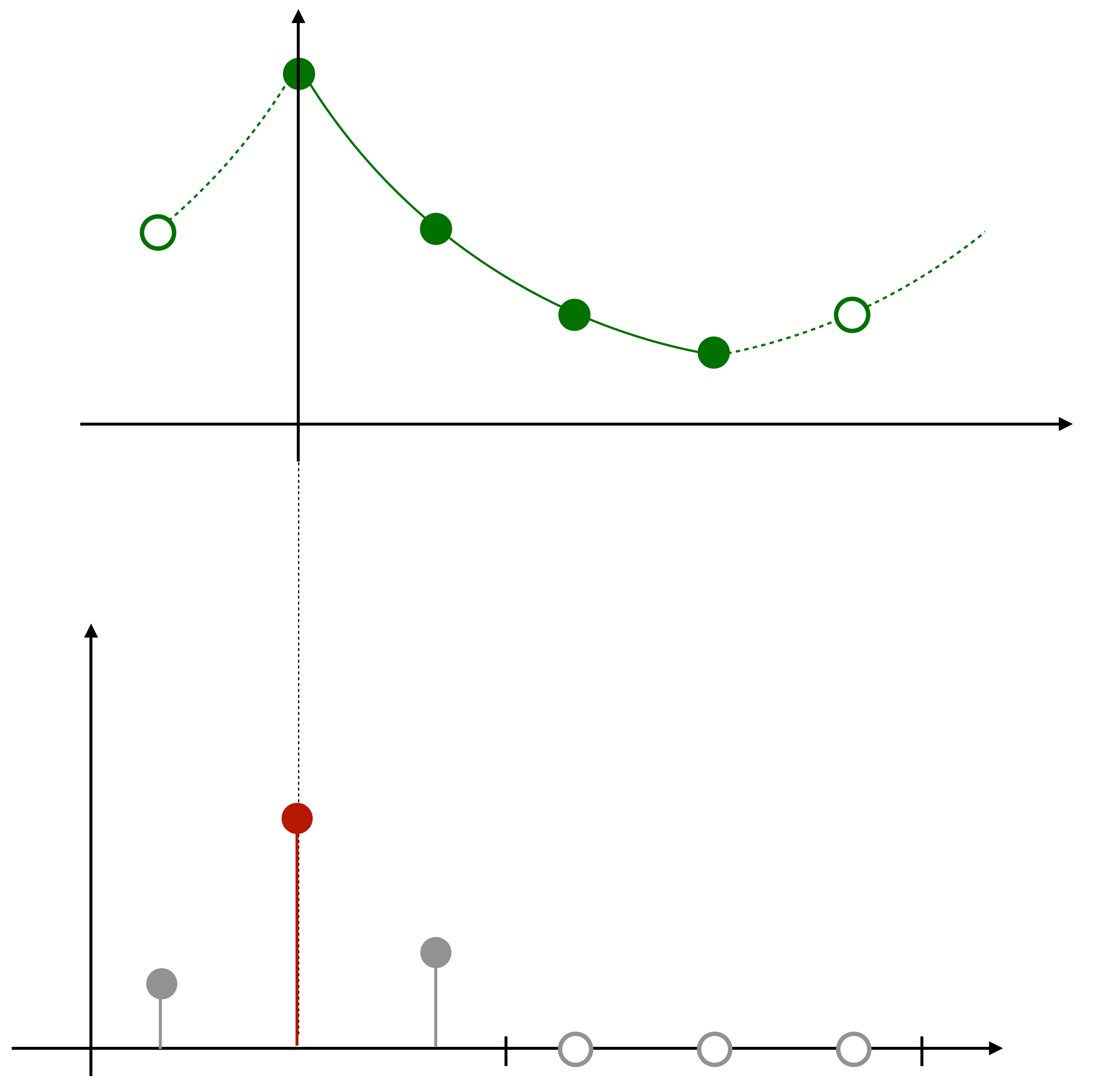}
\subcaption{convolution term $j=1$}
\label{fig:BC:full_unb_conv2}
\end{minipage}%
\begin{minipage}{0.25\textwidth}
\includegraphics[width=\textwidth]{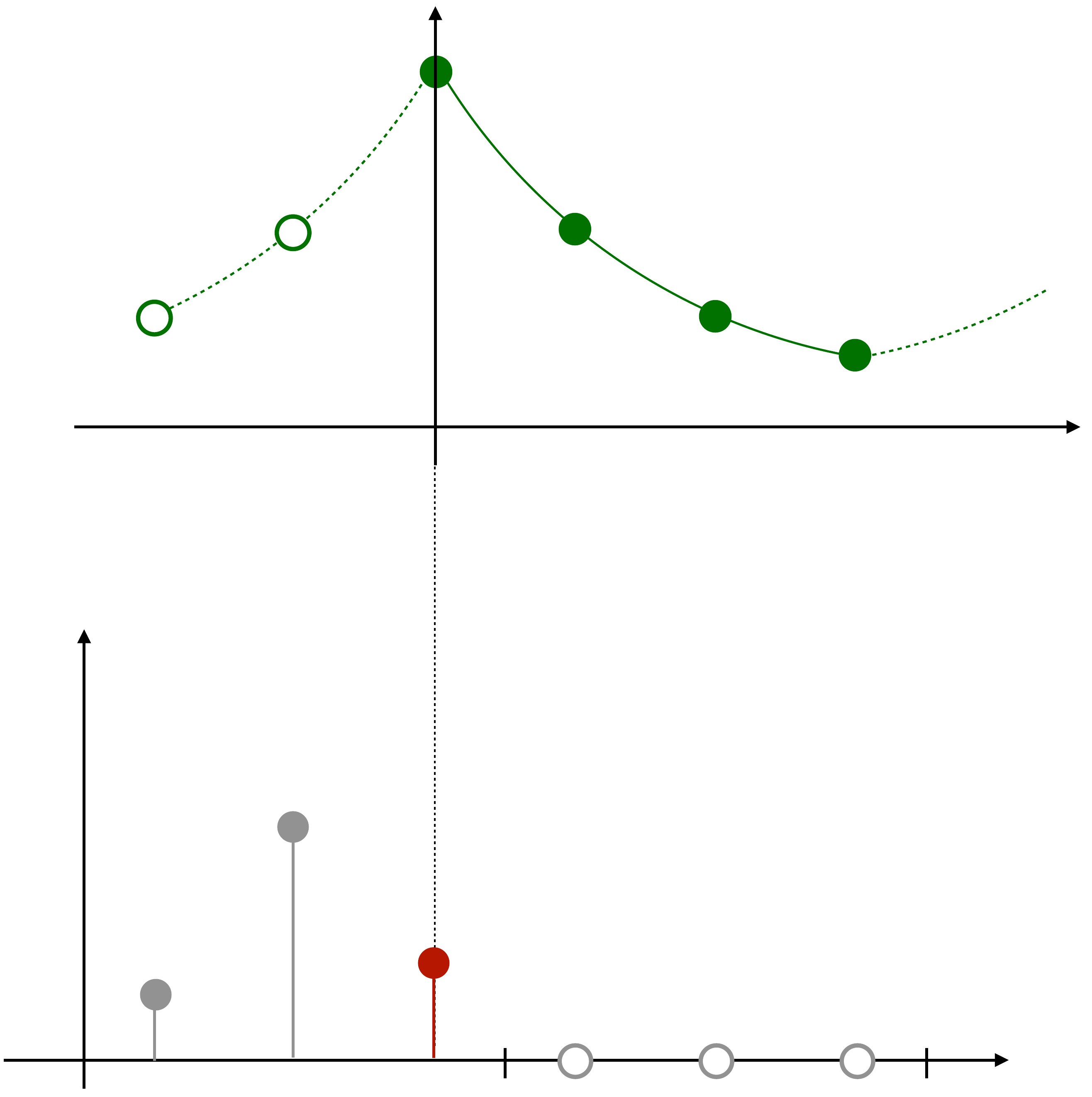}
\subcaption{convolution term $j=2$}
\label{fig:BC:full_unb_conv3}
\end{minipage}
\caption{Example of the steps involved in a convolution between the Green's function and a source term, $\phi(x_i) = \sum_{j=0}^{N-1} G(x_{j}- x_i) f(x_j)$. The filled bullets show the data in the initial domain $x=[0,L]$ of size $N=3$. The empty bullets denote the data in the extension of the domain, with symmetry on $G$ about $x=L$ and zero-padding on $f$. }
\end{figure}

Following the domain doubling technique by Hockney and Eastwood \cite{Hockney:1988}, $f$ is extended to a domain of size $2N$ with zero-padding\typo{;} the Green's function is extended to $2N$ and symmetrized about the N{th} data,  as in the example shown in \fig{fig:BC:full_unb_green} and \fig{fig:BC:full_unb_src}. The unbounded \revthree{(i.e. aperiodic)} convolution is  obtained through the convolution in the doubled domain with periodic BCs.
\fig{fig:BC:full_unb_conv1}, \fig{fig:BC:full_unb_conv2} and \fig{fig:BC:full_unb_conv3} illustrate the operations involved in that convolution, as if it was performed in spatial coordinates, with periodicity over the doubled domain of size $2L$. 
One easily verifies that performing the convolution between the padded fields with periodic BCs  yields the same results as if the domain was unbounded in the non-padded region ($x\in[0,L]$). The result in the padded region ($x \in ]L,2L]$) is spurious and is hence discarded.
One thus obtains $\tilde{f}$ and $\tilde{G}$ as the the forward transform of $f$ and $G$ using DFTs over the doubled domain, again of type 
real to complex or complex to complex, depending on the input data type.
%The backward transform of $\hat{\phi}$, all use 

As previously mentioned, the convolution in FLUPS is implemented through the multiplication in Fourier space.
It is to be noted that, because of the domain extension and the implicit periodicity, the Green's function and the field  eventually have the same size of $2N$ unknowns in the Fourier space. Hence, no special care needs to be taken for the computation of the point-to-point multiplication.

%\todo{change the subcaption stuff}
%\begin{figure}[h]
%\centering
%\begin{tikzpicture}[
%    x=\textwidth,
%    y=11/16*\textwidth,
%    every path/.style = {},
%  ]
%  \begin{scope}
%	\pgfmathsetmacro{\s}{.8}
%	
%	\node[rotate=0,above right] at (0,0) {\includegraphics[width=\s\textwidth, clip=true, trim=2cm 5cm 2cm 1cm]{semi_unbounded_dir_full.pdf}};
%
%	%Legends	
%	\node[rotate=0] at ({.28*\s},{.55*\s}) {a. Example Green's function};
%	\node[rotate=0] at ({.74*\s},{.55*\s}) {b. Example right hand side};
%	\node[rotate=0] at ({.5*\s},{.02*\s}) {\parbox{\s\textwidth}{ c. Decomposition of the terms involved in the convolution $\phi(x_i) = \sum_{j=1}^N G(x_{j}- x_i) f(x_j)$, with the shifted Green's function (top) and the RHS (bottom).}};
%	%
%	\node[rotate=0] at ({.2*\s},{.09*\s}) {$i=1$};
%	\node[rotate=0] at ({.5*\s},{.09*\s}) {$i=2$};
%	\node[rotate=0] at ({.805*\s},{.09*\s}) {$i=3$};
%	
%%	\node[rotate=0] at ({.36*\s},{.26*\s}) {$+$};
%   \end{scope}	
%\end{tikzpicture}

%\caption{Operations involved in the domain doubling technique by Hockney and Eastwood \cite{Hockney:1988}.}
%\label{fig:BC:full_unb}
%\end{figure}
%---------------------------------------------------
\subsubsection{Semi-unbounded direction}

A semi-unbounded direction is a direction with a symmetry condition at one end and an infinite BC at the other end.
Conceptually, this configuration is easily modeled with an extension of the domain.
As illustrated in \fig{fig:BC:odd_unb_conv}, the source term is first copied so as to explicitly satisfy the BC, hence extending the computational domain from $[0,L]$ to $[-L,L]$ in the case of a symmetry condition on the left side of the domain (or to $[0,2 L]$ in the case of a symmetry condition on the right side of the domain).
The resulting configuration is unbounded on a domain of size $2L$ and can be solved as described in the previous section: the Green's function is evaluated over $[0,2L]$ and symmetrized in the padded region, $[2L , 4L]$, and a DFT is used to obtain the data in the spectral space.
%
% If the sources are symmetrized in the extended domain, $[-L,L]$, in order to explicitly satisfy the symmetry condition, e.g. in $x=0$, as shown in \fig{fig:BC:oddOrEven_unb}, one can then apply the Hockney-Eastwood treatment to obtain the unbounded result of the convolution. The operations involved in the computation of the convolution performed on the extended domain are illustrated in \fig{fig:BC:odd_unb_conv}.
However, strictly applying this would require a domain of size $4N$ (because of two successive domain doublings), which is not acceptable for performance reasons.

\begin{figure}[!h]
\centering
\begin{minipage}[t]{0.3\textwidth}
\includegraphics[width=\textwidth,clip=true,trim=0 0 0 13.5cm]{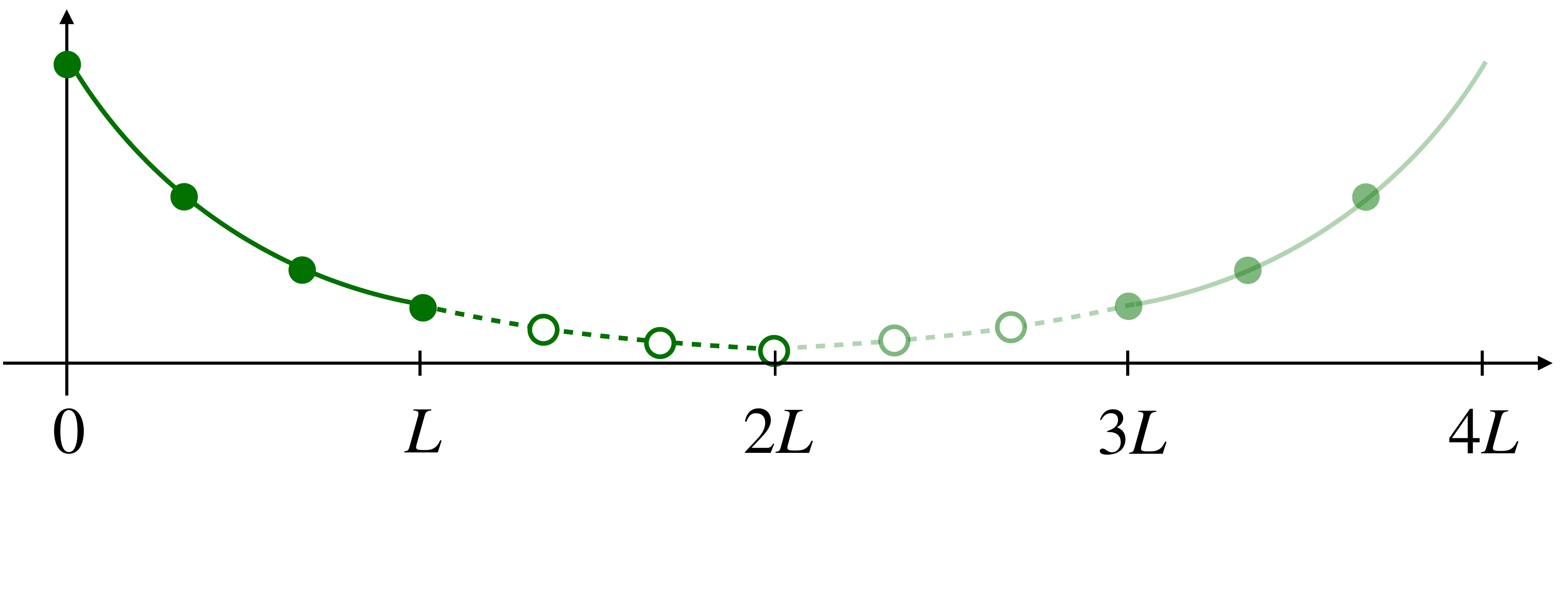}
\subcaption{\label{fig:BC:mixunb_Green}Green's function $G$ extended for a semi-unbounded direction}
\end{minipage}%
\hspace{0.015\textwidth}%
\begin{minipage}[t]{0.3\textwidth}
\includegraphics[width=\textwidth,clip=true,trim=0 0 0 13.5cm]{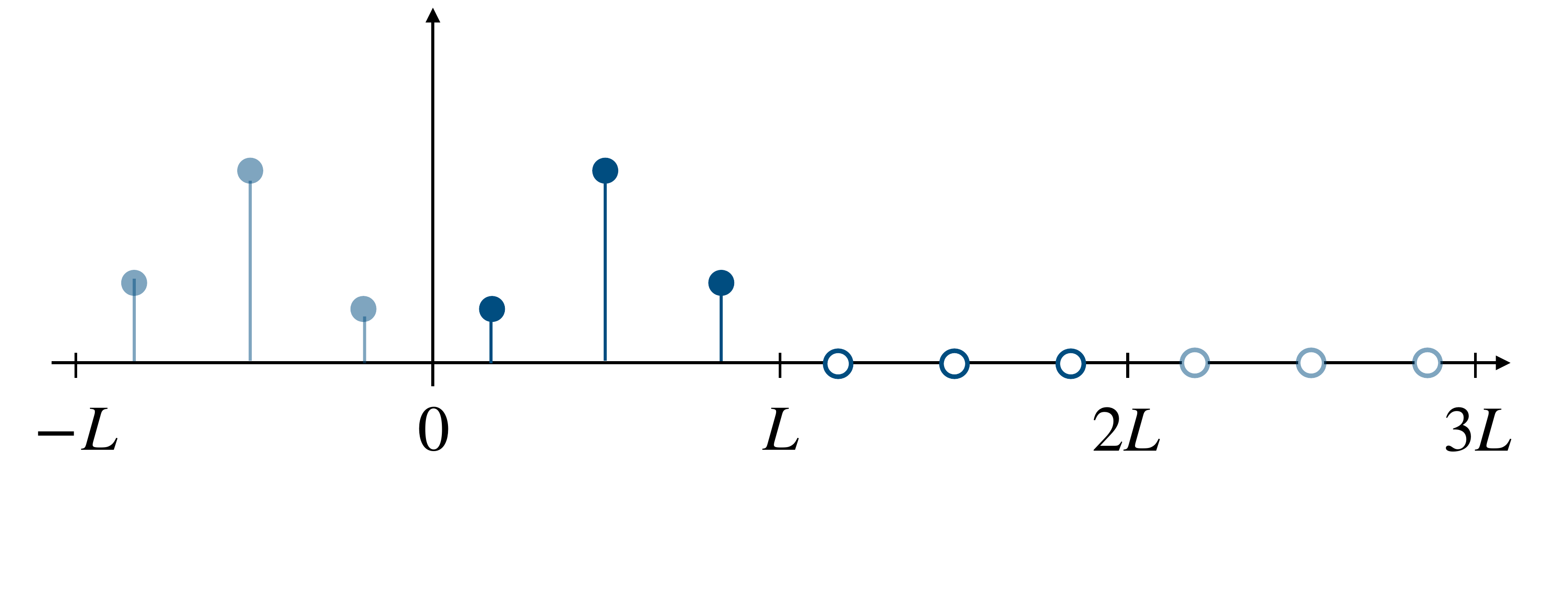}
\subcaption{\label{fig:BC:mixunb_FieldEven}right hand side $f$ with even symmetry on the left}
\end{minipage}%
\hspace{0.015\textwidth}%
\begin{minipage}[t]{0.3\textwidth}
\includegraphics[width=\textwidth,clip=true,trim=0 0 0 13.5cm]{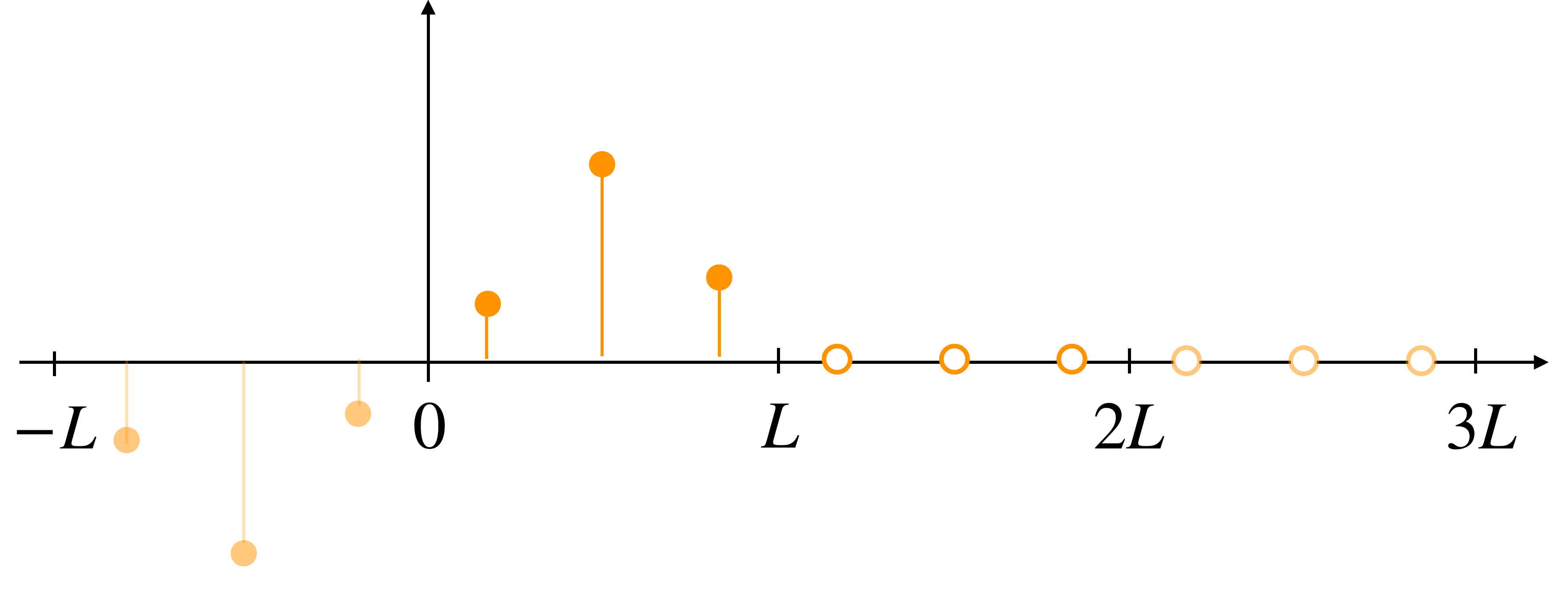}
\subcaption{right hand side $f$ with odd symmetry on the left}
\end{minipage}
\caption{Green's function and RHS initially on a domain $x=[0,L]$ of size $N=3$, and their virtual extension in a domain of size $4L$ for the resolution of a semi-infinite direction with symmetry on the left. Faded colors indicate ghost data that are not present when using a DCT or a DST to enforce the symmetry. Only the portion $[0,2L]$ is then effectively evaluated. }
\label{fig:BC:oddOrEven_unb}
\end{figure}

Instead, DSTs and DCTs are here exploited to reduce the computational cost, while still imposing the correct symmetry condition at the desired domain end. In this approach, the domain doubling is still performed as for the fully unbounded direction, but the zero-padding is here done on the side which is unbounded. In the example considered in \fig{fig:BC:mixunb_FieldEven}, it takes place on the right side, $\left]L,2L\right]$. The domain over which the DST/DCT will be executed is hence $[0,2L]$.
As for the Green's function (\fig{fig:BC:mixunb_Green}), it must be here evaluated in the extended domain and transformed using a DCT (whereas it was explicitly symmetrized in the extended domain for fully unbounded domains; see \fig{fig:BC:full_unb_green}). As shown in \fig{fig:BC:odd_unb_conv}, this guarantees that the convolution accounts for the proper BCs on the domain $\left[0,2L\right]$. 

\begin{figure}[h!]
\centering
\begin{minipage}{0.33\textwidth}
\includegraphics[width=\textwidth]{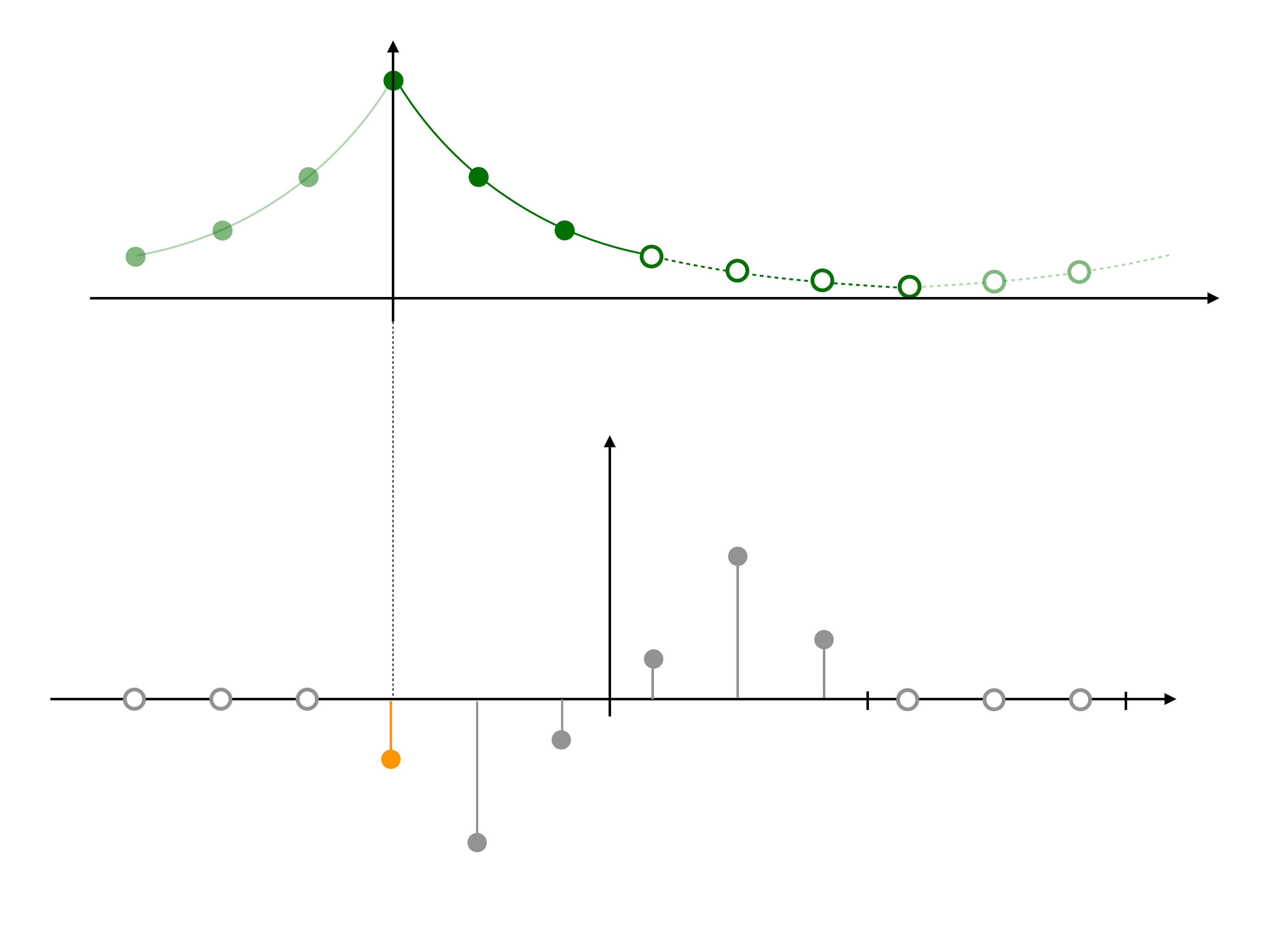}
\subcaption{convolution for $j=4$}
\label{fig:BC:odd_unb_conv1}
\end{minipage} %
\begin{minipage}{0.33\textwidth}
\includegraphics[width=\textwidth]{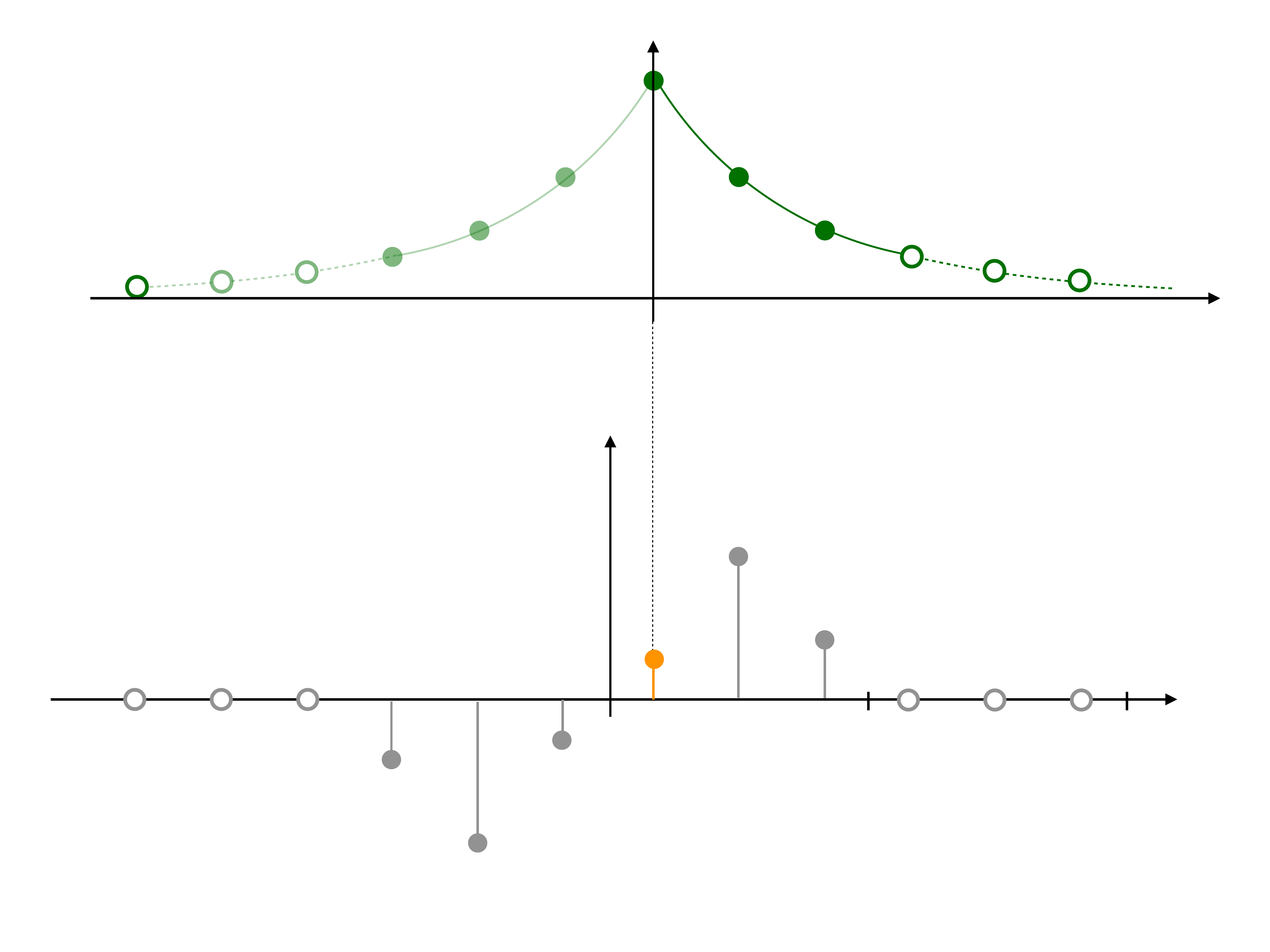}
\subcaption{convolution for $j=7$}
\label{fig:BC:odd_unb_conv2}
\end{minipage} %
\begin{minipage}{0.33\textwidth}
\includegraphics[width=\textwidth]{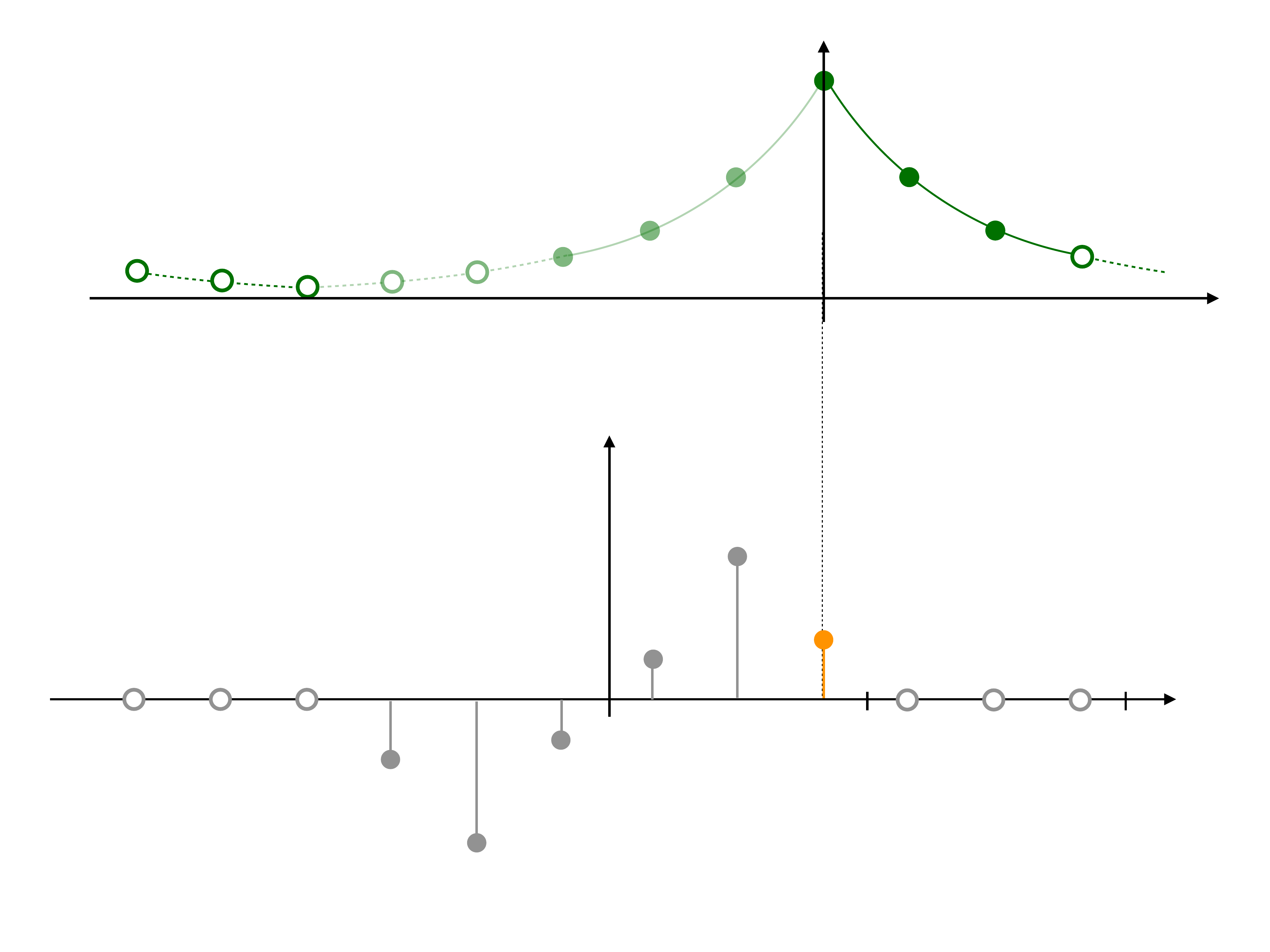}
\subcaption{convolution for $j=9$}
\label{fig:BC:odd_unb_conv3}
\end{minipage}
\caption{The convolution of $G$ (top) with $f$ (bottom) on a $4N$ size domain padded for a semi-unbounded transform with odd symmetry on the left. The correct solution is retrieved in $x=[-L,L]$.}
\label{fig:BC:odd_unb_conv}
\end{figure}

As shown in \fig{fig:BC:mixunb_Green}, the Green's function is evaluated in a ``vertex-centered'' framework, as it characterizes the influence of a source located in $\bx=\boldsymbol{0}$ on any other location in the domain. % and  in order to correctly enforce the symmetries at the edges of the domain. %Mmmh, we could use the different types of FFT to displace the symmetry plane of h/2...
Most importantly, $f$ is in a cell-centered layout. Therefore different types of Fourier transforms have to be used to compute $\tilde{f}$ and $\tilde{G}$ with the proper symmetry conditions.
%To ensure the compatibility between these two visions 
In order to guarantee that their product is consistent with the aimed convolution, the k{th} entry in the result of the FFT of $f$ must be related to the corresponding mode $k$ of $\tilde{G}$.

Practically, the field $f$ (with $2N$ cell-centered real data, as in \fig{fig:BC:mixunb_FieldEven}) will be transformed into $\tilde{f}$ using a type-$II$ DCT for an even symmetry, and using a type-$II$ DST for an odd symmetry, which both return $2N$ numbers (real or complex; the same kind as $f$). 
%As for the case of spectral directions, the Green's function has to be considered in a ``vertex-centered'' formalism. 
%However, the Green's function is here first evaluated as a function of the (doubled) spatial coordinate, and thus also needs to undergo an FFT. 
According to their definitions (\cref{eq:DCT2} for the DCT; \cref{eq:DST2} for the DST)
 when expressed in the case of a domain with $2N$ cells,
%\be
%%\hat{f}_k = 2 \sum_{j=1}^{n-1} f(x_j) \cos\left[ \pi \left(j+1/2\right) k / \left(n\right) \right]
%\tilde{f} = 2 \sum_{j=0}^{2N-1} f(x_j) \cos\left[ \pi \left(j+\frac12\right) \frac{\textrm{k}}{(2N)} \right]
%%\label{eq:DCT2}
%\eec
%or, using the type-$II$ DST, as
%\be
%\tilde{f} = 2 \sum_{j=0}^{2N-1} f(x_j) \sin\left[ \pi \left(j+\frac12\right) \frac{\textrm{k}+1}{(2N)} \right]
%%\label{eq:DST2_2}
%\eed
%One may notice that %, due to the cell-centered formalism, 
the flip-flop mode ($k=2N$) is trivially zero in the type-$II$ DCT, while the mode $k=0$ is trivially zero in the type-$II$ DST. 
%Indeed, when considering the definition of the cell-centered type-$II$ DCT, see \eqqref{eq:DCT2}, the flip-flop mode will can cancel the coefficients $\cos\left[ \pi \left(j+\frac12\right) \frac{2N}{(2N)} \right] = 0$. On the contrary, when considering the definition of the cell-centered type-$II$ DST, see \eqqref{eq:DST2}, the constant mode will also cancel the coefficients, $ \sin\left[ \pi \left(j+\frac12\right) \frac{0}{(2N)} \right] = 0$.
% Indeed, the flip-flop mode which is $k=2N$  $f \sim \cos\left[\pi \left(j+\frac12\right) \frac{2N+1}{(2N)}\right]$)
%   the mode $k=0$ of the type-$II$ DCT is non-zero as it corresponds to a constant in space and the flip-flop mode $k=2N+1$(i.e. $f \sim \cos\left[\pi \left(j+\frac12\right) \frac{2N+1}{(2N)}\right]$),  the result of the transform is zero by construction.
%Conversely, for the t, the mode $k=2N$ is the flip-flop in space while the result of the transform is null when $f$ is a constant.
%Hence in this transform the mean, i.e. the mode $k=0$ is null by definition, hence retrieving the missing information.
%One will refer to these comments in what follows.
%
On the other hand, the $2N+1$ numerical values of the Green's function (evaluated on the vertices of the interval $[0,2L]$, as in \fig{fig:BC:mixunb_Green}) are transformed using a type-$I$ DCT, which yields $2N+1$ values:
%The Green's function $\tilde{G}$, on the other hand, results from a type-$I$ DCT (with $2N+1$ vertex-centered data),
\be
%\hat{F} = F_{0} + \left( -1 \right)^k F_{n-1} + 2 \sum_{j=1}^{n-2} F_j \cos\left[ \pi j k / \left(n-1\right) \right]
\tilde{G} = G({x_0}) + \left( -1 \right)^\textrm{k} G(x_{2N}) + 2 \sum_{j=1}^{2N-1} G(x_j) \cos\left[ \pi j \left(\frac{\textrm{k}}{2N}\right) \right]
\eec
where the {k}{th} entry corresponds to mode $k$ and neither the constant mode $k=0$ nor the flip-flop mode $k=2N$ is null.
Hence particular care must be taken to match the entries of $\hat{f}$ and $\hat{G}$ with the same mode $k$ while performing the pointwise multiplication in Fourier space.
%
%The output of size $2N+1$ corresponds to the modes $k=0...2N$ of $G$, where 
%These correspond respectively to the  $0...2N$ modes for the DCT (because the $2N+1$ mode is implicitly null), and to the $1...2N+1$ modes for the DST (as the first mode is implicitly null).
%
By doing so, one finally obtains the field $\hat{\phi}$ of size $2N$, with the first or the last mode being trivially null and ready for the application of the backward \typo{transform}.
%from the implementation point of view, $\tilde{G}$ has to be resized to a size of $2N$ in order to perform the multiplication with $\tilde{f}$.  Considering the above comments, the mode $k=0$ is discarded in case of an even symmetry on the field, and the mode $k=2N$ in the case of an odd symmetry on the field, as they are not present in the output array $\tilde{f}$.
%Finally, the resulting $\tilde{f}$ and the resized $\tilde{G}$ have the same size and can be multiplied in Fourier space.

\subsubsection{Order of the transforms}
\label{sect:meth:order}
The order in which the directions are transformed to obtain the 3D FFT can be chosen arbitrarily. 
As the DCTs and DSTs are real to real transforms (for symmetric and semi-unbounded directions), one may want to avoid performing them on complex data, {i.e.} separately on the real and on the complex parts. 
Therefore, the symmetric directions are always treated first, in order to reduce the size and the cost of subsequent FFT evaluations (which can thus be real to real, or real to complex).
Then, the semi-unbounded directions are transformed, which yields real values on a doubled domain.
Finally, the FFTs in the periodic directions and in the unbounded directions are performed, using real to complex or complex to complex transforms, depending on the input data type.

FLUPS follows these directives to determine the actual order of executions of the FFTs. This order is stored in the \texttt{dimorder} array.

%---------------------------------------------------
\section{Additional notes on the parallel implementation}
\label{sec:implementation_note}

\subsection{Topologies and data layout}

As already mentioned, FLUPS uses topology objects in order to describe every field in the memory.
The topology specifies the useful information on the field, including % regarding its arrangement in memory
the global and local numbers of field data in each direction, the type of data ($\mathtt{n_f=1}$ for real data and $\mathtt{n_f=2}$ for complex data), the number of MPI ranks in each direction $N_c^x,N_c^y,N_c^z$, and the associated MPI communicator.
If the communicator is of type \texttt{MPI\_Cart}, the information on the mapping between the cores and the physical domain is retrieved from the communicator itself, otherwise it needs to be specified by the user.

Most importantly, every topology also has its own specific fastest rotating index (FRI), $\mathtt{ax_0}$, {i.e.} an integer indicating the axis ($X=0$, $Y=1$ or $Z=2$) along which data are contiguous in memory. The 3D data are stored sequentially in memory: data are decomposed in the order $\mathtt{ax_0},\mathtt{ax_1},\mathtt{ax_2}$, where the last two indices follow a natural permutation of indices, $\mathtt{ax_1 = \left( ax_0 + 1 \right) \% 3}$ and $\mathtt{ax_2 = \left( ax_0 + 2 \right) \% 3}$.
As an example, for a 3D data set of size ${N = \left[ N_x , N_y, N_z\right]}$ with an FRI aligned with $\mathtt{ax_0 = 1}$ (and thus with $\mathtt{ax_1 = 2}$ and $\mathtt{ax_2 = 0}$), the element located in ${\left(x=10, y=2, z=3\right)}$  will be located in memory
at $\mathtt{\left[ 2 \left( n_f \right) + 3 \left( N_y * n_f \right)  + 10 \left( N_z N_y * n_f \right) \right]}$.
%, where $\mathtt{n_f =1 }$ for real data and $\mathtt{n_f =2 }$ for complex datasets.

The size of the computational domain may be extended along the FRI in order to fulfil data alignment requirements. Indeed, only one FFTW plan is defined per pencil topology, and it is used to transform all the pencils in the domain. For this to be possible, the starting element of each pencil must be aligned in memory.
%the FFT is multi-threaded on the number of FFTs to perform and not on the FFT itself. 
Therefore, if the size of one pencil does not match a multiple of the memory chunk size, the domain is padded in the FRI dimension to ensure the correct alignment of all pencils.

In the other dimensions, the sizes of the pencils are chosen in order to ensure load balancing: the width of all pencils in a given direction differs by 1 at most.
For example, if the $x$ direction with $N_x=20$ was to be split between three cores, each core would be in charge of, respectively, 7, 7, and 6 data.

\subsection{Memory management and computational complexity}
In order to avoid repetitive large allocations of memory, the whole solver is designed to work ``in place''.
The total amount of memory required for the solve operation is allocated once at initialization. The size of the allocation, which corresponds to the maximum size that the data will occupy, is determined by the execution of a \emph{dry run}. This procedure accounts for the directions that will need zero-padding, and also predicts the type of FFT transform (real to real, real to complex, complex to complex) in order to determine the input and output sizes.
This information is used to create three topologies with pencils in each direction that will enable the series of 1D FFTs, to determine the optimal order \texttt{dimorder} in which these topologies will be used, and to prepare the switches between the successive topologies. 

Even though the size of the memory allocation corresponds to the maximum size of the problem (accounting for domain doubling), FFTs are only performed on the significant part\typo{, {i.e.,} not on} zero-padded data. %, determined by the topology object.
For instance, a 3D unbounded problem of size $N\times N\times N$ requires a total memory allocation of $(2N+2) \times (2N+2) \times (2N+2)$ floating point values, but the computational complexity associated with the forward (or backward) transform\footnote{The leading complexity of an FFT over a doubled domain is $2N \log(2N) = 2N (\log(2)+\log(N))$ thus still $\O{N \log(N)}$.} is as small as $7N^2 \O{N \log(N)}$. If the FFTs were all heedlessly computed on the full size domain $(2N+2)^3$, the complexity would be $12N^2 \O{N \log(N)}$---almost twice as large. This can also be compared to the cost of a fully periodic transform without padding, $3N^2 \O{N \log(N)}$.

As another example, a domain with odd-odd symmetry conditions in the first two dimensions, and fully unbounded in the third direction, leads to a memory allocation of $N\times N \times (2N+2)$ floating point values. First, $f$ is transformed twice in the symmetric directions using $N\times N$ DSTs of size $N$ (hence excluding the zero-padded region). Then, the fully unbounded direction is transformed using $N$ DFTs with $2N$ real numbers as an input and with $N+1$ complex numbers as an output (thus corresponding to $2N+2$ values). By explicitly discarding the zero-padded region for the first two DSTs, the total cost of one transform is  $4N^2 \; \O{N \log\left(N\right)}$.

%The order of execution of the 1D FFTs is arranged to start with real to real transforms: first, the odd-odd and even-even symmetries, then the semi-unbounded directions. Periodic directions are transformed next, and finally, the fully unbounded directions. This order of the FFT, $\mathtt{dimorder}$ will be reused to improve the topology layouts.

\subsection{Creation of topologies and topology switches}

Again, during the initialization of FLUPS, three pencil topologies dedicated to each of the 1D FFTs are first created. These topologies have an FRI aligned with the direction of the pencils.
For an illustration of the pencil decompositions, the reader is referred to  Figure~2 in \cite{Chatterjee:2018}, which follows the same idea.
Three switches are also set up. The first one is dedicated to remapping data from the physical topology to the first pencil topology, and the two others work as pencil-to-pencil.

%Again, the topology objects describe the decomposition of the computational domain. 
% When doing so, there is one degree of freedom on the number of processes assigned to each direction of the 2D pencil grid. This number is here chosen in order to maximize the number of interfaces between processes that are common in the origin and the destination topologies, for a reason explained below. %to ensure that the origin and the destination topologies in all topology switches will share interfaces.
%When going from one pencil topology to another, we make sure that the decompositions are compatible. 
% As shown in \fig{fig:switch_block:Pto0} for example, the physical topology (provided by the user, in red) and the first pencil topology (in blue) share one interface in $X$ and one in $Z$.
% In \fig{fig:switch_block:0to1}, the pencil decomposition in the $Y$ direction (blue topology) and the one in the $X$ direction (green topology) are such that both topologies have the same number of ranks in the $Z$ direction.
% By conserving a certain number of common interfaces, it is then possible to take advantage of  ``sub-communicators'', as highlighted in red in \fig{fig:switch_block:0to1}. The subset of processes located in a sub-communicator are to communicate in a all-to-all fashion during the execution of the topology switch, and the reduced size of the sub-communicator helps in lowering the communication latency on very large partitions.
%---------------------------------------------------
%Before executing any Fourier transform, FLUPS will first switch the data to the correct topology. 
%
When creating the new pencil topologies, FLUPS is free to choose the decomposition regardless of the physical topology. In particular, the number of pencils in each direction can be optimized based on the predetermined order of the switch operations (as described in \sect{sect:meth:order}).
%to reach the best configuration possible taking the switch operation into account.
%
This number is here chosen in order to take advantage of the interfaces between MPI ranks that are common in the origin and the destination topologies. 
Indeed, the common interfaces partition the domain into subsets of ranks: if a rank belongs to a subset in the input topology, it belongs to the same subset in the output topology. Moreover, each of the subsets contains a smaller number of ranks than the total $N_c^x\,N_c^y\,N_c^z$. Inside the subset, they are to communicate in an \textit{all-to-all} fashion during the execution of the topology switch.
%In p2p switches, a given process only communicates with $\sqrt{N_c}$ other processes. 
One \textit{sub-communicator} can thus be created per subset, and the reduced size helps in lowering the communication latency on very large partitions.
%
%When switching from one topology to another, one direction can be chosen freely, we call it the \textit{common direction}.
%Therefore, by choosing carefully the number of ranks in the common direction, we are able to define sub-communicators and hence reduce the communication latency on very large partitions.
%
For this to work, one further needs to ensure that rank numbering is consistent in both input and output topology communicators.
% In FLUPS, the binding between a given rank index and its 3D location relies on the $\textit{rank split}$ operation, which determines how the rank indices are incremented in each direction. 
%  Here, for all pencil topologies, the $\textit{rank split}$ is done following the $\mathtt{dimorder}$: this satisfies the requirement for the sub-communictors to be effective. 
%  %By doing so, we make sure that a processor will remain in the same ``layer'' in both topologies hence some data will not have to be communicated.

\begin{figure}[h!]
\begin{minipage}[b]{0.475\textwidth}
\centering
\includegraphics[width=\textwidth]{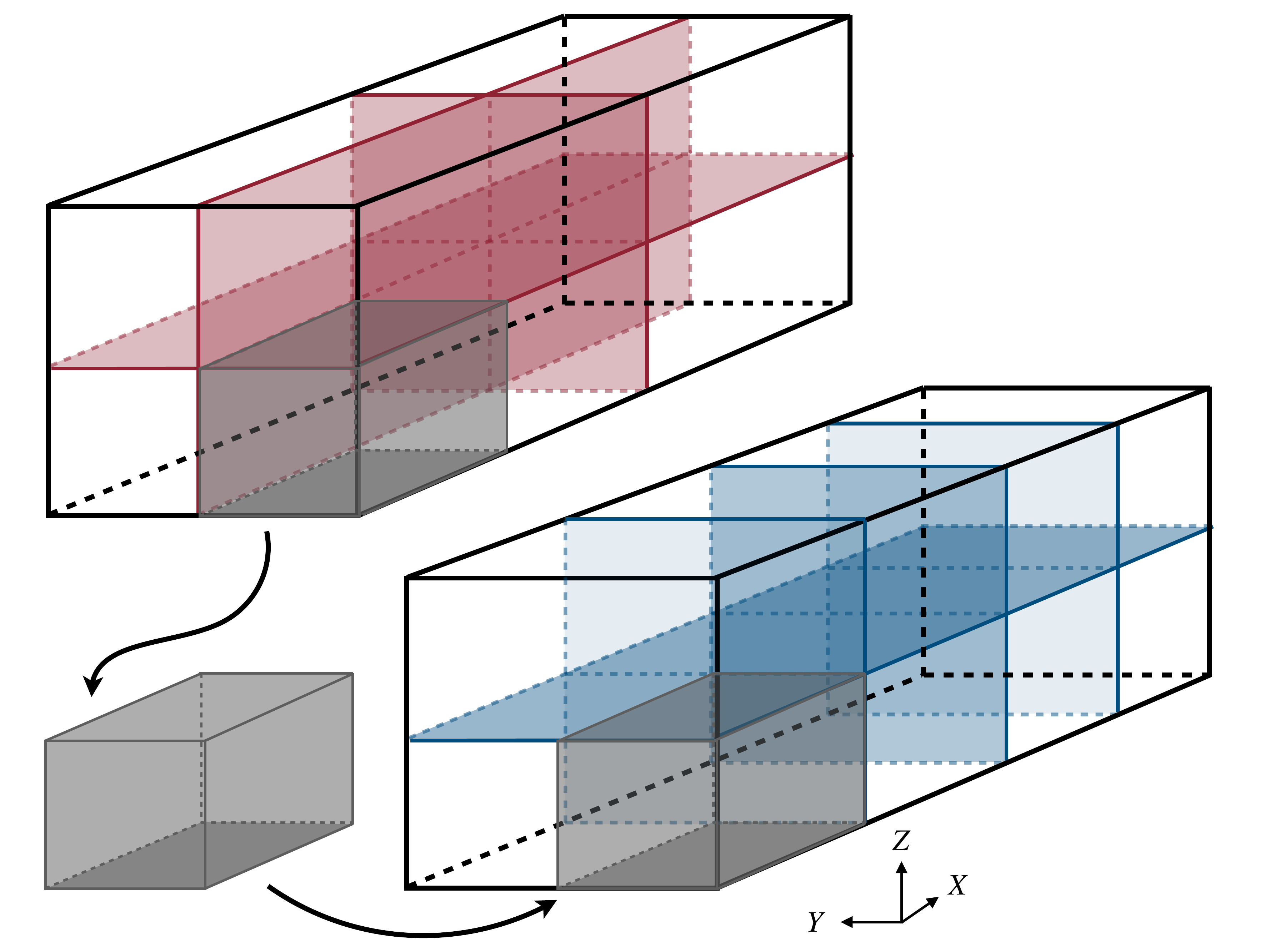}      
\subcaption{\label{fig:switch_block:Pto0} Switch from the physical topology (in red) to the first pencil topology aligned in the $Y$ direction (in blue). The grey area represents interface between ranks, highlighted for interfaces shared by both topologies.}
\end{minipage}%
\hspace{0.04\textwidth}
\begin{minipage}[b]{0.475\textwidth}
\centering
\includegraphics[width=\textwidth]{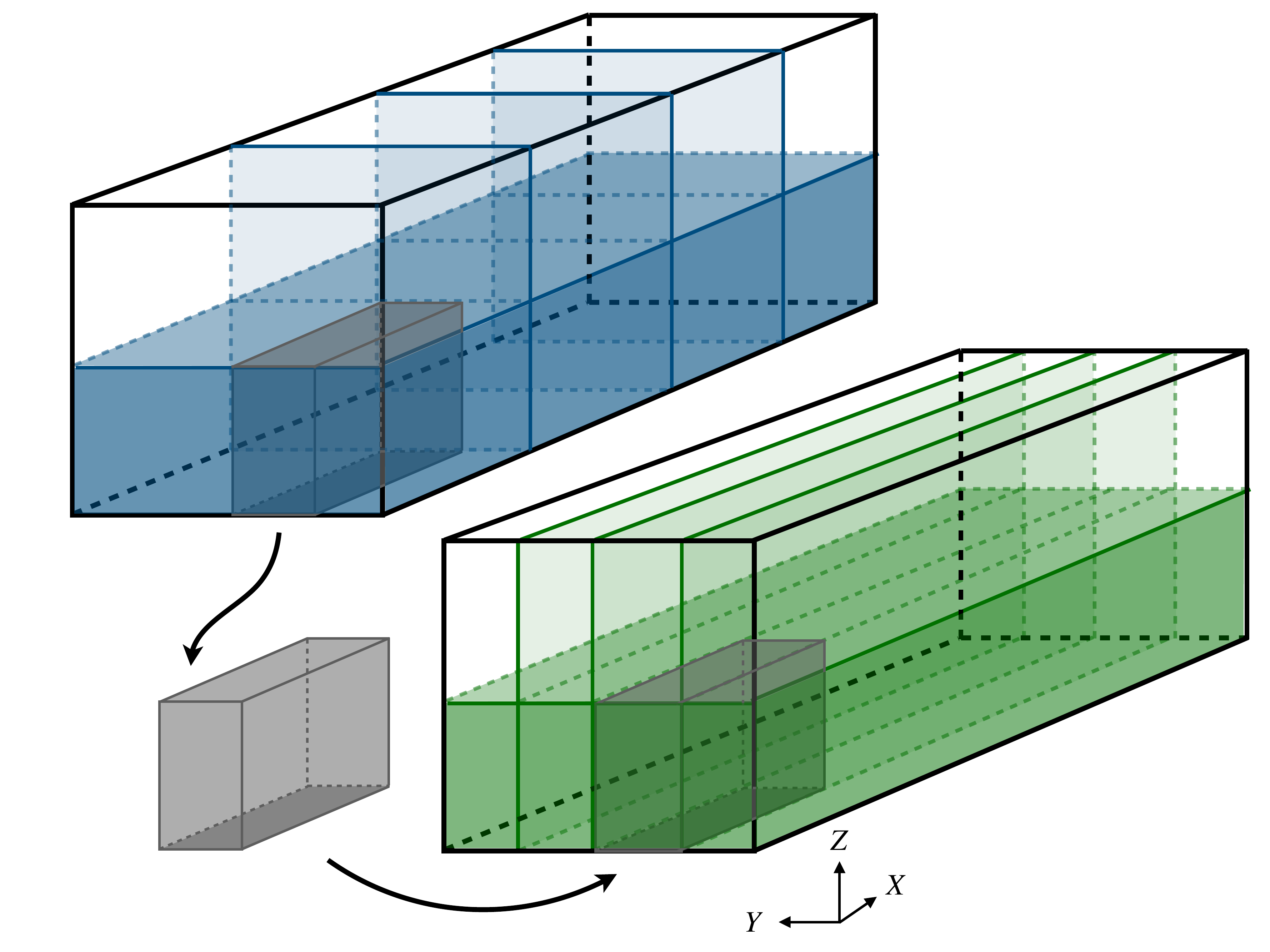}      
\subcaption{\label{fig:switch_block:0to1} Switch form the first pencil topology aligned in the $Y$ direction (in blue) to the second pencil topology aligned in the $X$ direction (in green). One of the two sub-communicators associated to this topology switch is highlighted in the corresponding topology's color.}
\end{minipage}
\caption{ Illustration of the topology switches that FLUPS would define starting from a physical topology with 8 MPI ranks, 2 in each direction (in red). Examples of communication blocks are shown in grey. }
\label{fig:switch_block}
\end{figure}

Consider a physical topology with the number of ranks per direction $\mathtt{[N_c^x=2,N_c^y=2,N_c^z=2]}$, and a $\mathtt{dimorder = [1 , 0 , 2]}$, as illustrated in \fig{fig:switch_block:Pto0}.
First, the rank IDs are split following the 3D decomposition given by $\mathtt{dimorder}$, which yields a 3D index. For example, the rank $2$ will have the 3D index $\mathtt{[1,0,0]}$, since $\mathtt{0 + 1*2 + 0 * 4 = 2}$. This is to ensure that a rank which is located in one sub-communicator will remain in the same sub-communicator in the next topology. Then, FLUPS sets up the pencil decomposition: the sizes of the first two directions in the $\mathtt{dimorder}$ array are multiplied, while the last one is left unchanged. In this case, $\mathtt{[N_c^x=4,N_c^y=1,N_c^z=2]}$, hence creating $4$ sub-communicators because of the 2 common interfaces.
Afterwards, FLUPS permutes the number of ranks to reach the second pencil decomposition, $\mathtt{[N_c^x=1,N_c^y=4,N_c^z=2]}$, creating $2$ sub-communicators, as depicted in \fig{fig:switch_block:0to1}.
Finally FLUPS would permute the last two directions of the $\mathtt{dimorder}$ array to obtain $\mathtt{[N_c^x=2,N_c^y=4,N_c^z=1]}$, creating $4$ sub-communicators (not presented in the figure). 
This strategy ensures that the last topology switch, which should involve the largest volume of transferred data due to the padding for unbounded directions, uses as many sub-communicators as possible (and thus decoupled communication patterns).

%While seting up the switch between topologies, we define sub-communicators between ranks as there is always a common boundary between the cores in each of the topologies.

MPI is used to communicate between the cores when switching from one topology to another. 
Prior to the communication, data in memory are grouped in blocks, defined as the largest possible subset of the domain that will be either sent to or received from a given rank (as shown in grey in \fig{fig:switch_block}). During the initialization of FLUPS, the size of those blocks and the communication patterns (origin and destination of each blocks) are determined and stored.
The maximum total size of the communication buffers is also precomputed so that only one memory allocation is performed, and they are reused for the three topology switches, akin to what is done for the field data.

The execution of a topology switch starts with the copy of the blocks into the send buffers. %, dedicated to the communication. 
As the blocks are made of several chunks of contiguous memory, this operation benefits from autovectorization along the FRI.
The start address of each chunk is not necessarily aligned in memory,  but when possible, instructions for aligned memory access are issued.
%Here also, we make sure that the total size of the communication buffer is a multiple of the alignement to ensure aligned memory access when copying from/to the field.
%Yet, since the length of one block in the FRI direction is not a-priori a multiple of the alignement, the starting pointer of the copying loop is not always aligned. As this would imply a very large memory overhead we do not pad the sizes, still we take advantage of it if the vector is aligned, due to a convenient domain decomposition among the ranks.
%
When the send buffers are filled, one proceeds to the communication. More details on this operation are provided hereafter. As a result, new data are available in receive buffers.
Since the FRI changes between the origin and the destination topologies, these data need to be transposed in memory to realign them in the correct direction. To do so, one option would consist in looping over the indices of the destination topology and simply copying back values from the buffer to the field. This, however, would result in slow, non-unit stride memory accesses. 
Instead, one here rather resorts to the FFTW library to efficiently ``shuffle'' the data in the buffer, through the functions \texttt{fftw\_plan\_guru\_r2r} for real data and %the function
\texttt{fftw\_plan\_guru\_dft} for complex data. 
The data can then be copied from the buffers back into the field, again benefiting from autovectorization and, potentially, aligned access. %The combination of shuffling and vertorized copy leads to a faster execution.

\subsection{Communication strategies}
For the communication in itself, two different methods are made available to the user, as the optimal communication scheme will depend on the implementation of the MPI standard as well as on the targeted hardware. The first communication method is implemented using calls to the \texttt{MPI\_alltoall} and \texttt{MPI\_alltoallv} routines. When all blocks across a whole sub-communicator have the same size, FLUPS automatically uses the \texttt{MPI\_alltoall} routine, as it is believed to sometimes outperform its variable-size counterpart thanks to more advanced internal optimizations. 
%We help FLUPS using the latest by padding the buffers to get the same size among processor to take advantage of the \texttt{MPI\_alltoall} routine. 
%as it was shown on Cray XXXX that its performance are substantially better than \texttt{MPI\_alltoallv}, because the latter is further optimized. XXX \todo{be maybe more subtle}.
%However, a call to \texttt{MPI\_alltoall} is only possible when the number of data sent to, and received by each process is the same (inside one sub-communicator).
%Practically, this is always true for pencil to pencil communications, even though it may require to pad some communication buffers to an identical size among each sub-communicator. Indeed, the last processor in a given direction may hold either one more plane of data compared to the others, or a few less.
%For example, if a topology has two (resp. three) pencils in a direction with 65 complex data, the first pencil holds 32 complex data while the second owns 33 complex data (respectively 22, 22 and 21 for the first, second and third pencils). The block size is increased to 33 complex values (resp. 22) on both processes (with one additional row of 0 on the first pencil), only for the purpose of communication, in order to use the \texttt{MPI\_alltoall} implementation.
In all other cases, the \texttt{MPI\_alltoallv} routine is used.

In the second method, all communications are handled using MPI \textit{non-blocking} persistent send and receive calls, detailing explicitly the origin, the destination and the size of each block.
The communication patterns are precomputed once and stored at initialization, hence sparing some time during the execution of the switches.
Even though point-to-point send and receive operations are generally expected to be slower than the \textit{all-to-all} communication (which benefits from global network optimization), a slight performance gain is here expected from the \textit{non-blocking} aspect. Indeed, as soon as a block is received by a core ({i.e.,} when the \texttt{MPI\_waitany} routine returns), the copy from the corresponding receive buffer to the main memory can start, while the process is still waiting to receive the other blocks. 
In practice, as soon as the communication is initiated, each process starts by copying the block that remains on that rank (if there is one).

\medskip

FLUPS is able to exploit shared-memory parallelism, when the user requires threads.
If so, all the copy operations (from/to buffers for instance) are accelerated with OpenMP \textit{parallel for} instructions.
In addition, OpenMP is also used to parallelize the FFTs. The total number of 1D transforms to be performed in a pencil topology is split between the available threads.

As FLUPS supports hybrid parallelism, for a given number of cores, an optimum is likely to be found between the number of threads and the number of MPI ranks. Generally, for a decreasing number of MPI ranks (\textit{i.e.} an increasing number of threads), the size of the blocks increases and one expects the communication latency to decrease, up to a point where the transfer speed does not depend on the size of the message.
The communication latency mainly depends on the network architecture of the targeted computer, and on the network congestion (see \sect{sec:scalability} for timings).

% \medskip

%Finally, given the order of execution, we also adapt the mapping between one MPI-rank and its locations on the 3D mesh, i.e. its 3D rank index. By doing so, we make sure that during the pencil transposition one rank will always send as much data to himself as possible, hence decreasing the communication cost. F

\subsection{Rank reordering}
An additional optimization of FLUPS was designed to potentially adapt the distribution of the MPI rank indices to the actual network architecture. %As shown in \todo{cite Chan:2008}, matching these lead to substantial increase in performance.
As the communication pattern between the ranks is known {a-priori} for each topology switch, one may want to optimize globally the position of each rank in the hardware communication network, in order to reduce the congestion of the network during the communications, and to avoid any bottleneck. Such an optimization reduces to permuting the rank indices between cores in the physical topology adequately, resulting in a given rank (associated to a given portion of the 3D domain) located at the desired location in the network. 
The determination of the optimal permutation---which should account for a specified communication graph, for the actual network architecture and for the resources allocated---is beyond the scope of this paper.
However, according to the MPI specifications, this task is supposedly achieved by the \texttt{MPI\_Dist\_graph\_create\_adjacent} routine when the reordering of the ranks is requested.
To exploit it, the communication graph between the MPI ranks is established at the initialization of  FLUPS: an edge exists between ranks $i$ and $j$ if they need to communicate during one of the pencil to pencil topology switches, and the associated edge weight is computed as the total amount of data that will be exchanged during the forward and backward transforms.
Based on the weighed graph, \texttt{MPI\_Dist\_graph} returns a new communicator where the rank indices are ``smartly'' permuted. 
FLUPS then recomputes the communication pattern of the first topology switch in order to ensure that the data in the user topology will be routed to the correct core with the new numbering in the first pencil topology. As a result, a bit more data needs to be exchanged during the execution of the first topology switch, between the user-specified physical topology with the original communicator and the first pencil topology with the new communicator.
The potential increase in communication delay related to this operation should be balanced by the gain over the subsequent topology switches.
Optionally, the optimized communicator can be retrieved by the user so that it can be used from the beginning of the computation on the user side, thus avoiding the data rerouting of the first topology switch.

FLUPS is thus ready to exploit such an optimization. % when it will be actually operational.
In practice, however, it was not possible to demonstrate the efficiency of this feature. At the time of this writing, the authors could not find an MPI implementation with documentation on the reordering. Several tests were executed on different machines using Intel's implementation of MPI (see \sect{sec:scalability}), but none of them led to rank indices actually permuted at the output of \texttt{MPI\_Dist\_graph}.
%Yet, it is not excluded that reordering be part of the implementation in later (or hardware-specific) versions.
The result presented in the following sections are hence obtained with the canonical rank ordering.

\section{Validation}
\label{sec_validation}

In this section, an extensive validation of FLUPS is presented for the Poisson equation (reproduced from \eq{eq:intro:Poisson})
\be
\nabla^2 \phi = f
\nonumber
\eec
with various combinations of BCs, in 3D.
The convergence of the solver is measured using the infinite norm of the error \typo{(as the infinite norm bounds the 2-norm)}, defined as
\be
%E_{\infty} = \max \abs{\phi(x,y,z) - \phi_{ref}(x,y,z)}
E_{\infty} = \sup_{x,y,z} \left\lbrace \abs{\phi(x,y,z) - \phi_{ref}(x,y,z)} \right\rbrace
\eec
where $\phi_{ref}$ is an analytical solution. For the latter, a composite expression is employed:
\revtwo{
\be
\phi_{ref}(x,y,z) = \phi_x(x)\phi_y(y)\phi_z(z)
\eed}%
The 1D functions \revtwo{$\phi_x$, $\phi_y$, and $\phi_z$} are chosen \revthree{to satisfy the set of BCs. Simple analytic expressions are preferred so that meaningful numerical solutions can be obtained even with small resolutions, hence allowing us to verify the slopes of the convergence more easily%
%circumventing the problems associated with under-resolved behavior
}.
For instance, sine and cosine are used in the periodic and symmetric directions, with an appropriate wavelength.
Fully unbounded solutions are guaranteed with a compact Gaussian function, {i.e.,} \revtwo{$\phi_x(x) = \exp { \cC \left(1 - \frac{1} { \left(1 - \left(\frac{x-x_c}{\sigma}\right)^2\right)} \right) }$} for $-1<\frac{x-x_c}{\sigma}<1$, and \revtwo{$\phi_x=0$} elsewhere. 
The dimensional parameter $\sigma$ and the center of the blob $x_c$ are set so that the value of the RHS at the unbounded boundaries is null. %$\frac{\sigma}{L} = 10$
\revtwo{The constant $\cC$ is arbitrarily set to $10$.}
For semi-infinite directions, a linear combination of compact Gaussian functions is used. The parameter $x_c$ and the sign of each term are chosen in order to explicitly impose the symmetric BC. Given these analytical expression\typo{s}, the \typo{RHS} $f$ is computed as
\revtwo{
\be
f(x,y,z) = \dfrac{d^2\phi_x}{dx^2}(x) \; \phi_y (y) \phi_z (z) +  \phi_x (x)\; \dfrac{d^2\phi_y}{dy^2} (y)  \; \phi_z(z) + \phi_x(x) \phi_y(y) \; \dfrac{d^2 \phi_z}{dz^2}(z) 
\eed}

In FLUPS, the BCs in a single direction can be set to periodic, or to a pair of options among even symmetry, odd symmetry, and unbounded.
This leads to a thousand possible combinations of BCs in 3D.
For the sake of concision, only three cases of validation are presented here.
All the other configurations were tested and validated as part of an automated process during the development of the library.
For the interested reader, FLUPS provides a C++ sample code dedicated to the test and the validation of the library.  When built, the associated executable can be easily used to reproduce the convergence results.

\subsection{Domain with symmetric and periodic BCs}%001033

The Poisson equation is here solved on a cubic domain of spatial extent $[0,L]$ in all directions, with even-even symmetry conditions in $X$, odd-even symmetry conditions in $Y$,  and periodic in $Z$.
The chosen reference solution is 
\be
\phi_{ref}(x,y,z) = \cos\left(2\pi \frac{x}{L}\right) \cos\left(\frac{9\pi}{2} \frac{y}{L}\right) \sin\left(8\pi \frac{z}{L}\right)
\eed

\begin{figure}[!h]
%\begin{minipage}{0.48\textwidth}
\centering
\pgfplotsset{xtick=data}
\pgfplotsset{xticklabel={
    \pgfkeys{/pgf/fpu=true}
    \pgfmathparse{int(10^\tick +1)}
    \pgfmathprintnumber[fixed]{\pgfmathresult}$^3$
}, xticklabel style={rotate=-45,anchor=west}}
% This file was created by tikzplotlib v0.8.6.
\begin{tikzpicture}

\definecolor{color0}{rgb}{0.12156862745098,0.466666666666667,0.705882352941177}
\definecolor{color1}{rgb}{1,0.498039215686275,0.0549019607843137}
\definecolor{color2}{rgb}{0.172549019607843,0.627450980392157,0.172549019607843}
\definecolor{color3}{rgb}{0.83921568627451,0.152941176470588,0.156862745098039}
\definecolor{color4}{rgb}{0.580392156862745,0.403921568627451,0.741176470588235}
\definecolor{color5}{rgb}{0.549019607843137,0.337254901960784,0.294117647058824}
\definecolor{color6}{rgb}{0.890196078431372,0.466666666666667,0.76078431372549}

\begin{axis}[
log basis x={10},
log basis y={10},
tick align=outside,
tick pos=left,
x grid style={white!69.01960784313725!black},
xlabel={N},
xmin=18.830018349522, xmax=3915.45024712477,
xmode=log,
xtick style={color=black},
y grid style={white!69.01960784313725!black},
ylabel={\(\displaystyle E_\infty\)},
ymajorgrids,
ymin=3.17221134457424e-16,
ymax=9.14240132575,
ymode=log,
ytick style={color=black}
]
\addplot [semithick, color0, mark=*, mark size=2, mark options={solid}]
table {%
32 2.442490654175e-15
64 1.7763568394e-15
128 2.22044604925e-15
256 1.817990202824e-15
512 2.22044604925e-15
1024 2.22044604925e-15
2048 2.275957200482e-15
};
\addplot [semithick, color1, mark=x, mark size=3, mark options={solid}]
table {%
32 0.04416171665067
64 0.01150336024537
128 0.00290453229755
256 0.0007279226500416
512 0.0001820924536398
1024 4.553009946129e-05
2048 1.138296147984e-05
};
\addplot [semithick, white!75.29411764705883!black, dashed]
table {%
24 0.0572611709991751
3072 3.49494451899262e-06
};
\addplot [semithick, color2, mark=o, mark size=2, mark options={solid}]
table {%
32 0.6888865984988
64 0.2849285094917
128 0.08185603319222
256 0.02120553912629
512 0.005349052056858
1024 0.0013402633303
2048 0.0003352536840311
};
\addplot [semithick, white!75.29411764705883!black, dashed]
table {%
24 1.41831080280313
3072 8.65668214601519e-05
};
\addplot [semithick, color3, mark=square, mark size=2, mark options={solid}]
table {%
32 0.3680443817142
64 0.04612317698417
128 0.003463043051366
256 0.0002267242035039
512 1.43361655337e-05
1024 8.986234374397e-07
2048 5.620487697477e-08
};
\addplot [semithick, white!75.29411764705883!black, dashed]
table {%
24 1.632646580852
3072 6.08208246846498e-09
};
\addplot [semithick, color4, mark=square*, mark size=2, mark options={solid}]
table {%
32 0.1477130975034
64 0.005124609124962
128 9.837747603336e-05
256 1.618947743309e-06
512 2.562666945316e-08
1024 4.017191024275e-10
2048 6.282530051749e-12
};
\addplot [semithick, white!75.29411764705883!black, dashed]
table {%
24 1.28994495763762
3072 2.9329952613753e-13
};
\addplot [semithick, color5, mark=triangle, mark size=2, mark options={}]
table {%
32 0.04684134546591
64 0.0004321327639693
128 2.102078505595e-06
256 8.676418028486e-09
512 3.436317896899e-11
1024 1.353361867018e-13
2048 2.553512956638e-15
};
\addplot [semithick, white!75.29411764705883!black, dashed]
table {%
24 0.773508505601733
3072 1.07345869082805e-17
};
\addplot [semithick, color6, mark=triangle*, mark size=2, mark options={solid}]
table {%
32 0.01220562500222
64 2.932599141148e-05
128 3.598487119394e-08
256 3.72130104509e-11
512 3.752553823233e-14
1024 2.275957200482e-15
2048 2.22044604925e-15
};
\addplot [semithick, white!75.29411764705883!black, dashed]
table {%
24 0.373282874496428
3072 3.16182893344267e-22
};
\end{axis}

\end{tikzpicture}
%\subcaption{All to all communication}
%\end{minipage}%
%\hspace{0.04\textwidth}%
\caption{\label{fig:conv:001033}Convergence of FLUPS  with symmetric and periodic BCs: %
\protect\ThinLineCircle{pyBlue}{pyBlue} \texttt{CHAT2} and \texttt{HEJ0} spectral, %
\protect\ThinLineCross{pyOrange} \texttt{LGF2} $\O{h^2}$,%
\protect\ThinLineCircle{pyGreen}{none} \texttt{HEJ2} $\O{h^2}$,%
\protect\ThinLineSquare{pyRed}{none} \texttt{HEJ4} $\O{h^4}$, %
\protect\ThinLineSquare{pyPurple}{pyPurple} \texttt{HEJ6} $\O{h^6}$, %
\protect\ThinLineTrianUp{pyBrown}{none} \texttt{HEJ8} $\O{h^8}$, %
\protect\ThinLineTrianUp{pyRose}{pyRose} \texttt{HEJ10} $\O{h^{10}}$.}
\end{figure}

The error obtained with FLUPS is measured in terms of infinite norm for all the implemented Green\typo{'s} functions:  the singular expression (\texttt{CHAT2}, \revthree{which is in the present case equivalent to \texttt{HEJ0}}), the regular expressions with order up to $m=10$ (\texttt{HEJ2},\texttt{HEJ4},\texttt{HEJ6}, \texttt{HEJ8} and \texttt{HEJ10}), and the lattice Green\typo{'s} function \texttt{LGF2}. Their mathematical expressions can be found in \app{sect:app:3dirspe}.
The results are shown in \fig{fig:conv:001033}, as a function of the resolution per direction of the domain.
Convergence is indeed observed, at the order corresponding to the theoretical order of each method. As the present test case is fully spectral, the singular Green's function used by \texttt{CHAT2} provides the exact (spectral) solution, and the error thus amounts to the machine precision regardless of the resolution.

\subsection{Domain with three fully unbounded BCs} %444444

Unbounded BCs are now used in all directions, on the same domain.
The reference solution is 
\be
\phi_{ref}(x,y,z) = \exp { 10 \left( 3 - \frac{1} { \left(1 - \left(\frac{2x}{L}-1\right)^2\right)} - \frac{1} { \left(1 - \left(\frac{2y}{L}-1\right)^2\right)} - \frac{1} { \left(1 - \left(\frac{2z}{L}-1\right)^2\right)} \right)  }
\eed

The convergence results obtained with the Green\typo{'s} functions (recalled in \app{sect:app:0dirspe})  are shown in \fig{fig:conv:444444}. Again, we retrieve the expected convergence behavior. \revthree{Moreover, the additional spectrally truncated \texttt{HEJ0} kernel indeed exhibits spectral-like convergence (saturating at the machine precision). }

\begin{figure}[!h]
\begin{minipage}{0.48\textwidth}
\centering
\pgfplotsset{xtick=data}
\pgfplotsset{xticklabel={
    \pgfkeys{/pgf/fpu=true}
    \pgfmathparse{int(10^\tick +1)}
    \pgfmathprintnumber[fixed]{\pgfmathresult}$^3$
}, xticklabel style={rotate=-45,anchor=west}}
% This file was created by tikzplotlib v0.8.6.
\begin{tikzpicture}

\definecolor{color0}{rgb}{0.12156862745098,0.466666666666667,0.705882352941177}
\definecolor{color1}{rgb}{1,0.498039215686275,0.0549019607843137}
\definecolor{color2}{rgb}{0.172549019607843,0.627450980392157,0.172549019607843}
\definecolor{color3}{rgb}{0.83921568627451,0.152941176470588,0.156862745098039}
\definecolor{color4}{rgb}{0.580392156862745,0.403921568627451,0.741176470588235}
%\definecolor{color5}{rgb}{0.580392156862745,0.43,0.27}
\definecolor{color5}{rgb}{0.549019607843137,0.337254901960784,0.294117647058824}
\definecolor{color6}{rgb}{0.890196078431372,0.466666666666667,0.76078431372549}

\begin{axis}[
log basis x={10},
log basis y={10},
tick align=outside,
tick pos=left,
x grid style={white!69.01960784313725!black},
xlabel={N},
xmin=18.830018349522, xmax=3915.45024712477,
xmode=log,
xtick style={color=black},
y grid style={white!69.01960784313725!black},
ylabel={\(\displaystyle E_\infty\)},
ymajorgrids,
%ymin=7.25941560455858e-15,
ymin=2.0e-15,
ymax=2.42746735762173,
ymode=log,
ytick style={color=black}
]
\addplot [semithick, color0, mark=*, mark size=2, mark options={solid}]
table {%
32 0.05041207253504
64 0.01307047437423
128 0.00329741387624
256 0.0008262246986102
512 0.0002066732633652
1024 5.167563588149e-05
2048 1.291936649972e-05
};
\addplot [semithick, white!75.29411764705883!black, dashed]
table {%
24 0.065061916885056
3072 3.97106426300391e-06
};
\addplot [semithick, color1, mark=x, mark size=3, mark options={solid}]
table {%
32 0.01084115333534
64 0.002776516073706
128 0.0006983047462505
256 0.0001748377430271
512 4.372579200684e-05
1024 1.093247040862e-05
2048 2.733181504899e-06
};
\addplot [semithick, white!75.29411764705883!black, dashed]
table {%
24 0.0138208800113365
3072 8.43559571004427e-07
};
\addplot [semithick, color2, mark=o, mark size=2, mark options={solid}]
table {%
32 0.3301075527807
64 0.1066411041943
128 0.02860155157511
256 0.00728016419432
512 0.00182829182276
1024 0.0004575908445825
2048 0.0001144301139446
};
\addplot [semithick, white!75.29411764705883!black, dashed]
table {%
24 0.530835718656071
3072 3.23996410312543e-05
};
\addplot [semithick, color3, mark=square, mark size=2, mark options={solid}]
table {%
32 0.09135896990668
64 0.008679550799164
128 0.000606213640304
256 3.897028542887e-05
512 2.452905218164e-06
1024 1.535777173212e-07
2048 9.602849404544e-09
};
\addplot [semithick, white!75.29411764705883!black, dashed]
table {%
24 0.307234667300778
3072 1.14453832544676e-09
};
\addplot [semithick, color4, mark=square*, mark size=2, mark options={solid}]
table {%
32 0.02194920130034
64 0.0005800217852716
128 1.03706335135e-05
256 1.67623956937e-07
512 2.641426632977e-09
1024 4.135947140327e-11
2048 6.487033132885e-13
};
\addplot [semithick, white!75.29411764705883!black, dashed]
table {%
24 0.146000633216415
3072 3.31967005914556e-14
};

\addplot [semithick, color5, mark=triangle, mark size=2, mark options={solid}]
table {%
32 0.004587666251222
64 3.190015644527e-05
128 1.439328423114e-07
256 5.878966814234e-10
512 2.386509826563e-12
1024 1.132427485118e-14
2048 3.441691376338e-15
};
\addplot [semithick, white!75.29411764705883!black, dashed]
table {%
24 0.0571006051792807
3072 7.92430082375849e-19
};
%\addplot [semithick, white!75.29411764705883!black, dashed]
%table {%
%32 0.0025
%512 5.820766091E-13
%};

\addplot [semithick, color6, mark=triangle*, mark size=2, mark options={solid}]
table {%
32 0.0008152503057831
64 3.174198293019e-06
128 8.483970323472e-09
256 1.256469452577e-11
512 1.396300609857e-14
1024 3.441691376338e-15
2048 3.552713678801e-15
};
\addplot [semithick, white!75.29411764705883!black, dashed]
table {%
24 0.0404035398638205
3072 3.42231294503585e-23
};
%\addplot [semithick, white!75.29411764705883!black, dashed]
%table {%
%32 0.0005
%256 4.656612873E-13
%};

\addplot [semithick, black, mark=diamond*, mark size=2, mark options={solid}]
table {%
%32 3.719969232286e-01 
%64 4.041981334386e-03 
%128 5.714149349179e-06 
%256 3.085912145818e-09 
%512 7.019371931858e-14 
%1024 3.441691376338e-15 
%2048 3.441691376338e-15 
32 2.808120694742e-08 
64 1.506400559466e-12 
128 2.331468351713e-15 
256 3.552713678801e-15 
512 3.552713678801e-15 
1024 3.441691376338e-15 
2048 3.441691376338e-15 
};
\end{axis}

\end{tikzpicture}
%\subcaption{All to all communication}
%\end{minipage}%
%\hspace{0.04\textwidth}%
\caption{\label{fig:conv:444444}Convergence of FLUPS  with fully unbounded BCs: %
\protect\ThinLineCircle{pyBlue}{pyBlue} \texttt{CHAT2} $\O{h^2}$, %
\protect\ThinLineCross{pyOrange} \texttt{LGF2} $\O{h^2}$,%
\protect\ThinLineCircle{pyGreen}{none} \texttt{HEJ2} $\O{h^2}$, %
\protect\ThinLineSquare{pyRed}{none} \texttt{HEJ4} $\O{h^4}$, %
\protect\ThinLineSquare{pyPurple}{pyPurple} \texttt{HEJ6} $\O{h^6}$, %
\protect\ThinLineTrianUp{pyBrown}{none} \texttt{HEJ8} $\O{h^8}$, %
\protect\ThinLineTrianUp{pyRose}{pyRose} \texttt{HEJ10} $\O{h^{10}}$, %
\protect\ThinLineDiamondD{black}{black} \texttt{HEJ0} spectral-like.
}
\end{minipage}
\hfill
%\begin{figure}[!h]
\begin{minipage}{0.48\textwidth}
\centering
\pgfplotsset{xtick=data}
\pgfplotsset{xticklabel={
    \pgfkeys{/pgf/fpu=true}
    \pgfmathparse{int(10^\tick +1)}
    \pgfmathprintnumber[fixed]{\pgfmathresult}$^3$
}, xticklabel style={rotate=-45,anchor=west}}
% This file was created by tikzplotlib v0.8.6.
\begin{tikzpicture}

\definecolor{color0}{rgb}{0.12156862745098,0.466666666666667,0.705882352941177}
\definecolor{color1}{rgb}{1,0.498039215686275,0.0549019607843137}
\definecolor{color2}{rgb}{0.172549019607843,0.627450980392157,0.172549019607843}
\definecolor{color3}{rgb}{0.83921568627451,0.152941176470588,0.156862745098039}
\definecolor{color4}{rgb}{0.580392156862745,0.403921568627451,0.741176470588235}
\definecolor{color5}{rgb}{0.549019607843137,0.337254901960784,0.294117647058824}
\definecolor{color6}{rgb}{0.890196078431372,0.466666666666667,0.76078431372549}

\begin{axis}[
log basis x={10},
log basis y={10},
tick align=outside,
tick pos=left,
x grid style={white!69.01960784313725!black},
xlabel={N},
xmin=18.830018349522, xmax=3915.45024712477,
xmode=log,
xtick style={color=black},
y grid style={white!69.01960784313725!black},
ylabel={\(\displaystyle E_\infty\)},
ymajorgrids,
%ymin=7.30325273352186e-15,
ymin=2.0e-15,
ymax=2.43826808122802,
ymode=log,
ytick style={color=black}
]
\addplot [semithick, color0, mark=*, mark size=2, mark options={solid}]
table {%
32 0.0518235237425
64 0.01313059668256
128 0.003303087620114
256 0.0008264614306819
512 0.0002066954661387
1024 5.167656107852e-05
2048 1.291945324067e-05
};
\addplot [semithick, white!75.29411764705883!black, dashed]
table {%
24 0.0653611923754098
3072 3.98933058931944e-06
};
\addplot [semithick, color1, mark=x, mark size=3, mark options={solid}]
table {%
32 0.0111335471897
64 0.002789357310967
128 0.0006996245756501
256 0.0001749302587223
512 4.374138332996e-05
1024 1.093568124999e-05
2048 2.733938714972e-06
};
\addplot [semithick, white!75.29411764705883!black, dashed]
table {%
24 0.0138848008368135
3072 8.47460988575043e-07
};
\addplot [semithick, color2, mark=o, mark size=2, mark options={solid}]
table {%
32 0.3387511211165
64 0.1071232888768
128 0.0286505539575
256 0.007282247883082
512 0.001828488182131
1024 0.0004575990366926
2048 0.0001144308822054
};
\addplot [semithick, white!75.29411764705883!black, dashed]
table {%
24 0.533235926853404
3072 3.25461381136111e-05
};
\addplot [semithick, color3, mark=square, mark size=2, mark options={solid}]
table {%
32 0.09429902077119
64 0.008727131283673
128 0.0006074661120887
256 3.898371594702e-05
512 2.453222315402e-06
1024 1.535810280062e-07
2048 9.602928896513e-09
};
\addplot [semithick, white!75.29411764705883!black, dashed]
table {%
24 0.308918898969719
3072 1.15081257734343e-09
};
\addplot [semithick, color4, mark=square*, mark size=2, mark options={solid}]
table {%
32 0.02274303800378
64 0.0005834824061727
128 1.039376373035e-05
256 1.676861950406e-07
512 2.64179422782e-09
1024 4.136224696083e-11
2048 6.490363801959e-13
};
\addplot [semithick, white!75.29411764705883!black, dashed]
table {%
24 0.146871726088636
3072 3.33947641794647e-14
};

\addplot [semithick, color5, mark=triangle, mark size=2, mark options={solid}]
table {%
32 0.004760002300366
64 3.679855857256e-05
128 2.492407118679e-07
256 1.120407302525e-09
512 4.530169642192e-12
1024 1.7962194232e-14
2048 4.218847493576e-15
};
\addplot [semithick, white!75.29411764705883!black, dashed]
table {%
24 0.065868641359906
3072 9.14111027981807e-19
};
\addplot [semithick, color6, mark=triangle*, mark size=2, mark options={solid}]
table {%
32 0.0008446445453053
64 5.889094175973e-06
128 1.095800687342e-08
256 1.526516413275e-11
512 1.665334536938e-14
1024 3.996802888651e-15
2048 4.218847493576e-15
};
\addplot [semithick, white!75.29411764705883!black, dashed]
table {%
24 0.0749607394799559
3072 6.34942160900688e-23
};
\addplot [semithick, black, mark=diamond*, mark size=2, mark options={solid}]
table {%
%32 0.3804530581775
%64 0.005031706251324
%128 8.585746954115e-06
%256 4.114618201712e-09
%512 8.643086246707e-14
%1024 3.996802888651e-15
%2048 4.218847493576e-15
32 2.891997286092e-08 
64 1.579737793511e-12 
128 3.219646771413e-15 
256 3.552713678801e-15 
512 4.440892098501e-15 
1024 4.107825191113e-15 
2048 4.329869796038e-15 
};
\end{axis}

\end{tikzpicture}
%\subcaption{All to all communication}
%\end{minipage}%
%\hspace{0.04\textwidth}%
\caption{\label{fig:conv:404414}Convergence of FLUPS  with semi-unbounded BCs: %
\protect\ThinLineCircle{pyBlue}{pyBlue} \texttt{CHAT2} $\O{h^2}$, %
\protect\ThinLineCross{pyOrange} \texttt{LGF2} $\O{h^2}$,%
\protect\ThinLineCircle{pyGreen}{none} \texttt{HEJ2} $\O{h^2}$, %
\protect\ThinLineSquare{pyRed}{none} \texttt{HEJ4} $\O{h^4}$, %
\protect\ThinLineSquare{pyPurple}{pyPurple} \texttt{HEJ6} $\O{h^6}$, %
\protect\ThinLineTrianUp{pyBrown}{none} \texttt{HEJ8} $\O{h^8}$, %
\protect\ThinLineTrianUp{pyRose}{pyRose} \texttt{HEJ10} $\O{h^{10}}$, %
\protect\ThinLineDiamondD{black}{black} \texttt{HEJ0} spectral-like.}
\end{minipage}
\end{figure}

\subsection{Domain with two semi-infinite directions and one fully unbounded BC } %404414

Semi-unbounded BCs are finally used.  An even symmetry is imposed on the right side of the domain in the $X$ direction, and an odd symmetry is imposed on the left side in the $Z$ direction.
The BCs in the $Y$ direction are kept fully unbounded.
The reference solution is 
\be
\begin{split}
\phi_{ref}(x,y,z) = &\left[ \exp { 10 \left(1 - \frac{1} { \left(1 - \left(\frac{2x-1.4}{L}\right)^2\right)} \right)  } + \exp { 10 \left(1 - \frac{1} { \left(1 - \left(\frac{2x-2.6}{L}\right)^2\right)} \right)  } \right] \\%
		& \qquad \exp { 10 \left(1 - \frac{1} { \left(1 - \left(\frac{2y}{L}-1\right)^2\right)} \right) } %
		    \left[ \exp { 10 \left(1 - \frac{1} { \left(1 - \left(\frac{2z-0.6}{L}\right)^2\right)} \right)  } - \exp { 10 \left(1 - \frac{1} { \left(1 - \left(\frac{2z+0.6}{L}\right)^2\right)} \right)  } \right].
%\phi_{ref}(x,y,z) = &\left[ \exp { 10 \left(1 - \frac{1} { \left(1 - \left(\frac{2x}{L}-1.4\right)^2\right)} \right)  } + \exp { 10 \left(1 - \frac{1} { \left(1 - \left(\frac{2x}{L}-2.6\right)^2\right)} \right)  } \right] \\%
%		& \qquad \exp { 10 \left(1 - \frac{1} { \left(1 - \left(\frac{2y}{L}-1\right)^2\right)} \right) } %
%		    \left[ \exp { 10 \left(1 - \frac{1} { \left(1 - \left(\frac{2z}{L}-0.6\right)^2\right)} \right)  } - \exp { 10 \left(1 - \frac{1} { \left(1 - \left(\frac{2z}{L}+0.6\right)^2\right)} \right)  } \right].
\end{split}
\ee

The Green's functions are those of \app{sect:app:1dirspe}, and the obtained convergence results are shown in \fig{fig:conv:404414} and matches the expected convergence behaviors.

%!TEX root = mixPoissonSolver_main_siam.tex
%!TEX encoding = UTF-8 Unicode

\section{Scalability}
\label{sec:scalability}
This section exposes the performance of the \textit{non-blocking} implementation compared to the \textit{all-to-all} version and the influence of the partition of the resources between  MPI ranks and threads.
A speed comparison with the P3DFFT library is also presented.

The mean time required for the execution of the solve operation is tested on three different clusters, listed in \cref{tab:clusters}.
%
%\begin{enumerate}
%\item \textit{Zenobe}: a tier-1 cluster\footnote{More specifications available at \url{http://www.top500.org/system/178439} (last visited November 2019).}, located in Belgium and equipped with \textit{Intel Xeon E5-2697v2} and \textit{InfiniBand QDR}. FLUPS and its dependencies is compiled using Intel's compilers and Intel's MPI.
%%FLUPS and his dependencies were compiled with Intel's compilers v17.4.196 and IMPI v17.3.196.
%\item \textit{MareNostrum}: a tier-0 cluster\footnote{\textit{Marenostrum} in the top 500 supercomputers, more specifications available at \url{www.top500.org/system/179067} (last visited November 2019)} located in Spain and equipped with \textit{Xeon Platinum 8160} and \textit{Intel Omni-Path}. Compiled using Intel's compilers and Intel's MPI.
%\item \textit{Juwels}: a tier-0 cluster\footnote{\textit{Juwels} in the top 500 supercomputers, more specifications available at \url{www.top500.org/system/179424} (last visited October 2019)} located in Germany equipped with \textit{Xeon Platinum 8168} and \textit{Infiniband EDR}.Compiled with Intel's compilers and Parastation MPI\footnote{Parastation MPI is developped by \textit{ParTec Cluster Competence Center GmbH}, see \url{www.par-tec.com/products/parastation-mpi.html}}.
%\end{enumerate}
%
For the sake of concision, a selection of plots representative of the behaviors of FLUPS is here presented. 
% A more thorough performance check available in the online documentation of FLUPS.

\begin{table}[h]
%\begin{center}
\begin{minipage}{\textwidth}
\centering
    \begin{tabular}{lllllllll}
           & \textbf{Name}  & \textbf{Location} & \textbf{Type} & $N_{\textrm{\textbf{CPU}}}$ & \textbf{CPU} & \textbf{Interconnect} & \textbf{Compiler} & \textbf{MPI}  \\
           \hline
        1. & \textit{Zenobe}\footnote{\textit{Zenobe} is in the top 500 supercomputers, more specifications available at \url{www.top500.org/system/178439} (last visited November 2019).} & Belgium & tier-1 & $8\,208$ & \textit{Intel Xeon E5-2697v2} & \textit{InfiniBand QDR} & Intel & Intel \\
        2. & \textit{MareNostrum}\footnote{\textit{MareNostrum} is in the top 500 supercomputers, more specifications available at \url{www.top500.org/system/179067} (last visited November 2019).} & Spain & tier-0 & $165\,888$ & \textit{Xeon Platinum 8160} & \textit{Intel Omni-Path} & Intel & Intel \\
        3. & \textit{Juwels}\footnote{\textit{Juwels} is in the top 500 supercomputers, more specifications available at \url{www.top500.org/system/179424} (last visited October 2019).} & Germany & tier-0 & $122\,768$ & \textit{Xeon Platinum 8168} & \textit{Infiniband EDR} & Intel & Parastation\footnote{Parastation MPI is developped by \textit{ParTec Cluster Competence Center GmbH}, see \url{www.par-tec.com/products/parastation-mpi.html} (last visited October 2019).}
    \end{tabular}
\end{minipage}
\captionof{table}{\label{tab:clusters}List of supercomputers used for testing the scalability of FLUPS.}
%\end{center}   
\end{table}

% \todo{ORDER OF THE STUFF:
% JUWELS, STRONG, 1152, a2a vs nb, -> 18k, eta ref nb
% JUWELS, weak, $64^3$, a2a vs nb, ->18k, eta ref nb + t(hist)
% Zenobe, weak, $128^3$, a2a, thread 1 vs 4, eta ref weak
% JUWELS vs Mare, weak, $128^3$, a2a, -> 9k, eta
% }

\subsection{Strong scaling}
Let $T\left( N_{c},M \right) $ be the time required to solve a problem of $M$ unknowns using $N_{c}$ cores.
Using this definition, the strong efficiency, defined by
\be
\eta_{strong} = \dfrac{N_{c}^{ref} \; T\left( N_{c}^{ref} , M^{ref} \right)}{N_{c} \; T\left( N_{c} , M^{ref} \right)}
\eec
describes the ability of the software to decrease the computational time while increasing the number of resources for a given problem size $M_{ref}$.
In this study, the reference size is chosen as $1152^3$ cells for a fully unbounded problem, hence leading to a final size of $2304^3$ due the domain doubling algorithm.
%This size was selected taking into account the maximum memory available on each cluster.
%
For each data point, the presented timing is obtained as the mean of $20$ consecutive solve operations, in order to reduce timing uncertainties and to isolate any external influence of a possible network congestion. 

\begin{figure}[!t]
\begin{minipage}{0.48\textwidth}
\centering
%ask to set the 
\pgfplotsset{xtick=data}
\pgfplotsset{xticklabel={
    \pgfkeys{/pgf/fpu=true}
    \pgfmathparse{int(10^\tick +1)}
    \pgfmathprintnumber[fixed]{\pgfmathresult}
},xticklabel style={rotate=-45,anchor=west}}
% This file was created by tikzplotlib v0.8.6.
\begin{tikzpicture}

\definecolor{color0}{rgb}{0.419607843137255,0.682352941176471,0.83921568627451}
\definecolor{color1}{rgb}{0.992156862745098,0.552941176470588,0.235294117647059}

\begin{axis}[
log basis x={10},
tick align=outside,
tick pos=left,
x grid style={white!69.01960784313725!black},
xlabel={\(\displaystyle N_{{core}}\)},
xmin=935.714760602383, xmax=90769.8153070941,
xmode=log,
xtick style={color=black},
y grid style={white!69.01960784313725!black},
ylabel={\(\displaystyle t_{strong}\) [s]},
ymajorgrids,
ymin=0, ymax=2.504893355,
ytick style={color=black}
]
\addplot [semithick, color0, mark=*, mark size=2, mark options={solid}]
table {%
1152 2.3918649
2304 1.29645175
4608 0.89154495
9216 0.49829825
18432 0.3211081
36864 0.205442
73728 0.1312958
};
\addplot [semithick, color1, mark=o, mark size=2, mark options={solid}]
table {%
1152 2.01762035
2304 1.5704817
4608 0.67917465
9216 0.37685075
18432 0.29865655
36864 0.18025375
73728 0.16773845
};
\end{axis}

\end{tikzpicture}
\subcaption{Mean execution time $[s]$ per solve}
\label{fig:strong:scaling:spd}
%\end{figure}
\end{minipage}%
\hspace{0.04\textwidth}%-------------
\begin{minipage}{0.48\textwidth}
%\begin{figure}
\centering
\pgfplotsset{xtick=data}
\pgfplotsset{xticklabel={
    \pgfkeys{/pgf/fpu=true}
    \pgfmathparse{int(10^\tick +1)}
    \pgfmathprintnumber[fixed]{\pgfmathresult}
},xticklabel style={rotate=-45,anchor=west}}
% This file was created by tikzplotlib v0.8.6.
\begin{tikzpicture}

\definecolor{color0}{rgb}{0.419607843137255,0.682352941176471,0.83921568627451}
\definecolor{color1}{rgb}{0.992156862745098,0.552941176470588,0.235294117647059}

\begin{axis}[
log basis x={10},
tick align=outside,
tick pos=left,
x grid style={white!69.01960784313725!black},
xlabel={\(\displaystyle N_{{core}}\)},
xmin=935.714760602383, xmax=90769.8153070941,
xmode=log,
xtick style={color=black},
y grid style={white!69.01960784313725!black},
ylabel={\(\displaystyle \eta_{strong}\)},
ymajorgrids,
ymin=0, ymax=1.04060283495861,
ytick style={color=black}
]
\addplot [semithick, color0, mark=*, mark size=2, mark options={solid}]
table {%
1152 1
2304 0.922465838007469
4608 0.670707881862827
9216 0.600008353430902
18432 0.465549004369557
36864 0.363829100792438
73728 0.284646493356985
};
\addplot [semithick, color1, mark=o, mark size=2, mark options={solid}]
table {%
1152 1
2304 0.642357166594173
4608 0.742673607591214
9216 0.669237207966284
18432 0.422228381982582
36864 0.349788206556036
73728 0.187943300827866
};
\end{axis}

\end{tikzpicture}
\subcaption{Efficiency}
\label{fig:strong:scaling:eta}
\end{minipage}
\caption{Strong scalability of the solve operation on \textit{Juwels}, for a fully unbounded problem of size $1152^3$ using \textit{all-to-all} (\protect\ThickLineCircle{pyBlue2}{pyBlue2}) and \textit{non-blocking} (\protect\ThickLineCircle{pyOrange2}{none}) communications.
}
\label{fig:strong:scaling}
\end{figure}
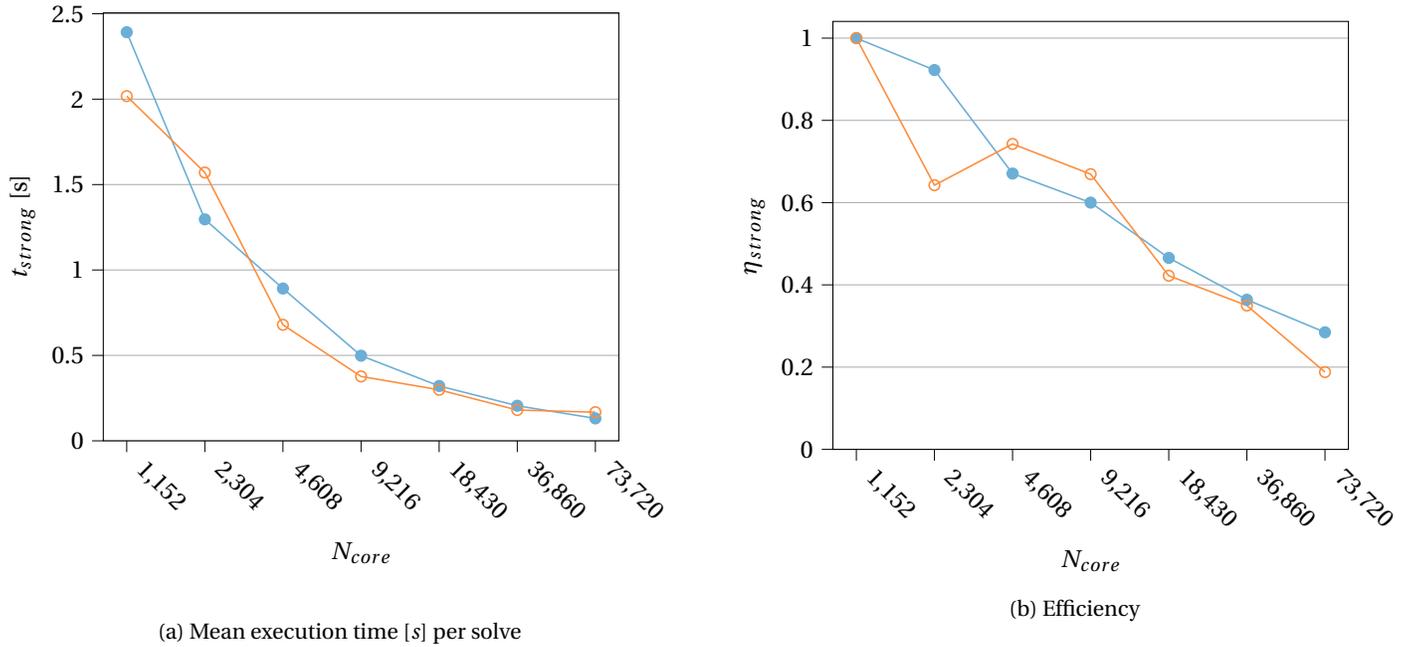

\fig{fig:strong:scaling:spd} shows the execution time. The \textit{non-blocking} implementation is \revtwo{usually} faster than the \textit{all-to-all} version, but the advantage \revtwo{significantly} shrinks when the number of cores increases.  \revtwo{For the largest number of cores, the \textit{all-to-all} version is even slightly faster.}
% This is due to the fact that, as expected, the copy operation scales better than the communication.
As expected, the gain of starting the memory copies while waiting for the other communications decreases \revtwo{when the number of unknowns per core decreases}, and becomes insignificant for a few thousands unknowns per rank. \revtwo{Correspondingly, the size of the MPI messages has become so small that the communication latency globally increases significantly.}
%\todo{Shall we show the timing histogram?}
%
As even more clearly observed in \fig{fig:strong:scaling:eta}, FLUPS exhibits the typical strong scalability of communication bounded applications.
In the present case, a loss of performance of $70\%$ is obtained when the number of cores is multiplied by 16 on a unbounded problem of size $1152^3$.
%To serve as a comparison, \todo{what P3DFFT WOULD DO}.

\subsection{Weak scalability}
The weak efficiency, defined as
\be
\eta_{weak} = \dfrac{T\left( N_{c}^{ref} , M^{ref} \right)}{ T\left( N_{c} , \dfrac{N_{c}}{N_{c}^{ref} } M^{ref}\right)}
\eec
represents the ability of the software to keep its initial efficiency while increasing the number of cores together with the size of the problem. \revtwo{We highlight that this scalability analysis is the most relevant for the interested user as, practically, the problem size usually scales with the number of resources.}

\begin{figure}[!t]
\centering
\begin{minipage}{0.48\textwidth}
\centering
\pgfplotsset{xtick=data}
\pgfplotsset{xticklabel={
    \pgfkeys{/pgf/fpu=true}
    \pgfmathparse{int(10^\tick +1)}
    \pgfmathprintnumber[fixed]{\pgfmathresult}
},xticklabel style={rotate=-45,anchor=west}}
% This file was created by tikzplotlib v0.8.6.
\begin{tikzpicture}

\definecolor{color0}{rgb}{0.419607843137255,0.682352941176471,0.83921568627451}
\definecolor{color1}{rgb}{0.992156862745098,0.552941176470588,0.235294117647059}

\begin{axis}[
log basis x={10},
tick align=outside,
tick pos=left,
x grid style={white!69.01960784313725!black},
xlabel={\(\displaystyle N_{{core}}\)},
xmin=935.714760602383, xmax=90769.8153070941,
xmode=log,
xtick style={color=black},
y grid style={white!69.01960784313725!black},
ylabel={\(\displaystyle t_{weak}\) [s]},
ymajorgrids,
ymin=0, ymax=1.8552191675,
ytick style={color=black}
]
\addplot [semithick, color0, mark=*, mark size=2, mark options={solid}]
table {%
1152 0.52782935
2304 0.61370865
4608 0.6703084
9216 0.7107974
18432 0.7743778
36864 0.9298072
73728 1.2441847
};
\addplot [semithick, color1, mark=o, mark size=2, mark options={solid}]
table {%
1152 0.44797415
2304 0.5725158
4608 0.6349171
9216 0.69923415
18432 0.74808675
36864 1.0566719
73728 1.7882075
};
\end{axis}

\end{tikzpicture}
\subcaption{Mean execution time}
\label{fig:weak:speedup:eta_ju}
%\end{figure}
\end{minipage}%
\hspace{0.04\textwidth}%
\begin{minipage}{0.48\textwidth}
\centering
\pgfplotsset{xtick=data}
\pgfplotsset{xticklabel={
    \pgfkeys{/pgf/fpu=true}
    \pgfmathparse{int(10^\tick +1)}
    \pgfmathprintnumber[fixed]{\pgfmathresult}
},xticklabel style={rotate=-45,anchor=west}}
% This file was created by tikzplotlib v0.8.6.
\begin{tikzpicture}

\definecolor{color0}{rgb}{0.419607843137255,0.682352941176471,0.83921568627451}
\definecolor{color1}{rgb}{0.992156862745098,0.552941176470588,0.235294117647059}

\begin{axis}[
log basis x={10},
tick align=outside,
tick pos=left,
x grid style={white!69.01960784313725!black},
xlabel={\(\displaystyle N_{{core}}\)},
xmin=935.714760602383, xmax=90769.8153070941,
xmode=log,
xtick style={color=black},
y grid style={white!69.01960784313725!black},
ylabel={\(\displaystyle \eta_{weak}\)},
ymajorgrids,
ymin=0, ymax=1.03747421230478,
ytick style={color=black}
]
\addplot [semithick, color0, mark=*, mark size=2, mark options={solid}]
table {%
1152 1
2304 0.860065032487321
4608 0.787442541373493
9216 0.742587620607504
18432 0.681617357832314
36864 0.567676126835757
73728 0.424237132959439
};
\addplot [semithick, color1, mark=o, mark size=2, mark options={solid}]
table {%
1152 1
2304 0.78246600355833
4608 0.705563214473197
9216 0.640664003610236
18432 0.598826472999288
36864 0.423948200004183
73728 0.250515753904399
};
\end{axis}

\end{tikzpicture}
\subcaption{Efficiency}
\label{fig:weak:scaling:eta_ju}
%\end{figure}
\end{minipage}%
\caption{\label{fig:weak:scaling}Weak scaling efficiency of the two implementations on \textit{Juwels}, for a fully unbounded problem and  $64^3$ unknowns per core using \textit{all-to-all} (\protect\ThickLineCircle{pyBlue2}{pyBlue2}) and \textit{non-blocking} (\protect\ThickLineCircle{pyOrange2}{none}) communications.}
\end{figure}
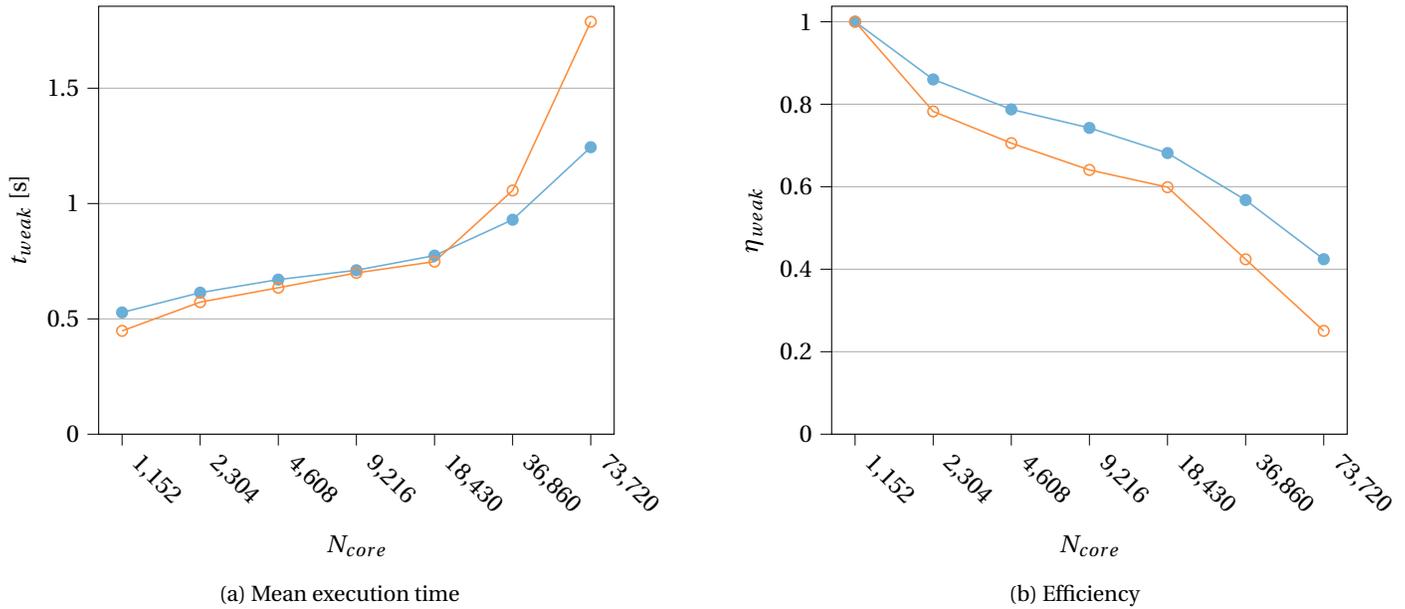

%------------------------------------------------------
% JUWELS BIG STUFF
%------------------------------------------------------
%\subsubsection{All-to-all vs Non-Blocking}
As expected from a communication bounded software, the weak scaling efficiency (\cref{fig:weak:scaling}) decreases with an increasing number of cores, since the number of communications through the network increases as $N_{\textrm{rank}}^{3/2}$, where $N_{\textrm{rank}}$ is here the number of MPI ranks.
%Thus, when $N_\textrm{core}$ increases,  the amount of transferred data increases and so does the congestion of the network. 
This results in the congestion of the network and explains the decreasing efficiency. 
Indeed, all the clusters used here have a network architecture in tree, with several nodes being connected to switches and switches being connected to racks (unlike \textit{BlueGene} and \textit{Cray} computers which feature a multidimensional grid network architecture).
Consequently, for calculations on more than $18430$ cores, the efficiency is even more reduced, probably due to bottlenecks in this tree network architecture.

%Therefore, as observed in \fref{fig:weak:bandwidth:zen},  the bandwidth scales as $N_{\mathrm{rank}}^{-3/2}$.

\begin{figure}[!t]
\begin{minipage}{0.48\textwidth}
\centering
\pgfplotsset{
legend image code/.code={\draw[#1, draw=none] (0cm,-0.1cm) rectangle (0.6cm,0.1cm);}
}
\pgfplotsset{xtick=data}
\pgfplotsset{xticklabel={
    \pgfkeys{/pgf/fpu=true}
    \pgfmathparse{int(10^\tick +1)}
    \pgfmathprintnumber[fixed]{\pgfmathresult}
},xticklabel style={rotate=-45,anchor=west}}
% This file was created by tikzplotlib v0.8.6.
\begin{tikzpicture}

\definecolor{color0}{rgb}{0.529411764705882,0.807843137254902,0.92156862745098}
\definecolor{color1}{rgb}{0,0.501960784313725,0.501960784313725}
\definecolor{color2}{rgb}{0.184313725490196,0.309803921568627,0.309803921568627}

\begin{axis}[
tick align=outside,
tick pos=left,
x grid style={white!69.01960784313725!black},
xlabel={\(\displaystyle N_{{core}}\)},
xmin=-0.795, xmax=6.795,
xtick style={color=black},
xtick={0,1,2,3,4,5,6},
xticklabels={1152,2304,4608,9216,18432,36864,73728},
y grid style={white!69.01960784313725!black},
ylabel={\(\displaystyle t\) [s]},
ymajorgrids,
ymin=0, ymax=2,
yminorgrids,
ytick style={color=black}
]
\draw[draw=white,fill=white!50.19607843137255!black] (axis cs:-0.45,0.42749635) rectangle (axis cs:0.45,0.52782935);
\draw[draw=white,fill=white!50.19607843137255!black] (axis cs:0.55,0.5136868) rectangle (axis cs:1.45,0.61370865);
\draw[draw=white,fill=white!50.19607843137255!black] (axis cs:1.55,0.5700823) rectangle (axis cs:2.45,0.6703084);
\draw[draw=white,fill=white!50.19607843137255!black] (axis cs:2.55,0.60809775) rectangle (axis cs:3.45,0.7107974);
\draw[draw=white,fill=white!50.19607843137255!black] (axis cs:3.55,0.66931165) rectangle (axis cs:4.45,0.7743778);
\draw[draw=white,fill=white!50.19607843137255!black] (axis cs:4.55,0.8263892) rectangle (axis cs:5.45,0.9298072);
\draw[draw=white,fill=white!50.19607843137255!black] (axis cs:5.55,1.1431502) rectangle (axis cs:6.45,1.2441847);
\draw[draw=white,fill=color0] (axis cs:-0.45,0.3670049) rectangle (axis cs:0.45,0.42749635);
\draw[draw=white,fill=color0] (axis cs:0.55,0.4560765) rectangle (axis cs:1.45,0.5136868);
\draw[draw=white,fill=color0] (axis cs:1.55,0.518741) rectangle (axis cs:2.45,0.5700823);
\draw[draw=white,fill=color0] (axis cs:2.55,0.5598887) rectangle (axis cs:3.45,0.60809775);
\draw[draw=white,fill=color0] (axis cs:3.55,0.62378655) rectangle (axis cs:4.45,0.66931165);
\draw[draw=white,fill=color0] (axis cs:4.55,0.78401305) rectangle (axis cs:5.45,0.8263892);
\draw[draw=white,fill=color0] (axis cs:5.55,1.10610535) rectangle (axis cs:6.45,1.1431502);
\draw[draw=white,fill=color1] (axis cs:-0.45,0.05960735) rectangle (axis cs:0.45,0.3670049);
\draw[draw=white,fill=color1] (axis cs:0.55,0.05636515) rectangle (axis cs:1.45,0.4560765);
\draw[draw=white,fill=color1] (axis cs:1.55,0.05515) rectangle (axis cs:2.45,0.518741);
\draw[draw=white,fill=color1] (axis cs:2.55,0.05683085) rectangle (axis cs:3.45,0.5598887);
\draw[draw=white,fill=color1] (axis cs:3.55,0.05587685) rectangle (axis cs:4.45,0.62378655);
\draw[draw=white,fill=color1] (axis cs:4.55,0.0574559) rectangle (axis cs:5.45,0.78401305);
\draw[draw=white,fill=color1] (axis cs:5.55,0.05723315) rectangle (axis cs:6.45,1.10610535);
\draw[draw=white,fill=color2] (axis cs:-0.45,0) rectangle (axis cs:0.45,0.05960735);
\draw[draw=white,fill=color2] (axis cs:0.55,0) rectangle (axis cs:1.45,0.05636515);
\draw[draw=white,fill=color2] (axis cs:1.55,0) rectangle (axis cs:2.45,0.05515);
\draw[draw=white,fill=color2] (axis cs:2.55,0) rectangle (axis cs:3.45,0.05683085);
\draw[draw=white,fill=color2] (axis cs:3.55,0) rectangle (axis cs:4.45,0.05587685);
\draw[draw=white,fill=color2] (axis cs:4.55,0) rectangle (axis cs:5.45,0.0574559);
\draw[draw=white,fill=color2] (axis cs:5.55,0) rectangle (axis cs:6.45,0.05723315);
\end{axis}

\end{tikzpicture}
\subcaption{\textit{All-to-all} communications}
\label{fig:weak:timepersolve:a2a:shm1}
\end{minipage}%
\hspace{0.04\textwidth}%
\begin{minipage}{0.48\textwidth}
%\begin{figure}
\centering
\pgfplotsset{xtick=data}
\pgfplotsset{xticklabel={
    \pgfkeys{/pgf/fpu=true}
    \pgfmathparse{int(10^\tick +1)}
    \pgfmathprintnumber[fixed]{\pgfmathresult}
},xticklabel style={rotate=-45,anchor=west}}
% This file was created by tikzplotlib v0.8.6.
\begin{tikzpicture}

\definecolor{color0}{rgb}{0.529411764705882,0.807843137254902,0.92156862745098}
\definecolor{color1}{rgb}{0,0.501960784313725,0.501960784313725}
\definecolor{color2}{rgb}{0.184313725490196,0.309803921568627,0.309803921568627}

\begin{axis}[
tick align=outside,
tick pos=left,
x grid style={white!69.01960784313725!black},
xlabel={\(\displaystyle N_{{core}}\)},
xmin=-0.795, xmax=6.795,
xtick style={color=black},
xtick={0,1,2,3,4,5,6},
xticklabels={1152,2304,4608,9216,18432,36864,73728},
y grid style={white!69.01960784313725!black},
ylabel={\(\displaystyle t\) [s]},
ymajorgrids,
ymin=0, ymax=2,
yminorgrids,
ytick style={color=black}
]
\draw[draw=white,fill=white!50.19607843137255!black] (axis cs:-0.45,0.33777825) rectangle (axis cs:0.45,0.44797415);
\draw[draw=white,fill=white!50.19607843137255!black] (axis cs:0.55,0.46511215) rectangle (axis cs:1.45,0.5725158);
\draw[draw=white,fill=white!50.19607843137255!black] (axis cs:1.55,0.5193348) rectangle (axis cs:2.45,0.6349171);
\draw[draw=white,fill=white!50.19607843137255!black] (axis cs:2.55,0.5887873) rectangle (axis cs:3.45,0.69923415);
\draw[draw=white,fill=white!50.19607843137255!black] (axis cs:3.55,0.63326995) rectangle (axis cs:4.45,0.74808675);
\draw[draw=white,fill=white!50.19607843137255!black] (axis cs:4.55,0.94418485) rectangle (axis cs:5.45,1.0566719);
\draw[draw=white,fill=white!50.19607843137255!black] (axis cs:5.55,1.6680264) rectangle (axis cs:6.45,1.7882075);
\draw[draw=white,fill=color0] (axis cs:-0.45,0.28943475) rectangle (axis cs:0.45,0.33777825);
\draw[draw=white,fill=color0] (axis cs:0.55,0.41824705) rectangle (axis cs:1.45,0.46511215);
\draw[draw=white,fill=color0] (axis cs:1.55,0.47324745) rectangle (axis cs:2.45,0.5193348);
\draw[draw=white,fill=color0] (axis cs:2.55,0.5527622) rectangle (axis cs:3.45,0.5887873);
\draw[draw=white,fill=color0] (axis cs:3.55,0.600903) rectangle (axis cs:4.45,0.63326995);
\draw[draw=white,fill=color0] (axis cs:4.55,0.91554575) rectangle (axis cs:5.45,0.94418485);
\draw[draw=white,fill=color0] (axis cs:5.55,1.64118355) rectangle (axis cs:6.45,1.6680264);
\draw[draw=white,fill=color1] (axis cs:-0.45,0.062064) rectangle (axis cs:0.45,0.28943475);
\draw[draw=white,fill=color1] (axis cs:0.55,0.05809105) rectangle (axis cs:1.45,0.41824705);
\draw[draw=white,fill=color1] (axis cs:1.55,0.056513) rectangle (axis cs:2.45,0.47324745);
\draw[draw=white,fill=color1] (axis cs:2.55,0.0531488) rectangle (axis cs:3.45,0.5527622);
\draw[draw=white,fill=color1] (axis cs:3.55,0.04696065) rectangle (axis cs:4.45,0.600903);
\draw[draw=white,fill=color1] (axis cs:4.55,0.04388995) rectangle (axis cs:5.45,0.91554575);
\draw[draw=white,fill=color1] (axis cs:5.55,0.0443969) rectangle (axis cs:6.45,1.64118355);
\draw[draw=white,fill=color2] (axis cs:-0.45,0) rectangle (axis cs:0.45,0.062064);
\draw[draw=white,fill=color2] (axis cs:0.55,0) rectangle (axis cs:1.45,0.05809105);
\draw[draw=white,fill=color2] (axis cs:1.55,0) rectangle (axis cs:2.45,0.056513);
\draw[draw=white,fill=color2] (axis cs:2.55,0) rectangle (axis cs:3.45,0.0531488);
\draw[draw=white,fill=color2] (axis cs:3.55,0) rectangle (axis cs:4.45,0.04696065);
\draw[draw=white,fill=color2] (axis cs:4.55,0) rectangle (axis cs:5.45,0.04388995);
\draw[draw=white,fill=color2] (axis cs:5.55,0) rectangle (axis cs:6.45,0.0443969);
\end{axis}

\end{tikzpicture}
\subcaption{\textit{Non-blocking} communications}
\label{fig:weak:timepersolve:nb:shm1}
\end{minipage}
\caption{Time spent in each part of the `solve' operation using $64^3$ unknowns by core, on \textit{Juwels}. From bottom to top, the shaded areas correspond to the computation of the 1D Fourier transforms (\protect\mymyLine{pyBar1}{solid}{6}), the time spent in the communication waiting for the data to be transferred (\protect\mymyLine{pyBar2}{solid}{6}), the data transfer from the memory to the send buffers and from the receive buffers to the transposed memory (\protect\mymyLine{pyBar3}{solid}{6}) and miscellaneous operations (\protect\mymyLine{pyBar4}{solid}{6}). The latter includes, among others, the multiplication in Fourier space for the convolution and the reset of the buffers to zero.}
\label{fig:weak:timepersolve:shm1}
\end{figure}
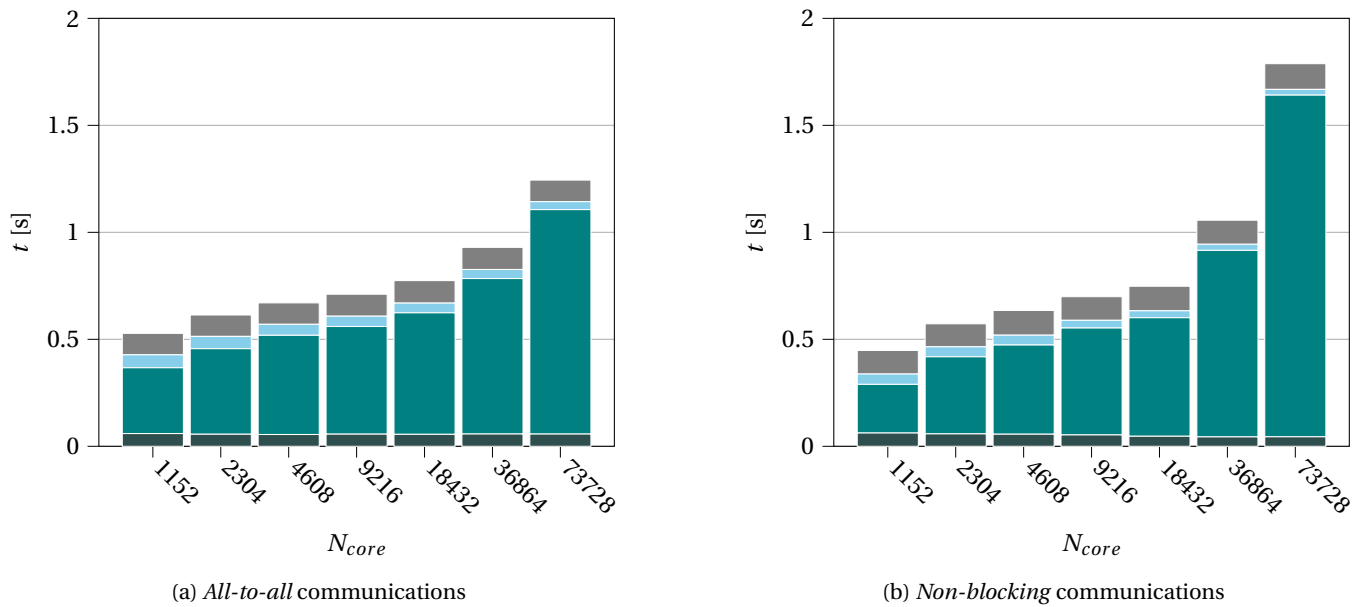

The computation time spent in each major step of the solver is shown in \fref{fig:weak:timepersolve:shm1}.
As expected, the communication is the most time-expensive part of the solver and drives the scalability, while the other operations almost scale perfectly.
Initially, the \textit{non-blocking} implementation is $10\%$ faster. For a large number of cores, the \textit{all-to-all} version becomes more advantageous as it is able to globally balance the communication over the tree network.

\begin{figure}[!t]
\begin{minipage}{0.48\textwidth}
\centering
%ask to set the 
\pgfplotsset{xtick=data}
\pgfplotsset{xticklabel={
    \pgfkeys{/pgf/fpu=true}
    \pgfmathparse{int(10^\tick +1)}
    \pgfmathprintnumber[fixed]{\pgfmathresult}
},xticklabel style={rotate=-45,anchor=west}}
% This file was created by tikzplotlib v0.8.6.
\begin{tikzpicture}

\definecolor{color0}{rgb}{0.419607843137255,0.682352941176471,0.83921568627451}
\definecolor{color1}{rgb}{0.992156862745098,0.552941176470588,0.235294117647059}

\begin{axis}[
log basis x={10},
tick align=outside,
tick pos=left,
x grid style={white!69.01960784313725!black},
xlabel={\(\displaystyle N_{{core}}\)},
xmin=1038.24053292768, xmax=10225.7922545773,
xmode=log,
xtick style={color=black},
y grid style={white!69.01960784313725!black},
ylabel={\(\displaystyle t_{weak}\) [s]},
ymajorgrids,
ymin=0, ymax=2.7409531325,
ytick style={color=black}
]
\addplot [semithick, color0, mark=*, mark size=2, mark options={solid}]
table {%
1152 1.9886995
2304 2.20778455
4608 2.50492325
9216 2.68157735
};
\addplot [semithick, color1, mark=o, mark size=2, mark options={solid}]
table {%
1152 1.4940617
2304 1.60123795
4608 1.9606152
9216 1.9770689
};
\end{axis}

\end{tikzpicture}
\subcaption{Mean execution time}
\label{fig:weak:speedup:mare}
\end{minipage}%
\hspace{0.04\textwidth}%
\begin{minipage}{0.48\textwidth}
\centering
\pgfplotsset{xtick=data}
\pgfplotsset{xticklabel={
    \pgfkeys{/pgf/fpu=true}
    \pgfmathparse{int(10^\tick +1)}
    \pgfmathprintnumber[fixed]{\pgfmathresult}
},xticklabel style={rotate=-45,anchor=west}}
% This file was created by tikzplotlib v0.8.6.
\begin{tikzpicture}

\definecolor{color0}{rgb}{0.419607843137255,0.682352941176471,0.83921568627451}
\definecolor{color1}{rgb}{0.992156862745098,0.552941176470588,0.235294117647059}

\begin{axis}[
log basis x={10},
tick align=outside,
tick pos=left,
x grid style={white!69.01960784313725!black},
xlabel={\(\displaystyle N_{{core}}\)},
xmin=1038.24053292768, xmax=10225.7922545773,
xmode=log,
xtick style={color=black},
y grid style={white!69.01960784313725!black},
ylabel={\(\displaystyle \eta_{weak}\)},
ymajorgrids,
ymin=0, ymax=1.01291922177818,
ytick style={color=black}
]
\addplot [semithick, color0, mark=*, mark size=2, mark options={solid}]
table {%
1152 1
2304 0.900767015513357
4608 0.793916340550554
9216 0.741615564436357
};
\addplot [semithick, color1, mark=o, mark size=2, mark options={solid}]
table {%
1152 1
2304 0.933066631352323
4608 0.762037191183665
9216 0.755695312388961
};
\end{axis}

\end{tikzpicture}
\subcaption{Efficiency}
\label{fig:weak:scaling:mare}
\end{minipage}
\caption{Weak scalability on \textit{MareNostrum} with \textit{all-to-all} (\protect\ThickLineCircle{pyBlue2}{pyBlue2}) and \textit{non-blocking} (\protect\ThickLineCircle{pyOrange2}{none}) communications, using $1$ thread per rank and $128^3$ unknowns per core .}
\label{fig:weak:mare}
\end{figure}
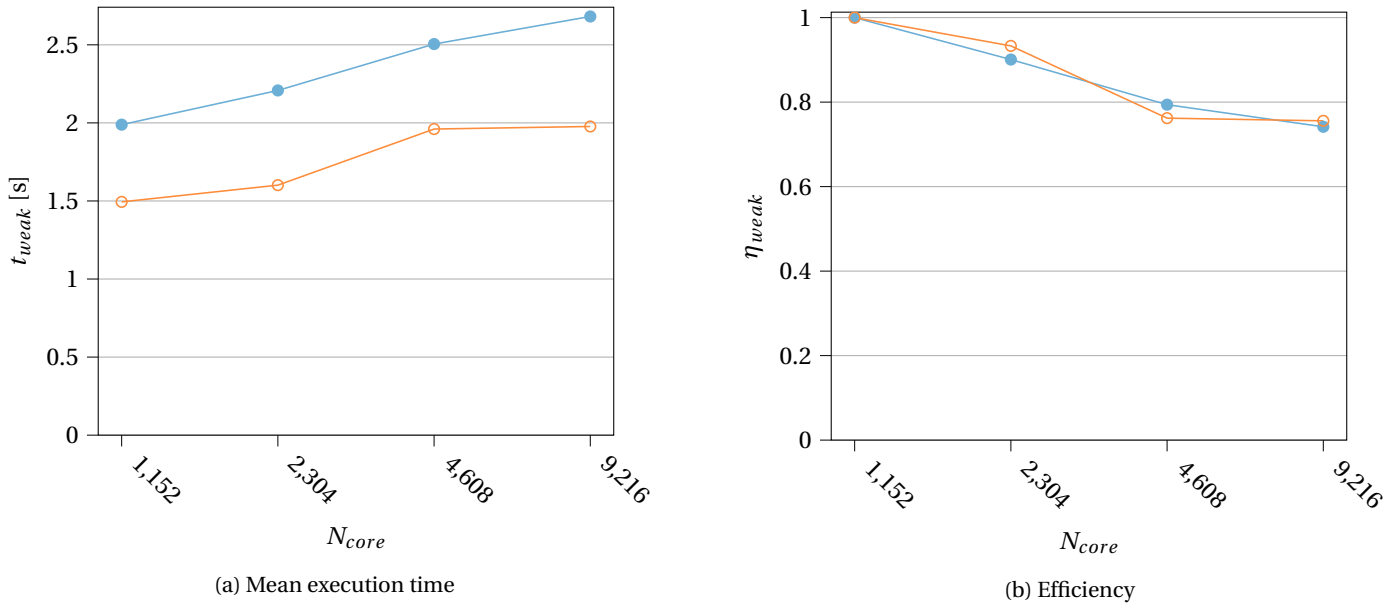

%------------------------------------------------------
% MARENOSTRUM, same even with a larger number of unkowns
%------------------------------------------------------
\medskip
A similar \textit{communication-bounded} behavior is also observed when increasing the computational load to $128^3$ unknowns per core (see \fig{fig:weak:mare}). However, for this size, the \textit{non-blocking} communication approach is \typo{clearly} faster by $\pm 20-25\%$ compared to the \textit{all-to-all} approach. This is due to the overlap between the communication and the copy of the data to the transposed memory. This technique turns out to be beneficial only when the number of data to copy is large enough to compensate the less efficient communication, compared to an \textit{all-to-all} implementation.

Incidentally, one could argue that other computational operations could possibly be executed during the communication. For instance, the computation of the FFTs could start as soon as the first complete row of data in the pencil has been received. However, since the FFTs represent less than $10\%$ of the computation time, this would likely not significantly improve the computational efficiency, while greatly increasing the implementation complexity. Even worse, it would overwhelm the network with a lot of very small messages, hence slowing down the solver.
% We note that, in practice, we observe the all to all communication pattern to be faster for a small number of unknowns per core, while the non-blocking communication is faster for a large number of unknowns per core.

%------------------------------------------------------
% ZENOBE, what about threads??
%------------------------------------------------------
%\medskip

\subsection{Hybrid MPI/OpenMP parallel performance}
\revtwo{%In the following paragraph, 
We compare the pure MPI and a hybrid MPI-OpenMP layout (without hyperthreading).
The former defines as many MPI ranks as the number of available cores,
%In the former, the available number of cores yields the number of MPI ranks 
$N_{\mathrm{rank}} = N_{\mathrm{core}}$.
In the latter, the cores that belong to the same socket can be grouped and execute a chosen number of threads
, such that $N_{\mathrm{rank}} = N_{\mathrm{core}}/N_{\mathrm{thread}}$. (We use the \texttt{MPI\_THREAD\_FUNNELED} mode in order to mitigate the communication cost.)}

\revtwo{
Let us first examine the expected influence of the addition of threads. On theoretical grounds, the simplified %analysis based on the 
PlogP model \cite{Alexandrov:1997} decomposes the communication time into a constant overhead, a linear dependency on the number of messages to send and a linear dependency on the message size. In the case of a pure MPI layout, one core sends and receives $P=\sqrt{N_{\mathrm{core}}}$ messages, each of size $N_{\mathrm{data}}$, hence leading to a communication time of
\be
T_{\textrm{comm}}^{MPI} = \alpha + P \; g + N_{\mathrm{data}} \; G \left(P\right)
\eec
where $g$ is the gap between messages (expressed in $[s]$) and $G$ is the gap per byte for long messages $[s/B]$. %which increases with $P$ due to the network contention.
% latency into a constant, $\alpha$ and a factor dependent on the number of messages to send/receive, $\beta$. In the case of a pure MPI layout, through the use of sub-communicators and considering that each core sends and receives $\sqrt{N_{\mathrm{core}}}$ messages, each of size $N_{\mathrm{data}}$, the communication cost for one core is approximated by 
% \be
% \alpha + \sqrt{N_{\mathrm{core}}} \left( \beta +  N_{\mathrm{data}}  \; b \right) = \alpha + \sqrt{N_{\mathrm{core}}} \; \beta + \sqrt{N_{\mathrm{core}}} N_{\mathrm{data}}  \; b
% \eec
% where $b$ stands for the theoretical bandwidth of the network.
%%%%%%%%%%% BEFORE
% Gathering $N_{\mathrm{thread}}$ cores together under the same MPI-rank will decrease the number of sent messages from $P$ to $P/N_{\mathrm{thread}}$, while increasing the size of each message. Doing so results in a communication time of 
% \be
% T_{\textrm{comm}}^{hybrid} = \alpha + P/N_{\mathrm{thread}} \; g + N_{\mathrm{thread}} N_{\mathrm{data}} \; G\left(P/N_{\mathrm{thread}} \right).
% \ee
% % This negative effects can however be balanced by an increase in the bandwidth of large message, \textit{i.e.} the rate at which data is transferred, also expressed as $1/G$.
% %On the other hand, 
% %by decreasing the number of sent messages, we expect an 
% However, the enlargement of the message length can result in an 
% increased bandwidth for large message (\textit{i.e.} the rate at which data is transferred), equivalent to $1/G$.
%%%%%% after
Grouping $N_{\mathrm{thread}}$ cores under the same MPI rank will decrease the number of sent messages from $P$ to $P/N_{\mathrm{thread}}$, which results in a communication time given by 
\be
T_{\textrm{comm}}^{hybrid} = \alpha + P/N_{\mathrm{thread}} \; g + N_{\mathrm{thread}} N_{\mathrm{data}} \; G\left(P/N_{\mathrm{thread}} \right).
\label{eq:Tcomm}
\ee
% This negative effects can however be balanced by an increase in the bandwidth of large message, \textit{i.e.} the rate at which data is transferred, also expressed as $1/G$.
%On the other hand, 
%by decreasing the number of sent messages, we expect an 
The increase in size of the messages has a beneficial effect through the second term of \eqqref{eq:Tcomm}, 
and a detrimental effect through the last term. The latter might however be compensated by the increased bandwidth  ({i.e.,} the rate at which data is transferred, equivalent to $1/G$) generally observed for larger messages. %This is what we need to verify here.
%%%%%%
%
Besides, the use of threads within a single MPI rank also adds a non-negligible overhead, due to the thread management in the copy and the shuffle operations. %Their cost can be modeled together as $\O{N_{\mathrm{data}} + \gamma \left(N_{\mathrm{thread}}-1\right)}$.
%
% All these  negative effects can however be balanced by an increase in the bandwidth of large message, \textit{i.e.} the rate at which data is transferred, also expressed as $1/G$.
%
Due to these antagonist effects, the use of threads may result in either improved or deteriorated performances, and dedicated tests must be performed\footnote{Even though the general concepts presented in this paragraph are true, the actual performance may vary depending on the targeted network architecture. In order to ease the determination by the user of the optimal thread setting for the machine he plans to use, we provide guidelines to run dedicated tests in the code documentation.}.
Notice that the computational cost associated with the FFT remains unchanged in a multithreaded environment, as the pencils are distributed among the threads and each thread treats them sequentially using single-thread FFTs.
%single-thread FFTs are distributed among the available threads, \textit{i.e.} the FFT is executed on a continuous pencil by one thread only.
}

\revtwo{
As observed in the weak scaling presented in \fref{fig:weak:zenobe}, only the small number of cores benefits from an acceleration when using 4 threads and \textit{all-to-all} communications, while it is significantly slower on large partitions. Indeed, in the present \textit{all-to-all} implementation, the communication and the copy are sequential. Hence, the copy overhead remains the same, but the communication cost drastically increases with the number of cores. Indeed, the communication time is driven by the cost associated with the message length, $N_{\mathrm{thread}} N_{\mathrm{data}} G$, which is here not compensated by a gain in bandwidth.  %due to the reduction of the number of messages.
% %
% This happens because there is no communication-copy overlap: the overhead associated with threads can only be balanced by a faster data transfer, which does not occur if the network is already saturated.
}

\revtwo{However, in the \textit{non-blocking} approach, the communication overlaps with the copy and with the shuffle operations. Even though the time-to-solution slightly increases for small $N_{{core}}$ with respect to the non-threaded implementation  (attributed to the overhead of threads), 
%it leaves more time for the communication of the next message to complete. By using this hybrid approach, 
the non-scalable communication times are thus partially hidden behind more scalable copy and shuffle operations, which explains the better scalability obtained with the hybrid MPI-OpenMP non-blocking approach.
Consequently, the threaded execution becomes faster than the non-threaded one at high $N_{\mathrm{core}}$.
 Furthermore, the \textit{non-blocking} version with threads is mostly faster than the \textit{all-to-all} version without threads.}

\begin{figure}[!t]
\begin{minipage}{0.48\textwidth}
\centering
%ask to set the 
\pgfplotsset{xtick=data}
\pgfplotsset{xticklabel={
    \pgfkeys{/pgf/fpu=true}
    \pgfmathparse{int(10^\tick +1)}
    \pgfmathprintnumber[fixed]{\pgfmathresult}
},xticklabel style={rotate=-45,anchor=west}}
% This file was created by tikzplotlib v0.8.5.
\begin{tikzpicture}

\definecolor{color0}{rgb}{0.419607843137255,0.682352941176471,0.83921568627451}
\definecolor{color1}{rgb}{0.619607843137255,0.792156862745098,0.882352941176471}
\definecolor{color2}{rgb}{0.992156862745098,0.552941176470588,0.235294117647059}
\definecolor{color3}{rgb}{0.992156862745098,0.682352941176471,0.419607843137255}

\begin{axis}[
log basis x={10},
tick align=outside,
tick pos=left,
x grid style={white!69.01960784313725!black},
xlabel={\(\displaystyle N_{{core}}\)},
xmin=245.686857635262, xmax=8102.41141573496,
xmode=log,
xtick style={color=black},
y grid style={white!69.01960784313725!black},
ylabel={\(\displaystyle t_{weak}\) [s]},
ymajorgrids,
ymin=0, ymax=10.5068334975,
ytick style={color=black}
]
\addplot [semithick, color0, mark=*, mark size=2, mark options={solid}]
table {%
288 4.17297265
576 5.1563639
1152 5.41308825
2304 6.20284505
4608 6.35686555
6912 6.2633176
};
\addplot [semithick, color1, mark=square*, mark size=2, mark options={solid}]
table {%
288 3.54859415
576 5.67114485
1152 6.70767325
2304 9.0931943
4608 8.41994735
6912 10.1745446
};
\addplot [semithick, color2, mark=o, mark size=2, mark options={solid}]
table {%
288 3.52876665
576 3.74565255
1152 4.6674952
2304 5.42634085
4608 6.18493265
6912 6.4921609
};
\addplot [semithick, color3, mark=square, mark size=2, mark options={solid}]
table {%
288 4.1896937
576 4.3861861
1152 5.09079305
2304 5.88007135
4608 6.04453715
6912 6.0734657
};
\end{axis}

\end{tikzpicture}
\subcaption{Mean execution time}
\label{fig:weak:speedup:zenobe}
%\end{figure}
\end{minipage}%
\hspace{0.04\textwidth}%
\begin{minipage}{0.48\textwidth}
\centering
%ask to set the 
\pgfplotsset{xtick=data}
\pgfplotsset{xticklabel={
    \pgfkeys{/pgf/fpu=true}
    \pgfmathparse{int(10^\tick +1)}
    \pgfmathprintnumber[fixed]{\pgfmathresult}
},xticklabel style={rotate=-45,anchor=west}}
% This file was created by tikzplotlib v0.8.5.
\begin{tikzpicture}

\definecolor{color0}{rgb}{0.419607843137255,0.682352941176471,0.83921568627451}
\definecolor{color1}{rgb}{0.619607843137255,0.792156862745098,0.882352941176471}
\definecolor{color2}{rgb}{0.992156862745098,0.552941176470588,0.235294117647059}
\definecolor{color3}{rgb}{0.992156862745098,0.682352941176471,0.419607843137255}

\begin{axis}[
log basis x={10},
tick align=outside,
tick pos=left,
x grid style={white!69.01960784313725!black},
xlabel={\(\displaystyle N_{{core}}\)},
xmin=245.686857635262, xmax=8102.41141573496,
xmode=log,
xtick style={color=black},
y grid style={white!69.01960784313725!black},
ylabel={\(\displaystyle \eta_{weak}\)},
ymajorgrids,
ymin=0, ymax=1.03256141041438,
ytick style={color=black}
]
\addplot [semithick, color0, mark=*, mark size=2, mark options={solid}]
table {%
288 1
576 0.809285909786158
1152 0.770904233826227
2304 0.672751393330388
4608 0.656451299335724
6912 0.66625595515067
};
\addplot [semithick, color1, mark=square*, mark size=2, mark options={solid}]
table {%
288 1
576 0.625728004461039
1152 0.529035034615021
2304 0.390247258875795
4608 0.421450871661329
6912 0.348771791712427
};
\addplot [semithick, color2, mark=o, mark size=2, mark options={solid}]
table {%
288 1
576 0.942096631466792
1152 0.756030054406912
2304 0.650303168847198
4608 0.570542453683792
6912 0.543542697778793
};
\addplot [semithick, color3, mark=square, mark size=2, mark options={solid}]
table {%
288 1
576 0.955201991999382
1152 0.822994307340779
2304 0.712524296155012
4608 0.693137223914655
6912 0.68983573909045
};
\end{axis}

\end{tikzpicture}
% \subcaption{bandwith on \textit{Zenobe}}
\subcaption{Efficiency}
\label{fig:weak:scaling:zenobe}
\end{minipage}
\caption{Weak scalability on \textit{Zenobe} with $1$ thread per rank (\textit{all-to-all}: \protect\ThickLineCircle{pyBlue2}{pyBlue2}, \textit{non-blocking}: \protect\ThickLineCircle{pyOrange2}{none})
and $4$ threads per rank (\textit{all-to-all}: \protect\ThickLineSquare{pyBlueLight}{pyBlueLight}, \textit{non-blocking}: \protect\ThickLineSquare{pyOrangeLight}{none}), using $128^3$ unknowns per core.}
\label{fig:weak:zenobe}
\end{figure}

%------------------------------------------------------
% COMPARISON OF JUWELS/MARE/ZENOBE
%------------------------------------------------------
\medskip
Finally, the comparison of the results obtained with different clusters reveals that the scalability is strongly dependent on the running environment, with a difference in efficiency larger than $10\%$ for large sizes, as seen in \fref{fig:weak:scaling:ju_ma}.
The driving factors are the different interconnects and the various MPI implementations.

%Similarly, we also define the speedup with respect to a reference number of cores, $N_{CPU}^{ref}$, for the strong and weak scaling respectively as
%\be
%s_{strong} = \dfrac{T\left( N_{CPU}^{ref} , M^{ref} \right)}{T\left( N_{CPU} , M^{ref} \right)}
%\eec
%and
%\be
%s_{weak} = \dfrac{T\left( N_{CPU}^{ref} , M^{ref} \right)}{T\left( N_{CPU} ,  \dfrac{N_{CPU}}{N_{CPU}^{ref} } M^{ref}\right)}
%\eed
% FIG\todo{XX} presents the scalability of the library when using hybrid distributed and shared memory parallelism. For a given number of CPUs, 

\subsection{Comparison with P3DFFT++}

A campaign of computations was dedicated to compare the performances of the P3DFFT library\footnote{The latest version of P3DFFT V3 available at the time of this study was used, with git SHA \textit{2150652831daab38b1028d16db7e4c7318df1133} from \url{github.com/sdsc/p3dfft.3}.} V3 (also known as P3DFFT++) with those of FLUPS (for the 3D transform only). To this end, a single executable is created, in which each library is used to perform the same fully periodic 3D FFT. As opposed to the previous section, the starting topology is here already set up with a pencil decomposition, such that the execution will only involve two topology switches and three computations of 1D FFT.
%which performs the same full periodic 3D FFT using each library.
%
Both libraries use the same binary from the compilation of FFTW. They also both use the \texttt{FFTW\_MEASURE} option when creating 1D FFT plans, such that the difference in timing %that are observed 
only originates in the implementation of the 3D FFT, specific to FLUPS and P3DFFT.

Similarly to the weak scaling study, a resolution of $128^3$ per core was selected, and the number of cores was increased (using 1 thread in FLUPS).
The results were obtained on \textit{Zenobe} and are shown in \fig{fig:flups_vs_p3d}.

\begin{figure}[!t]
\centering
\begin{minipage}[t]{0.48\textwidth}
\centering
\pgfplotsset{xtick=data}
\pgfplotsset{xticklabel={
    \pgfkeys{/pgf/fpu=true}
    \pgfmathparse{int(10^\tick +1)}
    \pgfmathprintnumber[fixed]{\pgfmathresult}
}}%,xticklabel style={rotate=-45,anchor=west}}
% This file was created by tikzplotlib v0.8.5.
\begin{tikzpicture}

\definecolor{color0}{rgb}{0.458823529411765,0.419607843137255,0.694117647058824}
\definecolor{color1}{rgb}{0.192156862745098,0.63921568627451,0.329411764705882}

\begin{axis}[
log basis x={10},
tick align=outside,
tick pos=left,
x grid style={white!69.01960784313725!black},
xlabel={\(\displaystyle N_{{core}}\)},
xmin=1038.24053292768, xmax=10225.7922545773,
xmode=log,
xtick style={color=black},
y grid style={white!69.01960784313725!black},
ylabel={\(\displaystyle \eta_{weak}\)},
ymajorgrids,
ymin=0, ymax=1.01843817315146,
ytick style={color=black}
]
\addplot [semithick, color0, mark=*, mark size=2, mark options={solid}]
table {%
1152 1
2304 0.840061039748426
4608 0.740551929452766
9216 0.631236536970851
};
\addplot [semithick, color1, mark=o, mark size=2, mark options={solid}]
table {%
1152 1
2304 0.900767015513357
4608 0.793916340550554
9216 0.741615564436357
};
\end{axis}

\end{tikzpicture}
\caption{Weak scaling efficiency of the \textit{all-to-all} implementation on \textit{Juwels} (\protect\ThickLineCircle{pyPurple}{pyPurple}) and \textit{MareNostrum} (\protect\ThickLineCircle{pyGreen}{pyGreen}), for a fully unbounded problem and $128^3$ unknowns per core.
}
\label{fig:weak:scaling:ju_ma}
\end{minipage}
\hfill
\begin{minipage}[t]{0.48\textwidth}
\centering
% \pgfplotsset{
% legend image code/.code={\draw[#1, draw=none] (0cm,-0.1cm) rectangle (0.6cm,0.1cm);}
% }
% \pgfplotsset{xticklabel style={rotate=-45,anchor=west}}
% This file was created by tikzplotlib v0.8.6.
\begin{tikzpicture}

\definecolor{color0}{rgb}{0.192156862745098,0.509803921568627,0.741176470588235}
\definecolor{color1}{rgb}{0.901960784313726,0.333333333333333,0.0509803921568627}

\begin{axis}[
tick align=outside,
tick pos=left,
x grid style={white!69.01960784313725!black},
xlabel={\(\displaystyle N_{{core}}\)},
xmin=-0.53935, xmax=4.53935,
xtick style={color=black},
xtick={0,1,2,3,4},
xticklabels={128,256,512,1024,2048},
y grid style={white!69.01960784313725!black},
ylabel={t [s]},
ymajorgrids,
ymin=0, ymax=0.5368021365,
ytick style={color=black}
]
\draw[draw=white,fill=color0] (axis cs:0.0115,0) rectangle (axis cs:0.3085,0.26731927);
\draw[draw=white,fill=color0] (axis cs:1.0115,0) rectangle (axis cs:1.3085,0.28066902);
\draw[draw=white,fill=color0] (axis cs:2.0115,0) rectangle (axis cs:2.3085,0.31037487);
\draw[draw=white,fill=color0] (axis cs:3.0115,0) rectangle (axis cs:3.3085,0.37758506);
\draw[draw=white,fill=color0] (axis cs:4.0115,0) rectangle (axis cs:4.3085,0.40536385);
\draw[draw=white,fill=color1] (axis cs:-0.3085,0) rectangle (axis cs:-0.0115,0.32796525);
\draw[draw=white,fill=color1] (axis cs:0.6915,0) rectangle (axis cs:0.9885,0.33568847);
\draw[draw=white,fill=color1] (axis cs:1.6915,0) rectangle (axis cs:1.9885,0.42827531);
\draw[draw=white,fill=color1] (axis cs:2.6915,0) rectangle (axis cs:2.9885,0.43130472);
\draw[draw=white,fill=color1] (axis cs:3.6915,0) rectangle (axis cs:3.9885,0.51124013);
\end{axis}

\end{tikzpicture}
%\subcaption{All to all communication}
%\label{fig:weak:timepersolve:a2a:shm1}
\caption{Wall time for a 3D FFT, using the \textit{all-to-all} version of FLUPS (\protect\mymyLine{pyBlue}{solid}{6}) and P3DFFT++ (\protect\mymyLine{pyRed}{solid}{6}), executed on \textit{Zenobe}, using $128^3$ unknowns per core.}
\label{fig:flups_vs_p3d}
\end{minipage}%
\end{figure}

On average, FLUPS is roughly 20\% faster than P3DFFT++, and exhibits a similar scalability. For this comparison, the \textit{all-to-all} version of FLUPS was used. An even larger difference is thus to be expected with the \textit{non-blocking} implementation.

%!TEX root = mixPoissonSolver_main_siam.tex
%!TEX encoding = UTF-8 Unicode

\section{Conclusions}%
\label{sec_conclu}

The Fourier-based Library of Unbounded Poisson Solvers (FLUPS) enables the fast resolution of the Poisson equation on 2D rectangle and 
3D parallelepiped-like domains with uniform resolution and with arbitrary boundary conditions on each boundary.
The library features a collection of solvers using a Green's function of various types, and with a verified convergence order from 2 to \revthree{spectral-like}, to solve the Poisson equation.
The convolution of the Green's function with the RHS of the Poisson equation is performed in Fourier space. To this end, data are transformed using fast Fourier transforms performed in 1D pencils and transposed efficiently.

The mathematical background for the resolution of the Poisson equation with the various boundary conditions was exposed, including the newly enabled semi-unbounded directions. In particular, the numerical treatment of data required for the RHS and for the Green's function was detailed in all cases.
The particularities of the numerical implementation were also addressed, especially related to the implementation of the 3D FFT. The comparison between two communication methods showed that point-to-point \textit{non-blocking} MPI send-receive instructions lead to a faster execution when they are interlaced with memory copy operations, compared to regular \textit{all-to-all} directives. 
Under some conditions on the communication latency and on the size of the MPI messages, the use of threads can even further improve the overall execution time.
The scalability of the library was measured on current top world-class supercomputers with tree-like network architectures. It features the behavior typical of communication-bounded software: when multiplying the number of cores by 16, the  
strong efficiency drops by $70\%$ (for a problem of size $1152^3$) and the weak efficiency by
$60\%$ (for $64^3$ unknowns per core).
Still, FLUPS has been demonstrated to be roughly $20\%$ faster than a reference third-party package for 3D FFTs.

The final implementation hence exhibits encouraging performances, propitious to massively parallel computational architectures, which proves that FLUPS is now ready for production use.

%\todo{a last word to kick asses, why we are better than others at Poisson}
% what was the problem

% what we did

% ongoing work
%\todo{missing features which are coming}

\section*{Acknowledgements}
The authors would like to acknowledge the insightful discussions with Prof. G. Winckelmans, Dr. M. Duponcheel and Dr. J. Lambrechts (\textit{UCLouvain}, Belgium).

D.-G. Caprace is funded by an Aspirant fellowship from the \textit{Fonds de la Recherche Scientifique de Belgique, F.R.S.-FNRS}.
Computational resources have been provided by the \textit{Consortium des \'Equipements de Calcul Intensif (C\'ECI)}, funded by the \textit{Fonds de la Recherche Scientifique de Belgique (F.R.S.-FNRS)} under Grant $n^{\circ} 2.5020.11$ and by the Walloon Region.
The present research also benefited from computational resources made available on the Tier-1 supercomputer of the \textit{Federation Wallonie-Bruxelles}, infrastructure funded by the Walloon Region under the grant agreement $n^{\circ} 1117545 $. 
Finally, we acknowledge PRACE for awarding us access to \emph{MareNostrum} at Barcelona Supercomputing Center (BSC), Spain; and \emph{JUWELS} at GCS@FZJ, Germany.

%\newpage
%\section*{References}
%\bibliographystyle{abbrvnat}
%\bibliographystyle{unsrtnat}
%\biboptions{sort&compress}
{\footnotesize
%\bibliographystyle{elsarticle-num-names-nourldoi}

% \IfFileExists{/Users/DeeGee/.bashrc}{% Adjust file name here to ~/user/.login or some other known file
% 	\bibliography{/Users/DeeGee/Documents/PhD/reference_all/AllPublications}
% }{
% 	\bibliography{/Users/tgillis/Dropbox/vortexbib/AllPublications.bib}
% }
\bibliography{extracted.bib}

}

\appendix
%!TEX root = mixPoissonSolver_main_SIAM.tex
%!TEX encoding = UTF-8 Unicode

\clearpage

\section{Analytical expressions of the Gaussian regularisation functions}
\label{sect:app:gaussian}

The regularization functions considered in this work are based on radially symmetric functions $\zeta_m(\rho)$,
\be
\zeta_{\varepsilon,m}(r) = \frac{1}{\varepsilon^d} \zeta_m \left(\frac{r}{\varepsilon} \right)
\eed
where $d$ is the dimensionality of the space.
For a given order $m$, $\zeta_\varepsilon^m$ satisfies the $m$ first moments, that is, $\forall \beta \in \{0,\dots,m-2\}$,
\be
0^\beta = 
    \begin{cases}
        2 \int_0^\infty r^\beta \zeta^m_\varepsilon(r) dr 			& \text{in 1D}, \\
       2\pi \int_0^\infty r^\beta \zeta^m_\varepsilon(r) r dr 		& \text{in 2D},\\
       4\pi \int_0^\infty r^\beta \zeta^m_\varepsilon(r) r^2 dr 		& \text{in 3D},
    \end{cases}
    \label{eq:zeta:moments}
\ee
Notice that the constraint is always satisfied for odd $\beta$ as the function $\zeta^m$ is radially symmetric.

\medskip

The Gaussian regularization functions were extensively \typo{used} in \cite{Hejlesen:2013} and by co-authors in other works, and their expressions are here reproduced for the sake of completeness.
In that particular case, 
\be
\zeta_m(\rho) = P_m(\rho) e^{\left( \frac{-\rho^2}{2} \right)}
\label{eq:zetam}
\eec
where $P_m(\rho) = a_1 + a_2 \rho^2 + \dots + a_{m/2} \rho^{m-2}$.
For a given $m$, the fulfilment of \eq{eq:zeta:moments} reduces to solving a system of $m/2$ equations for the coefficients $a_i$.
Alternatively, $\zeta_m$ can be obtained as the result of a recursive procedure \cite{Chatelain:2008e}.
For example, in 1D,
\be
P_2(\rho) = 1,\qquad
P_4(\rho) = \frac12\left(3-\rho^2\right),\qquad
P_6(\rho) = \frac18\left(15-10\rho^2+\rho^4 \right),\qquad
\revthree{P_8(\rho) = \frac{1}{48}\left(105-105\rho^2+21\rho^4 -\rho^6 \right),} %\qquad
\label{eq:zeta:P1D}
\ee
\begin{equation*}
\revthree{
P_{10}(\rho) = \frac{1}{384}\left(945-1\,260\rho^2+378\rho^4-36\rho^6 +\rho^{10} \right)
}
\end{equation*}

It is also convenient to express the Fourier transform of \eq{eq:zetam},
\be
\hat{\zeta}_m(s) = D_m(s) e^{-\frac{s^2}{2}}
\label{eq:zetamh}
\eec
where $D_m(s) = \sum_{n=0}^{m/2-1} \frac{(s^2/2)^n}{n!}$.

%=========================================================================================================
%=========================================================================================================

\section{Analytical expressions of the Green's functions in 3D}
\label{appendix:greenfunctions}

In this section, the expression of the Green\typo{'s} kernels $G$ used in FLUPS to solve the Poisson equation, $\nabla^2 \phi = f$, is reproduced. The solution is obtained as $\phi = G*f$.

%=========================================================================================================

\subsection{3 directions unbounded}
\label{sect:app:0dirspe}

\subsubsection{Singular}
The exact solution, and the cell-averaged value replacement, reproduced from \cite{Chatelain:2010}, are, respectively,
\be
G(\bx) = -  \frac{1}{4\pi |\bx|} \qquad \text{and} \qquad G(\boldsymbol{0}) = -  \frac{1}{2} \left(\frac{1}{2\pi}  \frac{3 h^3}{2}\right) ^{\frac{2}{3}}
\eed
\subsubsection{Regularized}
The regularized Green's functions, reproduced from \cite{Hejlesen:2013}, use $\rho = \frac{|\bx|}{\varepsilon} $, such that
%$\hat{\zeta}(s) = D_m(s) e^\frac{-s^2}{2} $ where $D_m(s) = \sum_{n=0}^{m/2-1} \frac{(s^2/2)^n}{n!}$. 
\be
G_m(|\bx|)=- \frac{1}{4\pi |\bx|} \left( Q_m(\rho) e^{-\frac{\rho^2}{2}} + \erf{ \frac{\rho}{\sqrt{2}} }  \right)
\eec
using the conventional definition of $\erf{z} = \frac{2}{\sqrt{\pi}} \int_0^z e^{-q^2} dq$, and where $Q_m$ is given by
\be
Q_2(\rho) = 0, \qquad %
Q_4(\rho) = \frac{1}{\sqrt{2\pi}} (\rho), \qquad %
Q_6(\rho) = \frac{1}{\sqrt{2\pi}} \left(\frac{7}{4} \rho - \frac14 \rho^3\right), \qquad %
\revthree{Q_8(\rho) = \frac{1}{\sqrt{2\pi}} \left(\frac{19}{8} \rho - \frac23 \rho^3 + \frac{1}{24} \rho^5\right) ,}
\ee
\begin{equation*}
\revthree{Q_{10}(\rho) = \frac{1}{\sqrt{2\pi}} \left(\frac{187}{64} \rho - \frac{233}{192} \rho^3 + \frac{29}{192} \rho^5 - \frac{1}{192} \rho^7 \right).}
\end{equation*}
The limit for $|\bx|\longrightarrow 0$ is given by
\be
G_2(\boldsymbol{0}) =-  \frac14 \frac{\sqrt{2}}{\pi^{3/2}\typo{\epsilon}}, \qquad G_4(\boldsymbol{0}) = - \frac38 \frac{\sqrt{2}}{\pi^{3/2}{\typo{\epsilon}}} , \qquad G_6(\boldsymbol{0}) = -\frac{15}{32} \frac{\sqrt{2}}{\pi^{3/2}\typo{\epsilon}} , \qquad
\revthree{
G_8(\boldsymbol{0}) = -\frac{35}{64} \frac{\sqrt{2}}{\pi^{3/2}\epsilon} , \qquad
G_{10}(\boldsymbol{0}) = -\frac{315}{512} \frac{\sqrt{2}}{\pi^{3/2}\epsilon}.}
\ee

\revthree{
A generalization to spectrally convergent kernels was proposed in \cite{Hejlesen:2019}, here noted as $G_{\infty}$, which leads to
\be
G_{\infty}(|\bx|) = \dfrac{1}{2 \pi^2 \sigma} \dfrac{Si(\rho)}{\rho},  \qquad G_{\infty}(\boldsymbol{0}) = - \dfrac{1}{2 \pi^2 \sigma}
\eec
where $\sigma = \frac{h}{\pi}$ is the spectral cutoff, and $Si(x) \triangleq \int_0^x \frac{\sin(t)}{t} d\!t$ is the sine integral function.
}

\subsubsection{LGF}
The lattice Green\typo{'s} function is defined as
\be
 G_h \left( \bx \right) = \dfrac{1}{\left( 2 \pi \right)^3} \iiint_{-\pi/h}^{\pi/h} \dfrac{ h^2 \;\; \exp{- \ic \; \bx \cdot \bk} 	}{4 \sin^2\left( k_x h /2 \right) + 4 \sin^2 \left( k_y h /2 \right) + 4 \sin^2 \left( k_z h /2 \right)} \quad d\bk 
\eec
where $h$ is the uniform grid size and \typo{$k_{x,y,z} \in \left[ \frac{-\pi}{h}, \frac{\pi}{h} \right] $}.\\
Here instead, it is evaluated using the \textit{indexed} form, $G \left( \bx \right) \triangleq \frac{1}{h}   \Theta \left( \bm \right)$,
where $\bm = \dfrac{\bx}{h} = \left\{ m_x \; ; \; m_y \; ; \; m_z \right\}$, $\bxi = \bk h$ and
\be
\Theta \left( \bm \right) = \dfrac{1}{\left( 2 \pi \right)^3} \iiint_{-\pi}^{\pi} \dfrac{ \exp{- \ic \bm \cdot \bxi } 	}{4 \sin^2\left( \xi_x /2 \right) + 4 \sin^2 \left( \xi_y /2 \right) + 4 \sin^2 \left( \xi_z /2 \right)}  \quad d\bxi = - \int_0^{\infty} e^{-6t} I_{m_x}\left( 2 t \right) I_{m_y}\left( 2 t \right) I_{m_z}\left( 2 t \right) dt \label{eq:appendix:lgf:integral}
\eed
The integral \eqqref{eq:appendix:lgf:integral} is precomputed using its second form\footnote{The integration was performed using the \texttt{quadgk} functions of Matlab along with the \texttt{besseli} function.} and stored  for every point satisfying $\max \left[ m_x , m_y, m_z \right] < 64$. For every other location, we use the following expansion such that the estimated error is lower that the machine precision \cite{Gillis:2018}, \textit{i.e.} with an estimated error of $64^{-9} \approx 10^{-17}$,
{\setstretch{2.0}
\begin{equation}
\begin{array}{lcl}
\Theta \left( \bm \right) = \dfrac{-1}{4 \pi \sabs{\bm}} &- \dfrac{1}{ 16 \pi \abs{\bm}^7} \Biggl[ & m_x^4 + m_y^4 + m_z^4 - 3 m_x^2 m_y^2 - 3 m_x^2 m_z^2 - 3 m_y^2 m_z^2 \Biggr]\\
& - \dfrac{1}{128 \pi \sabs{\bm}^{13}} \Biggl[ & \left( m_x^8 + m_y^8 + m_z^8 \right) - 244 \Bigl( m_x^6 \left( m_y^2 + m_z^2 \right) + m_y^6 \left( m_x^2 + m_z^2 \right) + m_z^6 \left( m_x^2 + m_y^2 \right) \Bigr) \\
& & - 228 m_x^2 m_y^2 m_z^2 \abs{\bm}^2 + 621 \left( m_x^4 m_y^4 + m_y^4 m_z^4 + m_x^4 m_z^4 \right) \Biggr] \\
& - \dfrac{1}{2048 \pi \sabs{\bm}^{19}} \Biggl[ & 2588 \left( m_x^{12} + m_y^{12} + m_z^{12} \right) \\
& & -65\,676 \left( m_x^{10} m_y^{2} + m_x^{2} m_y^{10} + m_x^{10} m_z^{2} + m_x^{2} m_z^{10} + m_y^{10} m_z^{2} + m_y^{2} m_z^{10} \right) \\
& &  +426\,144 \left( m_x^{8} m_y^{4} + m_x^{4} m_y^{8} + m_x^{8} m_z^{4} + m_x^{4} m_z^{8} + m_y^{8} m_z^{4} + m_y^{4} m_z^{8} \right) \\
& &  -712\,884 \left( m_x^{6} m_y^{6} + m_x^{6} m_z^{6} + m_y^{6} m_z^{6} \right) \\
& &  -62\,892 \left( m_x^{8} m_y^{2} m_z^2 + m_x^{2} m_y^{8} m_z^2+ m_x^{2} m_y^{2} m_z^8 \right) \\
& &  - 297\,876 \left( m_x^{6} m_y^{4} m_z^{2} + m_x^{6} m_y^{2} m_z^{4} + m_x^{4} m_y^{6} m_z^{2} + m_x^{2} m_y^{6} m_z^{4} + m_x^{2} m_y^{4} m_z^{6}+ m_x^{4} m_y^{2} m_z^{6} \right) \\
& & + 2\,507\,340 \left( m_x^4 m_y^4 m_z^4 \right) \Biggr] + \O{\dfrac{1}{\sabs{\bm}^9} } \; .
\end{array}
\label{eq_multipole_expansion_app}
\end{equation}}

%=========================================================================================================

\subsection{2 directions unbounded, 1 direction spectral}
\label{sect:app:1dirspe}

\subsubsection{Singular}
If $z$ is the spectral direction, the exact Green's function for the continuous problem is
\be
\tilde{G}(r,k_z) = 
    \begin{cases}
       \frac{1}{2\pi}  \log(r)  & \text{if $k_z=0$},\\
      -\frac{1}{2\pi} K_0(|k_z| r) & \text{otherwise},
    \end{cases}
\ee
where $r = \sqrt{x^2+y^2}$, and $K_0$ is the modified Bessel function of the second kind.
The singular value in $r = 0$ is replaced by the cell average value.
With the 2D equivalent radius $r_{eq} = \frac{h^2}{\sqrt{\pi}}$,
%r_eq2D = c_1osqrtpi * sqrt( hfact[ax0]*hfact[ax1]+hfact[ax1]*hfact[ax2]+hfact[ax2]*hfact[ax0] )

\be
\tilde{G}(0,k_z) = 
    \begin{cases}
       \frac{1}{8\pi}  \left(\pi - 6 + 2\log\left(\frac{\pi}{2} r_{eq}\right)\right)  & \text{if $k_z=0$},\\
       %.25 * c_1o2pi * (M_PI - 6.0 + 2. * log(.5 * M_PI * r_eq2D));
      -\frac{1 - k_z r_{eq} K_1\left(k_z r_{eq}\right)}{\pi\left( k_z r_{eq} \right)^2}  & \text{otherwise}.
      %return (1.0 - k * r_eq2D * besselk1(k * r_eq2D)) * c_1opi / ((k * r_eq2D) * (k * r_eq2D));
    \end{cases} 
\ee

\subsubsection{Regularized}

Defining $\rho =  \frac{\sqrt{x^2+y^2}}{\varepsilon}$ and $s_z =\varepsilon k_z $, the Green's function should be obtained as
\be
\tilde{G}_m(\rho,s_z) = - \frac{1}{2\pi} \int_0^\infty \frac{\hat{\zeta}_m \left( \sqrt{s^2 + s_z^2}\right)}{s^2+s_z^2} J_0 (s\rho) s \; ds
\label{eq:2dunb1ddct:spietz}
\eed
As noticed in \cite{Spietz:2018}, finding a closed form solution to \ref{eq:2dunb1ddct:spietz} is not trivial. The authors of the same reference suggest the use of an approximation under the form of a composite expression in the spectral space, which is here reproduced:
\be
\hat{G}_m(\bk) \approx %\simeq 
    \begin{cases}
       \hat{G}_m^{2D}(s)   & \text{if $s_z=0$},\\
       -\frac{\hat{\zeta}_m \left( \sqrt{s^2 + s_z^2}\right)}{|\bk|^2} & \text{otherwise},
    \end{cases} 
\ee
where $s = \varepsilon\sqrt{k_x^2+k_y^2}$. $\hat{G}^{m,2D}$ is obtained as the Fourier transform of the solution of a 2D unbounded regularized problem,
\be
 G_m^{2D}(\rho) = \frac{1}{2\pi} \left( \log(\varepsilon\rho) - R_m(\rho) e^{-\frac{\rho^2}{2}} + \frac12 E_1\left(\frac{\rho^2}{2}  \right)\right)
\eec
with the first exponential integral $E_1(z)=\int_1^\infty \frac{e^{-zq}}{q} dq$ , and
\be
R_2(\rho) = 0, \qquad %
R_4(\rho) = \frac12, \qquad %
R_6(\rho) = \frac34 - \frac18 \rho^2, \qquad %
\revthree{
R_8(\rho) = \frac{11}{12} - \frac{7}{24} \rho^2 + \frac{1}{48} \rho^4, \qquad %
R_{10}(\rho) = \frac{25}{24} - \frac{23}{48} \rho^2 + \frac{13}{192} \rho^4 - \frac{1}{384} \rho^6.}
\ee
\revthree{
The limit when $\rho \rightarrow 0$ is obtained as
\be
G_2^{2D}(0) = -\frac{1}{2\pi} \left(\frac{\gamma}{2} - \log(\sqrt{2}\varepsilon) \right), \qquad %
G_4^{2D}(0) = -\frac{1}{2\pi} \left(\frac{\gamma}{2} - \log(\sqrt{2}\varepsilon) + \frac12 \right), \qquad %
G_6^{2D}(0) = -\frac{1}{2\pi} \left(\frac{\gamma}{2} - \log(\sqrt{2}\varepsilon) + \frac34 \right), %
\ee
\begin{equation*}
G_8^{2D}(0) = -\frac{1}{2\pi} \left(\frac{\gamma}{2} - \log(\sqrt{2}\varepsilon) + \frac{11}{12}  \right), \qquad %
G_{10}^{2D}(0) = -\frac{1}{2\pi} \left(\frac{\gamma}{2} - \log(\sqrt{2}\varepsilon) + \frac{25}{24} \right).
\end{equation*}
with $\gamma = 0.5772156649$ being Euler's constant.
}

% \subsubsection{LGF} \todo{say we will not give it}

%=========================================================================================================

\subsection{1 direction unbounded, 2 directions spectral}
\label{sect:app:2dirspe}

\subsubsection{Singular}
If $x$ is the unbounded direction, the exact solution is
\be
\tilde{G}(x,\kappa) = 
    \begin{cases}
      \frac12 |x|  & \text{if $\kappa$ = 0},\\
	%        return -.5 * fabs(r);
      -\frac12 \frac{e^{-\kappa |x|}}{\kappa}  & \text{otherwise},
      	%    return .5 * exp(-k * r) / k;
    \end{cases}
\ee
where $\kappa = \sqrt{k_y^2 + k_z^2}$.

\subsubsection{Regularized}

Defining  $s=\varepsilon \sqrt{k_y^2+k_z^2}$ and $\rho=\frac{|x|}{\varepsilon}$, the regularized kernel is
\be
\tilde{G}_m(\rho,s) = -\frac{\varepsilon}{4s} \left( \left(1-\erf{\frac{s-\rho}{\sqrt{2}}}\right) e^{-s\rho} + \left(1-\erf{\frac{s+\rho}{\sqrt{2}}}\right) e^{s\rho} \right) - \frac{\varepsilon\sqrt{2}}{\sqrt{\pi}} \typo{U}_m(\rho,s) e^{-\frac{s^2+\rho^2}{2}}
\ee
and
\be
\typo{U}_2(\rho,s) = 0, \qquad %
\typo{U}_4(\rho,s) = \frac{1}{4}, \qquad %
\typo{U}_6(\rho,s) = \frac{5}{16} +
\frac{1}{16}(s^2-\rho^2), \qquad %
\revthree{
U_8(\rho,s) = \frac{11}{32} +
\frac{1}{12}s^2 - \frac{1}{8} \rho^2 - \frac{1}{48} s^2\rho^2 + \frac{1}{96}(s^4+\rho^4),} %
\ee
\begin{equation*}
\revthree{
U_{10}(\rho,s) = \frac{93}{256} +
\frac{73}{768}s^2-\frac{47}{256}\rho^2 - \frac{17}{384} s^2\rho^2 + \frac{11}{768} s^4 + \frac{23}{768} \rho^4 + \frac{1}{256} (s^2\rho^4 - s^4\rho^2) + \frac{1}{768}(s^6-\rho^6). %
}
\end{equation*}
When $s=0$, the solution of a 1D unbounded problem has to be used,
\be
\tilde{G}_m(\rho,0) = \frac{r}{2} \erf{\frac{\rho}{\sqrt{2}}} - \frac{\varepsilon}{\sqrt{2\pi}}   \typo{V}_m(\rho)
\eec
with
\be
\typo{V}_2(\rho) = \left( 1 - e^{-\frac{\rho^2}{2}} \right),\qquad %
\typo{V}_4(\rho) = \frac12 \left( 1 - e^{-\frac{\rho^2}{2}} \right) , \qquad %
\typo{V}_6(\rho) = \frac18 \left( 3 - \left( \rho^2 + 3 \right)e^{-\frac{\rho^2}{2}} \right), \qquad
\revthree{
V_8(\rho) = \frac{5}{16} -  \left(-\frac{1}{48} \rho^4 + \frac14 \rho^2 + \frac{5}{16} \right)e^{-\frac{\rho^2}{2}},
}
\ee
\be
\nonumber
\revthree{
V_{10}(\rho) = \frac{35}{128} -  \left(\frac{1}{384} \rho^6 -\frac{23}{384} \rho^4 + \frac{47}{128} \rho^2 + \frac{35}{128} \right)e^{-\frac{\rho^2}{2}}
}
\ee

% \subsubsection{LGF} \todo{say we will not give it}

%=========================================================================================================

\subsection{3 directions spectral}
\label{sect:app:3dirspe}

\subsubsection{Singular}

The well-known exact solution is 
\be
\hat{G}(\bk) = -\frac{1}{k^2} %\frac{1}{|\bk|^2}
\eec
with $k = \sqrt{k_x^2+k_y^2+k_z^2}$. By default, FLUPS arbitrarily sets $\hat{G}(\boldsymbol{0}) = 0$.

\subsubsection{Regularized}

Again with $k = \sqrt{k_x^2+k_y^2+k_z^2} $,  the regularized Green's function is given by 
\be
\hat{G}_m(\bk) = -\frac{\hat{\zeta}^m(\varepsilon k)}{k^2}
\eec
and again, $\hat{G}^m(\boldsymbol{0}) = 0$ by default. \revthree{The limit $m \rightarrow \infty$ corresponds exactly to the singular case.}

\subsubsection{Lattice Green's Function} 
The spectral LGF is computed as the spectral equivalent to the finite-difference Poisson equation, $\lapl_h \phi = f$,
\be
\hat{G}_h(\bk) = -\frac{h^2}{4 \sin^2(k_x \frac{h}{2}) + 4 \sin^2(k_y \frac{h}{2}) + 4 \sin^2(k_z \frac{h}{2})}
\eed

\end{document}